\documentclass[12pt]{article}
\usepackage{amsmath,scalerel}
\usepackage{amsfonts}
\usepackage{amssymb}
\usepackage{amsthm}
\usepackage{relsize}
\usepackage{dsfont}
\usepackage[latin1]{inputenc}
\usepackage[all]{xy}
\usepackage{graphicx}
\usepackage{latexsym}
\usepackage{color}
\usepackage{epsfig}
\usepackage{graphicx}
\usepackage{caption}
\usepackage{subcaption}
\usepackage{here}
\usepackage{float}
\usepackage{array}
\usepackage{tikz}
\usepackage{tikz-cd}
\usepackage{cite}
\usetikzlibrary{3d,arrows,calc,chains,matrix,positioning,scopes}

\def\l[{\left[}
\def\r]{\right]}
\newcommand{\eval}[1]{\left[\kern-0.15em\left[ #1 \right]\kern-0.15em\right]}

\def\ba{\begin{array}}
\def\ea{\end{array}}
\def\beq{\begin{equation}}
\def\eeq{\end{equation}}
\def\bea{\begin{eqnarray}}
\def\eea{\end{eqnarray}}

\DeclareMathOperator{\coker}{coker}
\DeclareMathOperator{\Ima}{Im}

\newcommand{\Hom}{\mathop{}\mathopen{}{\rm Hom}\!}

\newcommand{\egzz}{\, \stackrel{\mathbb{Z}}{{=}}\,}

\newtheorem{defi}{Definition}
\newtheorem{prop}{Property}
\newtheorem{coro}{Corollary}

\begin{document}
\pagestyle{empty}
\setcounter{page}{0}
\hspace{-1cm}

\begin{center}
{\Large {\bf 3D Topological Models and Heegaard Splitting II: Pontryagin duality and Observables}}%
\\ [1.5cm]

{\large F. Thuillier}

\end{center}

\vskip 0.7 truecm

{\it LAPTh, Université Savoie Mont Blanc, CNRS, B.P. 110, F-74941 Annecy Cedex, France}

\vspace{3cm}

\centerline{{\bf Abstract}}
In a previous article, a construction of the smooth Deligne-Beilinson cohomology groups $H^p_D(M)$ on a closed $3$-manifold $M$ represented by a Heegaard splitting $X_L \cup_f X_R$ was presented. Then, a determination of the partition functions of the $U(1)$ Chern-Simons and BF Quantum Field theories was deduced from this construction. In this second and concluding article we stay in the context of a Heegaard spitting of $M$ to define Deligne-Beilinson $1$-currents whose equivalent classes form the elements of $H^1_D(M)^\star$, the Pontryagin dual of $H^1_D(M)$. Finally, we use singular fields to first recover the partition functions of the $U(1)$ Chern-Simons and BF quantum field theories, and next to determine the link invariants defined by these theories. The difference between the use of smooth and singular fields is also discussed.

\vspace{2cm}


\newpage
\pagestyle{plain} \renewcommand{\thefootnote}{\arabic{footnote}}

\section{Introduction}

In a first article \cite{T19} we explained how to recover the smooth Deligne-Beilinson (DB) cohomology groups $H_D^p(M)$ by using a Heegaard splitting $X_L \cup_f X_R$ of an oriented closed smooth $3$-manifold $M$. Then, we identified a decomposition of DB $1$-cocycles of $X_L \cup_f X_R$ which turned out to be very efficient in the Chern-Simon (CS) and BF $U(1)$ topological theories. Doing so, we were able to check that the partition functions for these two theories coincide with those obtain in the so-called ``standard" approach as exposed in \cite{GT2013,GT2014,MT1,MT2}. Moreover, up to a reciprocity formula, the CS and BF partition functions are nothing but respectively the Reshetikhin-Turaev \cite{RT91} and Turaev-Viro \cite{TV} $U(1)$ invariants \cite{MOO92,GT2013,MT1}. It is remarkable that we can obtain partition functions by performing a functional integration on smooth fields (actually on $H_D^1(M)$) when it is well-known that in the only locally finite and translation-invariant Borel measure on the infinite dimensional Banach set of smooth fields is the trivial measure. In the $U(1)$ case, the use of smooth DB cohomology allows to identify some finite dimensional sub-integrals from the initial functional integral. Once the remaining infinite dimensional integral is factorized out and eliminated with the help of a normalization factor, the  desired partition functions emerge. This technique applies to the standard approach as well as to the one based on a Heegaard splitting. 

In this second and concluding article we study CS and BF observables, and compute their expectation values still in the context of a Heegaard splitting. The CS and BF observables are $U(1)$ holonomies and hence are given by exponentiating the line integral of gauge fields over closed curves, i.e. $1$-cycles. It turns out that integration along $1$-cycles goes to DB cohomology classes as an $\mathbb{R} / \mathbb{Z}$-valued linear functional, thus providing an example of the so-called Pontryagin duality. There are two ways to deal with the determination of the expectation values of these observables in the functional integration point of view. The first one is to stick with the smooth DB cohomology by replacing $1$-cycles with regularization whose support is as close to the $1$-cycles as we want. The holonomies are then given as integral of DB products. At the end of the computation, we take the limit which brings the regularized $1$-cycles back to their original singular position. The drawback of this method is that we must assume that functional integration commutes with the limit procedure. Another approach is to deal with $1$-cycle themselves and consequently with distributional fields. The drawback of this procedure is that we are faced with the usual problem of giving a meaning to products of distributions which appear all along the computation. Fortunately as we will see, this problem turns out to be quite easy to handle.

We can wonder about the differences between the standard and Heegaard splitting approaches to the $U(1)$ CS and BF theories. The most significant difference concern the degrees of freedom used in the functional integral. In the standard approach among the finite dimensional sub-integrals a sum over $T_1(M)$, the first torsion group of the $3$-manifold $M$ whose elements are integers, must be performed. In the Heegaard splitting approach the equivalent sum must be performed over $T_1(M)^\star$, the Pontryagin dual of $T_1(M)$ whose elements are angles. This difference is related to the following simple, although relatively unexpected, fact. When using a Heegaard splitting to describe DB classes, the exact sequence which naturally emerges from the construction is the one ending with the set of (generalized) curvatures and not the one ending with the second cohomology space of $M$, as in the standard approach. Beside a simple curiosity, let us point out that the Heegaard splitting approach could reveal interesting in the non-abelian case too where there does not seem to be an equivalent of DB cohomology. Hence, the Heegaard construction could help to identify relevant degrees of freedom in the functional integral, just as in the abelian case.

The notion of $3$-handlebody plays a central role in the construction proposed in this article and to our knowledge physicists are not very familiar with the geometry of smooth $3$-manifolds with boundary, the same remarks holding true for Heegaard splittings. This is why we thought it best to give enough details for this article to be as self contained as possible, even at the expense of its length.

\vspace{0.2cm}

\noindent The article is structured as follows:

$\bullet$ In Section 2 we exhibit some geometrical data provided by a $3$-handlebody $X$ and its boundary $\partial X \cong \Sigma$, concentrating on singular chains, forms and de Rham currents. The main purpose of the first subsection dedicated to the Riemann surface $\Sigma$ is Property \ref{pontdualcokerP}. The second subsection exposes some geometrical properties of $X$ relying on the fact that this manifold has a boundary. Both subsections end with few comments on DB cohomology.

$\bullet$ In section 3 we present a description of chains, forms and de Rham currents of a Heegaard splitting $X_L \cup_f X_R$. We end this section by briefly recalling the construction \cite{T19} of $H_D^1(X_L \cup_f X_R)$, the first smooth DB cohomology group of $X_L \cup_f X_R$. Beside fixing notations, this section tries to detail some useful points about Heegaard splitting and also provide an expression for the linking number of two $1$-cycles of $X_L \cup_f X_R$. 

$\bullet$ Section 4 is the core of the article and is dedicated to Pontryagin duality. As already mentioned, Pontryagin duality is closely related to integration of smooth DB classes along $1$-cycles of $X_L \cup_f X_R$. We first recall important facts concerning integration of $1$-forms and smooth DB $1$-cocycles along $1$-cycles of $X_L \cup_f X_R$. This yields the notion of DB $1$-cycles which themselves shed light on the construction of DB $1$-currents that we propose right after. The main difference with smooth DB 1-cocycles comes from the lack of gluing condition for DB $1$-currents. Then, and still in connection with DB $1$-cycles, an equivalence relation is exhibited which infers the notion of generalized DB classes the set of which is $H_D^1(X_L \cup_f X_R)^\star$, the Pontryagin dual of $H_D^1(X_L \cup_f X_R)$. Next, we show that $H_D^1(X_L \cup_f X_R)^\star$ is embedded into a canonical exact sequence which is the Pontryagin dual of an exact sequence into which $H_D^1(X_L \cup_f X_R)$ is embedded. This finally yields the central property of the article, i.e. Property \ref{decompsingularDBclasses}, which provides us with a parametrization of $H_D^1(X_L \cup_f X_R)^\star$. The rest of this section deals with an attempt to generalize the DB product to DB $1$-currents in order to extend the CS and BF (quantum) Lagrangians to singular DB classes. To that end, we first verify that smooth DB $1$-cocycles can be seen as regular DB $1$-currents, just like forms can be seen as regular de Rham currents. Then, a set of fundamental generalized DB products is exhibited and the zero regularization procedure is exposed in connection with the notion of linking number, self-linking and framing. As a collateral effect of the construction, an explicit realization of the so-called cycle map \cite{BGST05} is obtained.

$\bullet$ The last section is completely devoted to $U(1)$ CS and BF theories in the context of a Heegaard splitting. We first recall how the actions are related to DB product and its generalization to singular classes. Then, we discuss the functional measure which is supposed to allow us to compute partition functions and expectation values of holonomies. The so-called color periodicity and zero modes properties are exhibited. The next subsection more specifically deals with partition functions while the last one deals with expectation values. We eventually recover the expressions of the expectation values of $U(1)$ holonomies as given in \cite{GT2014,MT2}.

\section{Geometrical data of a 3-handlebody}

In this first section we recall some geometrical information which will prove useful in the other sections. As in this article we consider Heegaard splittings of oriented smooth closed $3$-manifolds, let $X$ be an oriented smooth $3$-handlebody whose boundary, denoted $\Sigma = \partial X$, is a Riemann surface of genus $g$. We assume that $\Sigma$ is endowed with the orientation induced by the one of $X$ and we denote by $i_\Sigma:\Sigma \rightarrow X$ the canonical inclusion. The first subsection deals with $\Sigma$ and the second with $X$. Each of these subsections start by considering chains as they are particularly easy to visualize in two and three dimensions with the help of simple drawings.

\subsection{Data for $\Sigma$}

Let $C_p(\Sigma)$ denote the set of $p$-chains on $\Sigma$, and let $\Omega^p(\Sigma)$ be the vector space of smooth $p$-forms on $\Sigma$. These spaces are respectively endowed with with the boundary operator $\partial$ and the de Rham exterior derivative $d$. The space of $p$-cycles of $\Sigma$ is denoted by $Z_p(\Sigma)$ while the space of closed $p$-forms is then denoted by $\Omega_0^p(\Sigma)$. The space of closed $p$-forms with integral periods by $\Omega^p_\mathbb{Z}(\Sigma)$. 

The de Rham $p$-currents of $\Sigma$ are the elements of the topological dual, $\Omega^p(\Sigma)^\diamond$, of $\Omega^p(\Sigma)$ \cite{dR55}. This space is endowed with the boundary operator $d^\dag$ defined by:
\bea
\label{bounddRcurrentsSigma}
\begin{aligned}
	\forall J \in \Omega^p(\Sigma)^\diamond \, , \, &\forall \alpha \in \Omega^p(\Sigma) \\
	\left( d^\dag J \right) \eval{\alpha} = & J \eval{d \alpha} \, ,
\end{aligned}
\eea
which gives rise to $\left( \Omega^\bullet(\Sigma)^\diamond , d^\dag \right)$, the dual complex of the differential complex $\left( \Omega^\bullet (\Sigma) , d \right)$. We denote by  $\Omega^p(\Sigma)^\diamond_0$ the space of closed $p$-currents of $\Sigma$ and by $\Omega^p(\Sigma)^\diamond_\mathbb{Z}$ the space of closed $p$-currents which are $\mathbb{Z}$-valued on $\Omega^p_\mathbb{Z}(\Sigma)$. The complex $\left( \Omega^\bullet(\Sigma)^\diamond , d^\dag \right)$ yields homology groups $H_p^\diamond(\Sigma)$ which fulfill Poincar\'e duality according to:
\bea
H_p^\diamond(\Sigma) \cong H_{dR}^{2-p}(\Sigma) \, ,
\eea
the de Rham cohomology groups $H_{dR}^q(\Sigma)$ being those of the complex $\left( \Omega^\bullet (\Sigma) , d \right)$.

To any $\omega \in \Omega^{2-p}(\Sigma)$ we associate the de Rham $p$-current $J_\omega$ defined by:
\bea
\label{productcurrentforms}
\kern-75pt \forall \alpha \in \Omega^{p}(\Sigma), \quad  J_\omega \eval{\alpha} = \oint_\Sigma \omega \wedge \alpha \, .
\eea 
This yields a canonical injection $J : \Omega^{2-p}(\Sigma) \to \Omega^p(\Sigma)^\diamond$. The elements of $\Ima J$ are commonly referred to as regular de Rham $p$-currents of $\Sigma$. Integration by parts implies that:
\bea
\kern-70pt \forall \omega \in \Omega^{2-p}(\Sigma) , \quad d^\dag J_\omega = (-)^{p+1} J_{d \omega} \, .
\eea
In particular, for $ \omega \in \Omega^{2-p}(\Sigma)$ and $\alpha \in \Omega^{p}(\Sigma)$, the $2$-form $\omega \wedge \alpha$ yields the regular 2-current $J_{\omega \wedge \alpha}$. Moreover, we can introduce the ``evaluation'':
\bea
\kern-20pt \left( J_\omega \wedge J_\alpha \right) \eval{1} = J_{\omega \wedge \alpha} \eval{1} = \oint_\Sigma \omega \wedge \alpha
\eea
Under certain conditions \cite{dR55}, it is possible to give a meaning to the evaluation $\left( j_1 \wedge j_2 \right) \eval{1}$ even if $j_1$ and $j_2$ are singular, i.e. not regular. For instance, when the singular support of $j_1$ (resp. $j_2$) does not intersect the singular support of $d^\dag j_2$ (resp. $d^\dag j_1$), then regularizations of $j_1$ and $j_2$ allow to define properly $\left( j_1 \wedge j_2 \right) \eval{1}$. In particular, $\left( j_1 \wedge d^\dag j_2 \right) \eval{1}$ is well defined when the singular supports of $d^\dag j_1$ and $d^\dag j_2$ do not intersect. Another example where $\left( j_1 \wedge j_2 \right) \eval{1}$ is well defined is when $j_2$ (or $j_1$) is regular since in this case its singular support is empty.

To any $c_\Sigma \in C_p(\Sigma)$ we associate the de Rham $p$-current $j_{c_\Sigma}$ defined by:
\bea
\label{chaintocurrent}
\kern-95pt \forall \alpha \in \Omega^p(\Sigma) , \quad j_{c_\Sigma} \eval{\alpha} = \int_{c_\Sigma} \! \! \alpha \, .
\eea
This generates a homomorphism $j : C_p(\Sigma) \to \Omega^p(\Sigma)^\diamond$. As in this article we mainly deal with integration properties along chains, we decide to identify $p$-chains which define the same de Rham $p$-current \cite{dR55}. Under this identification, the homomorphism $j$ becomes injective. Relation (\ref{chaintocurrent}) can be written as:
\bea
\kern-55pt \forall \alpha \in \Omega^p(\Sigma) , \quad j_{c_\Sigma} \eval{\alpha} = \left( j_{c_\Sigma} \wedge \alpha \right) \eval{1} = \left( j_{c_\Sigma} \wedge J_\alpha \right) \eval{1} \, .
\eea 
Indeed, $\left( j_{c_\Sigma} \wedge J_\alpha \right) \eval{1}$ as a meaning because the singular support of $J_\alpha$ is empty.  Moreover, for any $c \in C_p(\Sigma)$ we have:
\bea
j_{\partial c} = d^\dag j_c \, ,
\eea
This gives rise to an injection of the complex $\left( C_\bullet(\Sigma), \partial \right)$ into the complex $\left( \Omega^\bullet(\Sigma)^\diamond , d^\dag \right)$. Under this injection, $Z_p(\Sigma)$ is mapped into a subset of $\Omega^p(\Sigma)^\diamond_\mathbb{Z}$.

Let us consider a set of longitude and meridian $1$-cycles of $\Sigma$, $\lambda_a^\Sigma$ and $\mu_a^\Sigma$ ($a=1, \cdots , g$) respectively, the Kronecker indexes of which are:
\bea
\label{intersectionbasisSigma}
\lambda_a^\Sigma \odot \mu_b^\Sigma = \delta_{ab} \, ,
\eea
all other indexes being zero. The operator $\odot$ is the combination of the transverse intersection operator $\pitchfork$ with the degree operator $\sharp_0 : C_0(\Sigma) \rightarrow \mathbb{Z}$ which is a non-trivial extension of $\partial$ to the set $C_0(\Sigma)$ of $0$-chains of $\Sigma$ such that $\sharp_0 \circ \partial = 0$. Typically, $(\pm x)^{\sharp_0} := \pm 1$ for any point $x$ seen as a positively oriented $0$-chain of $\Sigma$. Note that $\pitchfork$ is only defined for chains which are in transverse position, hence its name. If $j_{\lambda_a^\Sigma}$ and $j_{\mu_b^\Sigma}$ are the de Rham $1$-currents of $\lambda_a^\Sigma$ and $\mu_b^\Sigma$ respectively, we have:
\bea
\label{interbasisSigmaCurrent}
\left( j_{\lambda_a^\Sigma} \wedge j_{\mu_b^\Sigma} \right) \eval{1} = \lambda_a^\Sigma \odot \mu_b^\Sigma =  \delta_{ab} \, ,
\eea
As an example, let us assume that $\Sigma = S^1 \times S^1$ endowed with angular coordinates $(\varphi_1 , \varphi_2)$, the orientation being specified by the volume form $d \varphi_1 \wedge d \varphi_2$. To get $\lambda^\Sigma \odot \mu^\Sigma = 1$ we can chose $\lambda^\Sigma$ and $\mu^\Sigma$ such that $j_{\lambda^\Sigma} = \delta(\varphi_1) d \varphi_1$ and $j_{\mu^\Sigma} = \delta(\varphi_2) d \varphi_2$ since then we have $\left( j_\lambda \wedge j_\mu \right) \eval{1} = \left( \delta(\varphi_1,\varphi_2) d \varphi_1 \wedge d \varphi_2 \right) \eval{1} = 1$ as expected.

Relations (\ref{intersectionbasisSigma}) induce an inner product in $\mathbb{Z}^g$ according to:
\bea
\label{innerproductZg}
\left\langle \vec{m} , \vec{n} \right\rangle = \left( \sum_{a=1}^g m^a \lambda_a^\Sigma \right) \odot \left( \sum_{b=1}^g n^b \mu_b^\Sigma \right) = \sum_{a,b=1}^g m^a \delta_{ab} n^b \, ,
\eea
Let us stress out that in this definition the longitude $1$-cycles appear to the left and the meridian $1$-cycles to the right of the operator $\odot$. Hence, this inner product is nothing but the Euclidean inner product on the free $\mathbb{Z}$-module $\mathbb{Z}^g$ which is the restriction to $\mathbb{Z}^g$ of the Euclidean inner product in $\mathbb{R}^g$. This last inner product will also be denoted $\left\langle \kern5pt , \kern5pt \right\rangle$.

The longitude and meridian $1$-cycles  generate the first homology group of $\Sigma$, and we have:
\bea
\left\{ \begin{aligned} 
& H_0(\Sigma) \cong \mathbb{Z} \\
& H_1(\Sigma) \cong \mathbb{Z}^{2g} \\
& H_2(\Sigma) \cong \mathbb{Z}
\end{aligned} \right. \, .
\eea

In anticipation of the Heegaard construction which will be considered in the next section, let $\Sigma_L$ and $\Sigma_R$ be two copies of the oriented closed surface $\Sigma$ and let $f:\Sigma_L \rightarrow \Sigma_R$ be an orientation \textit{reversing} diffeomorphism whose action on longitude and meridian $1$-cycles is given by:
\begin{equation}
\label{images}
\left\{ \begin{array}{l}
f(\lambda_a^{\Sigma_L}) = \sum\limits_{b=1}^g {{r_{ab}}} \, \lambda_b^{\Sigma_R} + \sum\limits_{b=1}^g {{s_{ab}}} \, \mu _b^{\Sigma_R} + \partial \varphi_a^{\Sigma_R} \\
f(\mu _a^{\Sigma_L}) = \sum\limits_{b=1}^g {{p_{ab}}} \, \lambda_b^{\Sigma_R} + \sum\limits_{b=1}^g {{q_{ab}}} \, \mu_b^{\Sigma_R} + \partial \psi_a^{\Sigma_R}
\end{array} \right. \, .
\end{equation}
where $\varphi_a^{\Sigma_R}$ and $\psi_a^{\Sigma_R}$ are 2-chains of $\Sigma_R$ with $a = 1 , \cdots , g$. This diffeomorphism induces an isomorphism $\hat{f}:H_1(\Sigma_L) \rightarrow H_1(\Sigma_R)$ which on its turn induces an automorphism of $\mathbb{Z}^{2g}$ as  $H_1(\Sigma) \cong \mathbb{Z}^{2g}$. The matrix of this automorphism is of the form:
\begin{equation}
\label{matrixMf}
\mathcal{M}_{\hat{f}} =
\left( {\begin{array}{*{20}{c}}
R & P\\
S & Q
\end{array}} \right) \, ,
\end{equation}
$P$, $Q$, $R$ and $S$ being integer $(g \times g)$ matrices defined by the integers $p_{ab}$, $q_{ab}$, $r_{ab}$ and $s_{ab}$, respectively. Since $f$ reverses orientation, it fulfills:
\bea
f(\lambda_a^{\Sigma_L}) \odot_{\Sigma_R} f(\mu_b^{\Sigma_L}) = - \, (\lambda_a^{\Sigma_L} \odot_{\Sigma_L} \mu_b^{\Sigma_L}) \, ,
\eea
or in terms of de Rham $1$-currents:
\bea
\left( f(j_{\lambda_a^{\Sigma_L}}) \wedge f(j_{\mu_b^{\Sigma_L}}) \right) [1] = - \, \left( j_{\lambda_a^{\Sigma_L}} \wedge j_{\mu_b^{\Sigma_L}} \right) [1] \, .
\eea
These relations imply that $det \mathcal{M}_{\hat{f}} = -1$ and that the matrix $\mathcal{M}_{\hat{f}}^{-1} = \mathcal{M}_{\hat{f}^{-1}}$ is:
\bea
\label{inversmatrixMf}
\left( {\begin{array}{*{20}{c}}
-Q^\dag & P^\dag \\
S^\dag & -R^\dag
\end{array}} \right) \, ,
\eea
which yields the following set of important relations \cite{CFH18,T19}:
\bea
\label{mainpropPQRS}
\begin{aligned}
Q^\dag P = P^\dag Q \; \; \; , \; \; \; P S^\dag - R Q^\dag = 1 \; \; \; , \; \; \; S^\dag R = R^\dag S \, , \\
R P^\dag = P R^\dag \; \; \; , \; \; \; P^\dag S - Q^\dag R = 1 \; \; \; , \; \; \; S Q^\dag = Q S^\dag \, ,  
\end{aligned} \, ,
\eea
the matrices $P^\dag$, $Q^\dag$, $R^\dag$ and $S^\dag$ being the transpose of respectively $P$, $Q$, $R$ and $S$, with respect to inner product (\ref{innerproductZg}). 

Each of the above matrices can be seen as the matrix of an endomorphism of $\mathbb{Z}^g$. The endomorphisms associated with $P$ and $P^\dag$ are embedded in the following exact sequences:
\bea
\label{Pexactsequence}
\begin{aligned}
	0 \rightarrow \ker P \xrightarrow{\subset} \mathbb{Z}^g & \xrightarrow{\kern2pt P \kern2pt} \mathbb{Z}^g \xrightarrow{[^{\kern5pt}]} \coker P \xrightarrow{} 0 \\
	0 \rightarrow \ker P^\dag \xrightarrow{\subset} \mathbb{Z}^g & \xrightarrow{P^\dag} \mathbb{Z}^g \xrightarrow{[^{\kern5pt}]_\dag} \coker P^\dag \xrightarrow{} 0
\end{aligned} \, ,
\eea
where $\coker P = \mathbb{Z}^g/\Ima P$ and $\coker P^\dag = \mathbb{Z}^g/\Ima P^\dag$. We then have the following property which stems from relations (\ref{mainpropPQRS}):

\begin{prop}\label{trivclassandQ}\leavevmode
\bea
\vec{n} \in \Ima P \Leftrightarrow Q^\dag \vec{n} \in \Ima P^\dag \, ,
\eea

\end{prop}
\noindent and an obvious corollary:
\begin{coro}\label{corototrivvlassandQ}\leavevmode
	\bea
	[\vec{n}] = 0 \Leftrightarrow Q^\dag [\vec{n}] = [Q^\dag \vec{n}]_\dag =  0 \, .
	\eea
	
\end{coro}

\noindent Indeed, if $\vec{n} = P \vec{m}$, then $Q^\dag \vec{n} = Q^\dag P \vec{m} = P^\dag Q \vec{m}$. Conversely, if $Q^\dag \vec{n} = P^\dag \vec{r}$ then $R Q^\dag \vec{n} = R P^\dag \vec{r} = P R^\dag \vec{r}$ on the one hand, and since $P S^\dag - R Q^\dag = 1$, $R Q^\dag \vec{n} = - \vec{n} + P S^\dag \vec{n}$ on the second hand. By combining these two relations we get $\vec{n} = P (S^\dag \vec{n} - R^\dag \vec{r})$ which ends up establishing the property. Since $Q^\dag (\vec{n} + P \vec{m}) = Q^\dag \vec{n} + Q^\dag P \vec{m} = Q^\dag \vec{n} + P^\dag Q \vec{m}$, we can extend $Q^\dag$ into a homomorphism $Q^\dag : \coker P \to \coker P^\dag$ by setting $Q^\dag [\vec{n}] = [Q^\dag \vec{n}]_\dag$, $[^{\kern+8pt}]$ and $[^{\kern+8pt}]_\dag$ denoting the passage to $\coker P$ and $\coker P^\dag$, respectively. This achieves the proof of the above property and of its corollary.

Since it is a finitely generated abelian group, $\coker P$ can be decomposed according to $\coker P = F_P \oplus T_P$, where $F_P$ is the free sector of $\coker P$ and is homomorphic to $\mathbb{Z}^b$ for some $b \in \mathbb{N}$, whereas $F_P$ is the torsion sector of $\coker P$ and is homomorphic to $\mathbb{Z}_{p_1} \oplus \cdots \oplus \mathbb{Z}_{p_N}$ with $p_i$ dividing $p_{i+1}$. If $\vec{n} \in \mathbb{Z}^g$ represents a torsion element of $\coker P$ then there exists $p \in \mathbb{N}^*$ such that $p \kern+1pt \vec{n} \in \text{Im } P$, and hence $\vec{n} = \hat{P} \vec{m}/p$ for some $\vec{m} \in \mathbb{Z}^g$, with $\hat{P}$ the canonical extension of $P$ to $\mathbb{R}^g$. Conversely, if $\vec{m}/p \in \mathbb{Q}^g$ is such that $\hat{P} \vec{m}/p \in \mathbb{Z}^g$, then $[\hat{P} \vec{m}/p] \in T_P$. As $P^\dag$ plays with respect to $f^{-1}$ the role that $P$ plays with respect to $f$, we straightforwardly conclude that $\coker P^\dag = F_{P^\dag} \oplus T_{P^\dag} \cong \mathbb{Z}^b \oplus \mathbb{Z}_{p_1} \oplus \cdots \oplus \mathbb{Z}_{p_N}$, and that $\vec{n}$ represents and element of $T_{P^\dag}$ if and only if there exist $(p,\vec{m}) \in \mathbb{N}^* \times \mathbb{Z}^g$ such that $\vec{n} = \hat{P}^\dag \vec{m}/p$. All these results can alternatively be obtained with the help of the Smith normal form of $P$, which then allows to easily check that $\ker P \cong \mathbb{Z}^b \cong \ker P^\dag$. For latter convenience, we introduce the following notations:
\bea
	H_P = \coker P = F_P \oplus T_P \quad , \quad H_{P^\dag} = \coker {P^\dag} = F_{P^\dag} \oplus T_{P^\dag} \, ,
\eea
with $F_P \cong F_{P^\dag} \cong \mathbb{Z}^b$ and $T_P \cong T_{P^\dag} \cong \mathbb{Z}_{p_1} \oplus \cdots \oplus \mathbb{Z}_{p_N}$.

Pontryagin duality playing a central role in this article, let us have a closer look at $(\mathbb{Z}^g)^\star = \Hom \, (\mathbb{Z}^g,\mathbb{R}/\mathbb{Z})$, the Pontryagin dual of $\mathbb{Z}^g$. It is isomorphic to $(\mathbb{R}/\mathbb{Z})^g$ since any $\vartheta \in (\mathbb{Z}^g)^\star$ can be generated with the help of some $\boldsymbol{\theta} \in (\mathbb{R}/\mathbb{Z})^g$ according to:
\bea
\kern-50pt \forall \vec{n} \in \mathbb{Z}^g, \quad  \vartheta (\vec{n}) = \left\langle \boldsymbol{\theta} , \vec{n} \right\rangle  = \overline{\left\langle \vec{\theta} , \vec{n} \right\rangle} \, ,
\eea
where $\vec{\theta} \in \mathbb{R}^g$ is a representative of $\boldsymbol{\theta}$ and $\overline{^{\kern10pt ^{}}}$ denotes the restriction to $\mathbb{R}/\mathbb{Z}$. The right-hand side of the second equality above obviously depends on $\boldsymbol{\theta}$ and not on the particular representative $\vec{\theta}$. From now on, $\boldsymbol{\theta}$ will refer to the homomorphism $\vartheta = \left\langle \boldsymbol{\theta} , \cdot \right\rangle$ it defines, thus yielding the identification of $(\mathbb{Z}^g)^\star$ with $(\mathbb{R}/\mathbb{Z})^g$.

The homomorphism $P$ induces a dual homomorphism $P^\star: (\mathbb{R}/\mathbb{Z})^g \to (\mathbb{R}/\mathbb{Z})^g$ defined as follows:
\bea
\kern-50pt \forall \vec{n} \in \mathbb{Z}^g , \quad ( P^\star \boldsymbol{\theta} ) (\vec{n}) = \boldsymbol{\theta} (P \vec{n}) \, .
\eea
In term of representatives the above relation takes the form:
\bea
\label{Pactiononrepres}
\kern-50pt \forall \vec{n} \in \mathbb{Z}^g , \quad   ( P^\star \boldsymbol{\theta} ) (\vec{n}) = \overline{\left\langle \hat{P}^\dag \vec{\theta} , \vec{n} \right\rangle} \, ,
\eea
which shows that on representatives $P^\star$ is nothing but $\hat{P}^\dag$. We similarly define $P^{\dag \star}$ by:
\bea
\label{Pdagactiononrepres}
\kern-50pt \forall \vec{n} \in \mathbb{Z}^g , \quad   ( P^{\dag \star} \boldsymbol{\theta} ) (\vec{n}) = \overline{\left\langle \hat{P} \vec{\theta} , \vec{n} \right\rangle} \, ,
\eea
which shows that on representatives $P^{\dag \star}$ is nothing but $\hat{P}$. Then, we have the following exact sequences associated with $P^\star$ and $P^{\dag \star}$:
\bea
\label{dualexactsequPPdag}
\begin{aligned}
0 \rightarrow \ker P^\star \rightarrow (\mathbb{R}/\mathbb{Z})^g & \xrightarrow{\kern2pt P^\star \kern2pt} (\mathbb{R}/\mathbb{Z})^g \xrightarrow{} \coker P^\star \rightarrow 0 \\
0 \rightarrow \ker (P^\dag)^\star \rightarrow (\mathbb{R}/\mathbb{Z})^g & \xrightarrow{P^{\dag \star}} (\mathbb{R}/\mathbb{Z})^g \xrightarrow{} \coker P^{\dag \star} \rightarrow 0
\end{aligned} \, .
\eea
The action of $\boldsymbol{\theta} \in (\mathbb{R}/\mathbb{Z})^g$ goes to $\coker P^\dag = \mathbb{Z}^g/\Ima P^\dag$ if and only if $\boldsymbol{\theta}$ is zero on $\Ima P^\dag$, which means that $\boldsymbol{\theta} (P^\dag \vec{m}) = (P^{\dag \star} \boldsymbol{\theta}) (\vec{m}) = 0 \in \mathbb{R}/\mathbb{Z}$ for any $\vec{m} \in \mathbb{Z}^g$, and hence if and only if $\boldsymbol{\theta} \in \ker P^{\dag \star}$. This shows that $\ker P^{\dag \star} \cong (\coker P^\dag)^\star$. Similarly, we have $\ker P^\star \cong (\coker P)^\star$ so that the exact sequences (\ref{dualexactsequPPdag}) turn out to be the Pontryagin dual of the exact sequences (\ref{Pexactsequence}).
Thus, we can write:
\bea
\ker P^{\dag \star} = H_{P^\dag}^\star = F_{P^\dag}^\star \oplus T_{P^\dag}^\star \cong (\coker P^\dag)^\star \cong (\mathbb{R}/\mathbb{Z})^b \oplus \mathbb{Z}_{p_1}^\star \oplus \cdots \oplus \mathbb{Z}_{p_N}^\star \, .
\eea
If $\boldsymbol{\theta}$ is an element of $H_{P^\dag}^\star$ then for any of its representative $\vec{\theta}$ there exists $\vec{L} \in \mathbb{Z}^g$ such that $P \vec{\theta} = \vec{L}$. For $\boldsymbol{\theta}$ fixed the set of such $\vec{L}$ is given as $\vec{L} + P \vec{l}$ with $\vec{l}$ spanning $\mathbb{Z}^g$. Let us pick one $\vec{L}_{\boldsymbol{\theta}}$ for each $\boldsymbol{\theta} \in \ker P^{\dag \star}$. Now, let $\boldsymbol{\theta}$ be an element of $F_{P^\dag}^\star$. On the first hand, if $\vec{\theta}$ is a representative of $\boldsymbol{\theta}$ then there exists $\vec{l} \in \mathbb{Z}^g$ such that $P \vec{\theta} = \vec{L}_{\boldsymbol{\theta}} + P \vec{l}$. On the second hand, since $P$ is an integer matrix and $\boldsymbol{\theta} \in F_{P^\dag}^\star \cong (\mathbb{R}/\mathbb{Z})^b$, this last equation is fulfilled if and only if $\hat{P} \vec{\theta} \egzz 0$, which means that $\boldsymbol{\theta}$ admits a representative $\vec{\theta}_f \in \ker \hat{P}$. This implies that $\boldsymbol{\theta} \in F_{P^\dag}^\star$ if and only if we can pick $\vec{L}_{\boldsymbol{\theta}} = \vec{0}$ or equivalently if and only if $[ \vec{L}_{\boldsymbol{\theta}} ] = 0$. Let us now assume that $\boldsymbol{\theta} \in T_{P^\dag}^\star$. Since $T_{P^\dag}^\star \cong \mathbb{Z}_{p_1}^\star \oplus \cdots \oplus \mathbb{Z}_{p_N}^\star$ there exists $(p,\vec{m}) \in \mathbb{Z} \times \mathbb{Z}^g$ such that $\vec{m}/p$ is a representative of $\boldsymbol{\theta}$, and then $\vec{L}_{\boldsymbol{\theta}}$ fulfills $p \vec{L}_{\boldsymbol{\theta}} = P (\vec{m} + p \vec{l})$ for some $\vec{l} \in \mathbb{Z}^g$. In other words $[\vec{L}_{\boldsymbol{\theta}}] \in T_P$.
 
Let us gather and complete the above results in the following property:

\begin{prop}\label{pontdualcokerP}\leavevmode

Let us introduce the notations:
\bea
H_P^\star = \ker P^\star = F_P^\star \oplus T_P^\star \quad , \quad H_{P^\dag}^\star = \ker P{^ \dag}^\star = F_{P^\dag}^\star \oplus T_{P^\dag}^\star \, .
\eea

1) We have the following set of abelian group isomorphisms:
\bea
\begin{aligned}
& H_P^\star \cong (\coker P)^\star \quad , \quad H_{P^\dag}^\star \cong (\coker P^\dag)^\star \\
& H_P^\star \cong (\mathbb{R}/\mathbb{Z})^b \oplus \mathbb{Z}_{p_1}^\star \oplus \cdots \oplus \mathbb{Z}_{p_N}^\star \cong H_{P^\dag}^\star
\end{aligned} \, .
\eea

2) Let $\vec{n}$ be an element of $\mathbb{Z}^g$. Then:
\bea
\left[\vec{n}\right] \in T_P \Leftrightarrow \left( \exists (p,\vec{m}) \in \mathbb{N}^* \times \mathbb{Z}^g , \; \; \vec{n} = \hat{P} \frac{\vec{m}}{p} \right) \, ,
\eea
and similarly for $T_{P^\dag}$.

3) Let $\boldsymbol{\theta} = \boldsymbol{\theta}_f + \boldsymbol{\theta}_\tau$ be an element of $H_{P^\dag}^\star = F_{P^\dag}^\star \oplus T_{P^\dag}^\star$. Then $\boldsymbol{\theta}$ has a representative of the form $\vec{\theta} = \vec{\theta}_f + \vec{\theta}_\tau$ such that:
\bea
\label{decompPontdualclass}
&\bullet& \hat{P} \vec{\theta}_f = \vec{0} \, , \\
&\bullet& \exists (p,\vec{m}) \in \mathbb{N}^* \times \mathbb{Z}^g , \quad \vec{\theta}_\tau = \frac{\vec{m}}{p} \, , \kern100pt 
\eea
$\hat{P} \vec{\theta} = \hat{P} \vec{\theta}_\tau = \hat{P} \vec{m}/p \in \mathbb{Z}^g$ then representing an element of $T_P$.

4) Let $\boldsymbol{\theta}$ and $\boldsymbol{\theta}'$ be two elements of $H_{P^\dag}^\star$ with representatives $\vec{\theta} = \vec{\theta}_f + \vec{\theta}_\tau$ and $\vec{\theta}' = \vec{\theta}'_f + \vec{\theta}'_\tau$. Then, the expression:
\bea
\label{linkingformHstar}
\Gamma (\vec{\theta} , \vec{\theta}') = \left\langle  \hat{P} \vec{\theta} , \hat{Q} \vec{\theta}' \right\rangle = \left\langle  \hat{P} \vec{\theta}_\tau , \hat{Q} \vec{\theta}'_\tau \right\rangle = \Gamma (\vec{\theta}_\tau , \vec{\theta}'_\tau)  \, ,
\eea
induces an $\mathbb{R}/\mathbb{Z}$-valued symmetric bilinear pairing on $T_{P^\dag}^\star$ which is referred to as the \textbf{linking form} of $T_{P^\dag}^\star$. The linking form of $T_P$ is then driven by setting:
\bea
\gamma ( \vec{n}_\tau , \vec{n}'_\tau) = \Gamma ( \vec{\theta}_\tau , \vec{\theta}'_\tau) \, ,
\eea
with $\vec{n}_\tau = P \vec{\theta}_\tau$ and $\vec{n}'_\tau = P \vec{\theta}'_\tau$.

\end{prop}

\noindent To prove the last point of the above property let us first note that $P^\dag Q \vec{\theta}' = Q^\dag P \vec{\theta}' = Q^\dag P \vec{\theta}'_\tau \egzz 0$ because $P^\dag Q = Q^\dag P$ and $P \vec{\theta}' = P \vec{\theta}'_\tau \in \mathbb{Z}^g$. This shows that $Q \vec{\theta}'$ represents an element of $H_P^\star$ and yields the following sequence of equalities:
\bea
\begin{aligned}
	\left\langle  \hat{P} \vec{\theta} , \hat{Q} \vec{\theta}' \right\rangle & = \left\langle  \hat{P} \vec{\theta}_\tau , \hat{Q} \vec{\theta}' \right\rangle = \left\langle  \vec{\theta}_\tau , \hat{P}^\dag \hat{Q} \vec{\theta}' \right\rangle = \left\langle  \vec{\theta}_\tau , \hat{Q}^\dag \hat{P} \vec{\theta}' \right\rangle = \left\langle  \vec{\theta}_\tau , \hat{Q}^\dag \hat{P} \vec{\theta}'_\tau \right\rangle \\ 
	& = \left\langle  \hat{P} \vec{\theta}_\tau , \hat{Q} \vec{\theta}'_\tau \right\rangle \, .
\end{aligned} \, .
\eea
Finally, symmetry and bilinearity being obvious, the relation:
\bea
\left\langle P (\vec{\theta}_\tau + \vec{l}) , Q (\vec{\theta}'_\tau  + \vec{l}') \right\rangle \egzz \left\langle P \vec{\theta}_\tau  , Q \vec{\theta}'_\tau \right\rangle \, ,
\eea
shows that $\Gamma$ induces an $\mathbb{R}/\mathbb{Z}$-valued symmetric bilinear pairing on $T_{P^\dag}^\star$, as claimed. 

A DB $1$-cocycle of $\Sigma$ is a way to represent a $U(1)$ principal bundle with $U(1)$-connection on $\Sigma$ with the help of local quantities like $1$-forms, functions and integers. By identified these principal bundles with connections under the action of $U(1)$ isomorphisms, we obtain an equivalence relation and hence classes of DB $1$-cocycles of $\Sigma$, the set of which is denoted $H_D^1(\Sigma)$. This is a $\mathbb{Z}$-module which stands in the following exact sequence:
\bea
\label{HD1Sigmaexactsequence}
0 \rightarrow \frac{\Omega^1(\Sigma)}{\Omega_\mathbb{Z}^1(\Sigma)} \rightarrow H_D^1(\Sigma) \rightarrow H^2(\Sigma) \rightarrow 0 \, ,
\eea
The Pontryagin dual of $H_D^1(\Sigma)$, $H_D^1(\Sigma)^\star := \Hom\,(H_D^1(\Sigma), \mathbb{R}/\mathbb{Z})$, is obtained by taking the Pontryagin dual of the above sequence, which yields:
\bea
\label{sequenceSigmaPontDual}
0 \rightarrow H^2(\Sigma)^\star \cong H^0(\Sigma,\mathbb{R}/\mathbb{Z}) \rightarrow H^1_D(\Sigma)^\star \xrightarrow{\delta_{-1}^\star} \left(\frac{\Omega^1(\Sigma)}{\Omega^1_\mathbb{Z}(\Sigma)}\right)^\star \cong \Omega^1(\Sigma)^\diamond_\mathbb{Z} \rightarrow 0 \, .
\eea
The construction giving rise to $H_D^1(\Sigma)$ can be generalized thus giving rise to a family of $\mathbb{Z}$-modules $H_D^p(\Sigma)$. As $H_D^0(\Sigma)$ is involved in the construction of $H_D^1(X_L \cup X_R)$, let recall that this particular space stands in the exact sequence:
\bea
\label{HDSigmaexactsequence}
0 \rightarrow H^1(\Sigma,\mathbb{R}/\mathbb{Z}) \rightarrow H_D^0(\Sigma) \xrightarrow{\bar{d}} \Omega^1_\mathbb{Z}(\Sigma) \rightarrow 0 \, ,
\eea
where $\bar{d}:H^1_D(\Sigma_R) \rightarrow \Omega^2_{\mathbb{Z}}(\Sigma_R)$ is a linear morphism that we refer to as the \textbf{curvature} operator of $H_D^0(\Sigma)$. The DB product which pairs $H_D^0(\Sigma)$ and $H_D^1(\Sigma)$ yields a conical injection $H^0_D(\Sigma) \xrightarrow{\mathfrak{J}} H^1_D(\Sigma)^\star$. This is consistent with the fact that $\Omega^1_\mathbb{Z}(\Sigma)$ is the subset of regular elements of $\Omega^1(\Sigma)^\diamond_\mathbb{Z}$. We refer to \cite{HLZ,BGST05} for details concerning all the groups $H_D^p(\Sigma)$.

\subsection{Data for $X$}

Let $C_p(X)$ denotes the abelian group of singular $p$-chains of X. For each $p$ the embedding $i_\Sigma: \Sigma \rightarrow \partial X$ induces a chain map $i_\Sigma: C_p(\Sigma) \rightarrow C_p(X)$ thanks to which me can identify $C_p(\Sigma)$ as a subgroup of $C_p(X)$. The quotient $C_p(X,\partial X) := C_p(X) / i_\Sigma(C_p(\Sigma))$ is an abelian group usually referred to as the group of relative singular $p$-chains of $(X,\partial X)$. The subgroup $i_\Sigma(C_p(\Sigma))$ can be seen as the free abelian group generated by standard $p$-simplices whose image are in $\partial X$ \cite{D72}. This induces a decomposition rule of any $p$-chain $c$ of $X$ according to:
\bea
\label{canondecompXchain}
c = c^\circ + c^\Sigma \, ,
\eea
where $c^\Sigma \in i_\Sigma(C_p(\Sigma))$ and $c^\circ = c - c^\Sigma$ corresponds to a unique element of $C_p(X,\partial X)$. The above decomposition is equivalent to the direct sum decomposition \cite{D72}
\bea
\label{directdecompchainX}
C_p(X) \cong C_p(\Sigma) \oplus C_p(X,\partial X) \, ,
\eea 
or to the statement that the exact sequence:
\bea
0 \rightarrow C_p(\Sigma) \rightarrow C_p(X) \rightarrow C_p(X,\partial X) \rightarrow 0 \, ,
\eea
splits. We refer to $c^\circ$ as the \textbf{relative component} of $c$ and to $c^\Sigma$ as its \textbf{$\Sigma$-component}, and with an irrelevant abuse we identify $C_p(X,\partial X)$ (resp. $C_p(\Sigma)$) with the subgroup of relative components (resp. $\Sigma$-components) of $p$-chains of $X$.

We provide each $C_p(X)$ with the standard boundary operator $\partial$ thus getting the singular complex $\left(C_\bullet(X) , \partial \right)$. With respect to the decomposition rule (\ref{canondecompXchain}), the boundary operator $\partial$ has the following behavior:
\bea
\partial c = \partial c^\circ + \partial c^\Sigma = (\partial c^\circ)^\circ + (\partial c^\circ)^\Sigma + \partial c^\Sigma \, 
\eea
which entails a decomposition of $\partial$ itself according to:
\bea
\label{decompboundary}
\partial = \partial_\circ^\circ + \partial_\circ^\Sigma + \partial_\Sigma^\Sigma \, ,
\eea
with:
\bea
\left\{ \begin{aligned}
	& \partial_\circ^\circ c = \partial_\circ^\circ c^\circ = (\partial c^\circ)^\circ \\
	& \partial_\circ^\Sigma c = \partial_\circ^\Sigma c^\circ = (\partial c^\circ)^\Sigma \\
	& \partial_\Sigma^\Sigma c = \partial_\Sigma^\Sigma c^\Sigma = \partial c^\Sigma
\end{aligned} \right. \, .
\eea
The nilpotency relation $\partial \partial = 0$ then implies:
\bea
\label{nilpotdecomp}
\left\{ \begin{aligned}
	& \partial_\circ^\circ \circ \partial_\circ^\circ = 0 \\
	& \partial_\Sigma^\Sigma \circ \partial_\circ^\Sigma + \partial_\circ^\Sigma \circ \partial_\circ^\circ = 0 \\
	& \partial_\Sigma^\Sigma \circ \partial_\Sigma^\Sigma = 0
\end{aligned} \right. \, .
\eea
A $p$-chain $c$ of $X$ which fulfills:
\bea
\label{relatcycle}
\partial_\circ^\circ c = (\partial c^\circ)^\circ = 0 \, .
\eea
is called a relative $p$-cycle of $(X,\partial X)$. This also means that $\partial c \in C_{p-1}(\Sigma)$. The subgroup of relative $p$-cycles of $(X,\partial X)$ is denoted $Z_p(X,\partial X)$. A $p$-chain $c$ of $X$ which fulfills : 
\bea
\label{relatbound}
c^\circ = \partial_\circ^\circ u = (\partial u^\circ)^\circ \, .
\eea
is called a relative $p$-boundary of $(X,\partial X)$. This also means that $c= \partial u + v^\Sigma$. The subgroup of relative $p$-boundaries of $(X,\partial X)$ is denoted $B_p(X,\partial X)$. The $p$-th relative homology group of $(X,\partial X)$ is then:
\bea
H_p(X,\partial X) = \frac{Z_p(X,\partial X)}{B_p(X,\partial X)} \, .
\eea
These groups are the homology groups of the complex $\left( C_\bullet(X,\partial X) , \partial_\circ^\circ \right)$. Of course, we still have the homology groups $H_p(X)$ associated with the complex $\left( C_\bullet(X) , \partial \right)$. Moreover, since $i_\Sigma$ is a chain map, for each $p$, we have an injection $i_\Sigma:H_p(\Sigma) \rightarrow H_p(X)$. It is quite obvious that any $p$-cycle of $X$ is a relative $p$-cycle of $(X,\partial X)$, the converse being untrue. Nevertheless, any relative $p$-cycle of $(X,\partial X)$ defines a $(p-1)$-cycle of $\partial X$. Indeed, let $c$ be relative $p$-cycle of $(X,\partial X)$. Then $\partial_\circ^\circ c = 0$ and from decomposition (\ref{decompboundary}) and the last of the relations (\ref{nilpotdecomp}) we deduce that:
\bea
\partial \partial_\circ^\Sigma c = \partial_\Sigma^\Sigma \partial_\circ^\Sigma c = - \partial_\circ^\Sigma \partial_\circ^\circ c = 0
\eea
In other words, if $c$ is a relative $p$-cycle of $(X,\partial X)$ then $\partial_\circ^\Sigma c$ is a $(p-1)$-cycle of $\partial X$.  Moreover, if $c$ is a relative $p$-boundary of $(X,\partial X)$, then $c = \partial u + v^\Sigma$ and hence $\partial_\circ^\Sigma c = - \partial_\Sigma^\Sigma (\partial_\circ^\Sigma u^\circ)$. So, if $c$ is a relative $p$-boundary of $(X,\partial X)$ then $\partial_\circ^\Sigma c$ is a boundary of $\Sigma = \partial X$. From all this we deduce that the operator $\partial_\circ^\Sigma$ gives rise to a connecting homomorphism $H_p(X,\partial X) \rightarrow H_{p-1}(\Sigma)$, so that we obtain the so-called long exact sequence in homology for $X$ \cite{D72}:
\bea
\label{longexacseqhomo}
\cdots \rightarrow H_p(\Sigma) \rightarrow H_p(X) \rightarrow H_p(X,\partial X) \rightarrow H_{p-1}(\Sigma) \rightarrow \cdots \, .
\eea

Let us extend the above construction to de Rham currents.  $\Omega^p(X)$ denotes the space of smooth $p$-forms of $X$, and let $\Omega^p(X,\partial X)$ denotes the subspace of smooth $p$-forms of $X$ that vanish on $\partial X$, i.e. the kernel of the pullback $i_\Sigma^*:\Omega^p(X) \rightarrow \Omega^p(\Sigma)$. These spaces have topological duals which are respectively denoted by $\Omega^p(X)^\diamond$ and $\Omega^p(X,\partial X)^\diamond$. Then, for each $p$, there is an exact short sequence:
\bea
\label{exactsequdeRhamX}
0 \rightarrow \Omega^p(\Sigma)^\diamond \xrightarrow{i_\Sigma} \Omega^p(X)^\diamond \xrightarrow{\pi_\circ^\diamond}   \Omega^p(X,\partial X)^\diamond \rightarrow 0 \, ,
\eea
which is the dual of the short exact sequence:
\bea
0 \rightarrow \Omega^p(X,\partial X) \xrightarrow{\pi_\circ} \Omega^p(X) \xrightarrow{i_\Sigma^*}  \Omega^p(\Sigma) \rightarrow 0 \, .
\eea
The first sequence allows to identify $\Omega^p(X,\partial X)^\diamond$ with the quotient $\Omega^p(X)^\diamond / i_\Sigma(\Omega^p(\Sigma)^\diamond)$. The space $i_\Sigma(\Omega^p(\Sigma)^\diamond)$ is then the space of de Rham $p$-currents of $X$ with support in $\partial X$, an element of which is generically denoted by $J^\Sigma$. By fixing a base for this subspace and by completing this base into a base of $\Omega^p(X)^\diamond$ we generate a decomposition rule of the form: 
\bea
\label{decompcurrents}
J = J^\circ + J^\Sigma \, ,
\eea 
for all $J \in \Omega^p(X)^\diamond$. We will refer to $J^\circ$ as the \textbf{relative component} of $J$ and to $J^\Sigma$ as its  $\Sigma$-\textbf{component}, as with an inconsequential abuse we can see $\Omega^p(X,\partial X)^\diamond$ and $\Omega^p(\Sigma)^\diamond$ as subspaces of $\Omega^p(X)^\diamond$. We endow $\Omega^p(X)^\diamond$ with the boundary operator $d^\dag$ defined by:
\bea
(d^\dag J) \eval{\alpha} = J \eval{d \alpha} \, ,
\eea 
for all $J \in \Omega^p(X)^\diamond$ and all $\alpha \in  \Omega^{p-1}(X)$. We have the decomposition:
\bea
\label{decompddag}
d^\dag = d^{\dag \circ}_\circ + d^{\dag \Sigma}_\circ + d^{\dag \Sigma}_\Sigma \, ,
\eea
which stems from the relation:
\bea
\label{decompderivcurrent}
d^\dag J = d^{\dag \circ}_\circ J^\circ + d^{\dag \Sigma}_\circ J^\circ + d^{\dag \Sigma}_\Sigma J^\Sigma \, .
\eea
The nilpotency property $d^\dag d^\dag = 0$ is then equivalent to:
\bea
\label{derivdRcurrent}
\left\{ \begin{aligned}
& d^{\dag \circ}_\circ d^{\dag \circ}_\circ = 0 \\
& d^{\dag \Sigma}_\Sigma d^{\dag \Sigma}_\circ + d^{\dag \Sigma}_\circ d^{\dag \circ}_\circ = 0 \\
& d^{\dag \Sigma}_\Sigma d^{\dag \Sigma}_\Sigma = 0
\end{aligned} \right. \, .
\eea
Hence, we have three complexes arising from the construction, the complex $\left( \Omega^\bullet(X)^\diamond , d^\dag \right)$, the relative complex $\left( \Omega^\bullet(X,\partial X)^\diamond , d^{\dag \circ}_\circ \right)$ and the complex $\left( i_\Sigma(\Omega^\bullet(\Sigma)^\diamond) , d^{\dag \Sigma}_\Sigma \right)$ which is canonically isomorphic to the complex $\left( \Omega^\bullet(\Sigma)^\diamond , d^{\dag}_\Sigma \right)$. The first two complexes generate families of homology groups, $H^\diamond_p(X)$ and $H^\diamond_p(X,\partial X)$, and the operator $d^{\dag \Sigma}_\circ$ induces a connecting homomorphism $H^\diamond_p(X,\partial X) \to H^\diamond_{p-1}(X)$. All this yields the so-called long exact sequence in currents homology:
\bea
\cdots \rightarrow H_p^\diamond(\Sigma) \rightarrow H_p^\diamond(X) \rightarrow H_p^\diamond(X,\partial X) \rightarrow H_{p-1}^\diamond(\Sigma) \rightarrow \cdots H_0^\diamond(X,\partial X) \rightarrow 0 \, .
\eea
Although they look similar, the modules occurring in the above long sequence are over real numbers \cite{dR55} while those appearing in the long sequence (\ref{longexacseqhomo}) are over integers.

To any $\omega \in \Omega^{3-p}(X)$ we associate the de Rham $p$-current $J_{\omega}$ defined by:
\bea
\kern-70pt \forall \alpha \in \Omega^p(X) \; , \; \; J_{\omega} \eval{\alpha} = \int_X \omega  \wedge \alpha  \, ,
\eea
which defines an injective homomorphism $J: \Omega^{3-p}(X) \to \Omega^p(X)^\diamond$. The elements of $\Ima J$ are called regular de Rham $p$-currents of $X$. If we apply decomposition (\ref{decompcurrents}) to a regular $p$-current $J_{\omega}$, then for support reasons $J_\omega^\Sigma = 0$. Hence, we have:
\bea
\label{decompregularcurrent}
\kern-80pt \forall \omega \in \Omega^{3 - p}(X) , \quad J_\omega = J_\omega^\circ \, .
\eea
The boundary of a regular $p$-current $J_\omega$ is then:
\bea
d^\dag J_\omega \eval{\alpha} = J_\omega \eval{d \alpha} = \int_X \omega \wedge d \alpha \, .
\eea
A simple integration by parts yields:
\bea
d^\dag J_\omega \eval{\alpha} = (-)^p \int_X d \omega \wedge \alpha + (-)^{p+1} \oint_\Sigma \omega^\Sigma \wedge \alpha^\Sigma \, ,
\eea
which implies that:
\bea
d^\dag J_\omega \eval{\alpha} = (-)^p J_{d \omega} + (-)^{p+1} J_{\omega^\Sigma} \, .
\eea
Recalling that $J_\omega = J_\omega^\circ$ and comparing this relation with (\ref{derivdRcurrent}), we deduce that:
\bea
\left\{ \begin{aligned}
& d^{\dag \circ}_\circ J_\omega = d^{\dag \circ}_\circ J_\omega^\circ  = (-)^p J_{d \omega} \\
& d^{\dag \Sigma}_\circ J_\omega = d^{\dag \Sigma}_\circ J_\omega^\circ = (-)^{p+1} J_{\omega^\Sigma} 
\end{aligned} \right. \, .
\eea
Hence, unlike $J_\omega$ the current $d^\dag J_\omega$ has a $\Sigma$-component. We gather this into a property that will prove helpful in subsection 4.3.

\begin{prop}\label{decompderivregular}\leavevmode

Let $J_\omega$ be a regular $p$-current of $X$. Then $J_\omega = J_\omega^\circ$ and:
\bea
d^\dag J_\omega = d^{\dag \circ}_\circ J_\omega + d^{\dag \Sigma}_\circ J_\omega = (-)^p J_{d \omega} + (-)^{p+1} J_{\omega^\Sigma} \, .
\eea

\end{prop}
\noindent The presence of a relative sign between the two terms forming the second equality above will be clarified a little further. From all this, we deduce that there is a Poincar\'e-Lefschetz duality between these homology groups and the corresponding de Rham cohomology groups according to:
\bea
\label{currentPoinLefsdual}
H^\diamond_p(X) \cong H_{dR}^{3-p}(X,\partial X) \quad , \quad H^\diamond_p(X,\partial X) \cong H_{dR}^{3-p}(X) \, .
\eea

As usual, a $p$-chain $c$ of $X$ defines a de Rham $p$-current $j_c$ according to:
\bea
\kern-50pt \forall \omega \in \Omega^p(X) , \quad j_c \eval{\omega} = \int_c \omega  \, .
\eea
As with $\Sigma$, $p$-chains of $X$ which have the same integrals on all $p$-forms define the same de Rham $p$-current and hence are considered as equivalent \cite{dR55}. Furthermore, if  $c$ is a chain of $X$ with decomposition $c= c^\circ + c^\Sigma$ and de Rham current $j_c$, then we have:
\bea
j_c = j_c^\circ + j_c^\Sigma \, .
\eea 
with $j_{c^\circ} = j_c^\circ$ and $j_{c^\Sigma} = j_c^\Sigma$. Moreover, we have:
\bea
d^\dag j_c = j_{\partial c} \, ,
\eea
with:
\bea
d^{\dag \circ}_\circ j_c = j_{\partial^\circ_\circ c} \quad , \quad d^{\dag \Sigma}_\circ j_c = j_{\partial^\Sigma_\circ c} \quad , \quad d^{\dag \Sigma}_\Sigma j_c = j_{\partial^\Sigma_\Sigma c} \, .
\eea
All this explains the similitude between relations (\ref{canondecompXchain}), (\ref{decompboundary}) and (\ref{nilpotdecomp}) on the one hand, and relations (\ref{decompcurrents}), (\ref{decompddag}) and (\ref{derivdRcurrent}) on the other hand.

Because $\Sigma$ is the oriented boundary of $X$, half of the first homology group of the latter is trivialized. Indeed, each meridian $1$-cycle $\mu_a^\Sigma$ can be seen as the boundary of a meridian disk $D_a$ of $X$. So, we fix once and for all a set of meridian disks $D_a$ such that:
\bea
\label{disks}
\partial D_a = - \mu_a^\Sigma \, ,
\eea
for each $a=1, \cdots , g$. Hence, we have:
\bea
\label{disks2}
\left\{ \begin{aligned}
& \partial_\circ^\circ D_a = 0 \\
& \partial_\circ^\Sigma D_a = - \mu_a^\Sigma \\
& \partial_\Sigma^\Sigma D_a = 0
\end{aligned} \right. \, ,
\eea
the first relation implying that each $D_a$ is a relative $2$-cycle: $D_a = D_a^\circ$.

Unlike meridian cycles, the longitude $1$-cycles $\lambda_a^\Sigma$ aren't trivial in $X$. Nevertheless, we can introduce a set of annuli $A_a^\Sigma$ in $X$ such that: 
\bea
\label{annuli}
\partial A_a = \lambda_a - \lambda_a^\Sigma \, .
\eea
The $1$-cycles $\lambda_a$ generate $H_1(X) \simeq \mathbb{Z}^{g}$ and are usually referred to as the core $1$-cycles of $X$. The above relation can also be written as:
\bea
\label{annuli2}
\left\{ \begin{aligned}
& \partial_\circ^\circ A_a = \lambda_a \\
& \partial_\circ^\Sigma A_a = - \lambda_a^\Sigma \\
& \partial_\Sigma^\Sigma A_a = 0
\end{aligned} \right. \, .
\eea
Note that relation (\ref{annuli}) implies that each non-trivial $1$-cycle $\lambda_a$ is actually a relative boundary. This yields:
\bea
H_1(X) = \mathbb{Z}^g \, \, \, ,  \, \, \, H_1(X,\partial X) = 0 \, .
\eea
For their part, the meridian disks are generators of $H_2(X,\partial X)$ so that we have:
\bea
H_2(X) = 0 \, \, \, ,  \, \, \, H_2(X,\partial X) = \mathbb{Z}^g \, .
\eea
We complete this with the following, quite trivial, relations:
\bea
\left\{ \begin{aligned}
& H_0(X) = \mathbb{Z} \, \, \, ,  \, \, \, H_0(X,\partial X) = 0 \\
& H_3(X) = 0 \, \, \, \, ,  \, \, \, H_3(X,\partial X) = \mathbb{Z} 
\end{aligned} \right. \, .
\eea

As the orientation of $\Sigma$ is inherited from the one of $X$, relations (\ref{intersectionbasisSigma}) induce the following ones in $X$:
\bea
\label{intersectionbasisX}
\lambda_a \odot D_b = \delta_{ab} \, ,
\eea
which define the Kronecker indexes of the $1$-chains $\lambda_a$ with the $2$-chains $D_b$. These relations imply that:
\bea
\label{krondisklambda}
\partial^\Sigma_\circ (D_a \pitchfork A_b) = - (\partial^\Sigma_\circ D_a) \pitchfork (\partial^\Sigma_\circ A_b) \, ,
\eea
and more generaly:
\bea
\label{sigmatorelativeboundary}
\partial^\Sigma_\circ (c^0 \pitchfork \tilde{c}^0) = - \partial^\Sigma_\circ c^0 \pitchfork \partial^\Sigma_\circ \tilde{c}^0 \, ,
\eea
where $c^0$ and $\tilde{c}^0$ are relative $2$-chains of $X$ the intersection of which is well-defined. Of course, we assume that $c^0$ and $\tilde{c}^0$  are such that all the intersections appearing in (\ref{sigmatorelativeboundary}) have a meaning. The above Kronecker indexes induces a pairing between $H_1(X)$ and $H_2(X,\partial X)$ and the same Euclidean inner product on $\mathbb{Z}^g$ as (\ref{innerproductZg}). In terms of de Rham currents, we have:
\bea
\label{disks&annulibound}
\left\{ \begin{aligned}
& d^{\dag \circ}_\circ j_{D_a} = 0 \\
& d^{\dag \Sigma}_\circ j_{D_a} = - j_{\mu_a^\Sigma} \\
& d^{\dag \Sigma}_\Sigma j_{D_a} = 0
\end{aligned} \right. \; \; \; \; \; \; , \; \; \; \; \; \; 
\left\{ \begin{aligned}
& d^{\dag \circ}_\circ j_{A_a} = j_{\lambda_a} \\
& d^{\dag \Sigma}_\circ j_{A_a} = - j_{\lambda_a^\Sigma} \\
& d^{\dag \Sigma}_\Sigma j_{A_a} = 0
\end{aligned} \right. \, ,
\eea
relations (\ref{intersectionbasisX}) thus reading:
\bea
\label{KroneckerWithCurrents}
\left( j_{\lambda_a} \wedge j_{D_b} \right) \eval{1} = \lambda_a \odot D_b = \delta_{ab} \, .
\eea
Finally, still in terms of currents, relation (\ref{sigmatorelativeboundary}) takes the form:
\bea
\label{sigmatorelativecurrents}
d^{\dag \Sigma}_\circ (J^0 \wedge \tilde{J}^0) = - d^{\dag \Sigma}_\circ J^0 \wedge d^{\dag \Sigma}_\circ \tilde{J}^0 \, ,
\eea
when the exterior product have a meaning. Note that we can recover (\ref{sigmatorelativeboundary}) or (\ref{sigmatorelativecurrents}) by considering two regular $2$-currents $J_\omega$ and $J_{\tilde{\omega}}$ obtained by regularizing $c^0$ and $\tilde{c}_0$ (or $J^0$ and $\tilde{J}_0$) and then by writing the following sequence of equalities:
\bea
\label{sigmatorelativeboundcurr}
d^{\dag \Sigma}_\circ ( J_\omega \wedge J_{\tilde{\omega}} ) = d^{\dag \Sigma}_\circ  J_{\omega \wedge \tilde{\omega}} = - J_{\omega^\Sigma \wedge \tilde{\omega}^\Sigma} = - J_{\omega^\Sigma} \wedge J_{\tilde{\omega}^\Sigma} = - d^{\dag \Sigma}_\circ J_\omega \wedge d^{\dag \Sigma}_\circ J_{\tilde{\omega}} \, ,
\eea
with the help of Property \ref{decompderivregular}.

Let us consider the case where $\partial X = S^1 \times S^1$, with $\omega_A$ and $\omega_D$ two regularizations of the chain currents $j_A$ and $j_D$, respectively. It is no hard to check that $\omega_A^\Sigma$, $\omega_D^\Sigma$ and $d \omega_A$ are regularizations of the chain currents $j_{\lambda^\Sigma}$, $j_{\mu^\Sigma}$ and $j_\lambda$, respectively, whereas $d \omega_D = 0$. Then, by using Property \ref{decompderivregular}, we deduce that:
\bea
\left\{ \begin{aligned}
& d^\dag J_{\omega_A} = d^{\dag \circ}_\circ J_{\omega_A} + d^{\dag \Sigma}_\circ J_{\omega_A} = J_{d \omega_A} - J_{\omega_A^\Sigma} \\
& d^\dag J_{\omega_D} = d^{\dag \circ}_\circ J_{\omega_D} + d^{\dag \Sigma}_\circ J_{\omega_D} = - J_{\omega_A^\Sigma}
\end{aligned} \right. \, ,
\eea
relations which agree with those of (\ref{disks&annulibound}). Note that $\omega_D^\Sigma$ is a regularization of $j_{-\partial D}$ and not of $j_{\partial D}$, which shed lights on the presence of the relative sign in the formula of Property (\ref{disks&annulibound}).

A $U(1)$ connection on $X_L \cup_f X_R$ necessarily gives rise to two connections, one on $X_L$ and one on $X_R$, that glue on $\Sigma$, via $f$, up to a gauge transformation. Since a $U(1)$ connection is equivalent to a DB $1$-cocycle, the set of classes of $U(1)$ connections of $X$ is nothing but the Deligne-Beilinson $\mathbb{Z}$-module $H_D^1(X)$ which is embedded into the exact sequence \cite{HL01}:
\bea
\label{HD1Xsequence}
0 \rightarrow \frac{\Omega^1(X)}{\Omega^1_\mathbb{Z}(X)} \rightarrow H_D^1(X) \rightarrow H^2(X) \rightarrow 0 \, .
\eea
From Poincar\'e-Lefschetz duality we have $H^2(X) \cong H_1(X,\partial X) = 0$ and hence:
\bea
H_D^1(X) \cong \Omega^1(X) / \Omega^1_\mathbb{Z}(X) \, .
\eea 
This already suggests to construct DB $1$-cocycles of $X_L \cup_f X_R$ by starting with $1$-forms on $X_L$ and $X_R$ that glue on $\Sigma$ up to a gauge transformation the set of which is $\Omega^1_\mathbb{Z}(\Sigma)$, this last space appearing in exact sequence (\ref{HDSigmaexactsequence}). 

Since $X$ has a boundary, it is also possible to consider the DB space $H_D^1(X,\partial X)$ which is embedded into the exact sequence:
\bea
\label{HD1Xsequence}
0 \rightarrow H^1(X,\partial X,\mathbb{R}/\mathbb{Z}) \rightarrow H_D^1(X,\partial X) \rightarrow \Omega^{2}_{\mathbb{Z}}(X,\partial X) \rightarrow 0 \, .
\eea
At the level of Pontryagin dual spaces, there is a natural injection \cite{HL01} $H_D^1(X) \rightarrow H_D^1(X,\partial X)^\star$ which suggests to consider this last space in the construction of $H_D^1(X_L \cup_f X_R)^\star$. We then have the following exact sequence:
\bea
\label{DualDBsequeforXdX}
0 \rightarrow \Omega^{2}_{\mathbb{Z}}(X,\partial X)^\star \xrightarrow{\bar{d}^*} H^1_D(X,\partial X)^\star \rightarrow H^{2}(X) = 0 \rightarrow 0 \, ,
\eea
which implies that $H^1_D(X,\partial X)^\star \cong \Omega^{2}_{\mathbb{Z}}(X,\partial X)^\star$. Finally, the following property shows that $H^1_D(X,\partial X)^\star$ can be seen as a quotient space, just like $H^1_D(X)$.

\begin{prop}\label{H1Dualisquotient}\leavevmode
	
	Let $\Omega^2(X,\partial X)^\diamond_\mathbb{Z}$ be the space of relative $2$-currents that are $\mathbb{Z}$-valued on $\Omega_\mathbb{Z}^1(X,\partial X)$, the space of closed $1$-forms with integral periods of $X$ that vanish on $\partial X$. Then, we have:
	\bea
	H_D^1(X,\partial X)^\star \cong \Omega^{2}_{\mathbb{Z}}(X,\partial X)^\star \cong \frac{\Omega^2(X,\partial X)^\diamond}{\Omega^2(X,\partial X)^\diamond_\mathbb{Z}}   \, .
	\eea
	
\end{prop}

\noindent The first isomorphism above is a simple consequence of the exact sequence (\ref{DualDBsequeforXdX}). The second is quite obvious since the elements of $\Omega^2(X,\partial X)^\diamond_\mathbb{Z}$ are by definition the relatively closed de Rham currents of $(X,\partial X)$ which are $\mathbb{Z}$-valued on $\Omega^{2}_{\mathbb{Z}}(X,\partial X)$. We refer the reader to \cite{HL01} for details.

\section{Geometrical data for $X_L \cup_f X_R$}

We consider here the following Heegaard construction. Let $X_L$ and $X_R$ be two copies of the handlebody $X$ and consider an orientation reversing diffeomorphism $f:\Sigma_L \rightarrow \Sigma_R$, where $\Sigma_L = \partial X_L$ and $\Sigma_R = \partial X_L$. Then, met us consider the oriented smooth $3$-manifold $X_L \cup_f X_R$ obtained by gluing $X_L$ and $X_R$ along their diffeomorphic boundaries. It is well-known that any oriented smooth closed $3$-manifold can be constructed this way. In this second section we explain how to represent chains, forms and currents on $X_L \cup_f X_R$ with the help of chains, forms and currents on $X_L$ and $X_R$. As before, we start with chains, then we consider currents and we finish with smooth DB cohomology groups, dedicating one subsection to each of these topics. An expression of the linking number is also provided in the second subsection.

\subsection{Chains and cycles}

We define a $p$-chain of $X_L \cup_f X_R$ as a couple $(c_L , c_R)$ of $p$ chains on $X_L$ and $X_R$. If we decompose these chains according to (\ref{canondecompXchain}) we obtain the representation $(c_L^\circ + c_L^\Sigma,c_R^\circ + c_R^\Sigma)$. However, such a representation is degenerated. Indeed, let us consider a couple of the form $(u^\Sigma , f(u^\Sigma))$. When gluing $X_L$ and $X_R$ along their boundary via the homeomorphism $f$, then $u^\Sigma$ obviously ``compensates" $f(u^\Sigma)$ so that $(u^\Sigma , f(u^\Sigma))$ actually represents the zero $p$-chain of $X_L \cup_f X_R$. Hence, we must consider the equivalence relation:
\bea
(c_L,c_R) \sim_f (c_L + u^\Sigma , c_R + f(u^\Sigma)) \, ,
\eea
for any $p$-chain $u^\Sigma$ of $\Sigma$, the set of $p$-chains of $X_L \cup_f X_R$ being then defined as the quotient:
\bea
\left( C_p(X_L) \times C_p(X_R) \right) / \sim_f \, .
\eea
This set is denoted by $C_p(X_L \cup_f X_R)$. To avoid the use of equivalence classes of chains, we can consider representatives of the form:
\bea
\label{leftrepofchains}
(c_L^\circ , c_R^\circ + c_R^\Sigma) \, ,
\eea
called \textbf{right representatives}. Obviously, any $p$-chain of $X_L \cup_f X_R$ has a unique right representative.

The boundary of $(c_L , c_R)$ is trivially defined by $\partial (c_L , c_R) = (\partial c_L , \partial c_R )$. Since $f$ induces a chain map between $C_p(\Sigma_L)$ and $C_p(\Sigma_R)$, this boundary operator is compatible with the equivalence relation $\sim_f$ and hence goes to classes. On right representatives $\partial$ is given as:
\bea
\label{rightreppartialaction}
\partial (c_L^\circ , c_R^\circ + c_R^\Sigma) = \left( \partial^\circ_\circ c_L^\circ , \partial^\circ_\circ c_R^\circ + \partial_\circ^\Sigma c_R^\circ - f\left( \partial_\circ^\Sigma c_L^\circ \right) + \partial_\Sigma^\Sigma c_R^\Sigma \right) \, .
\eea

\noindent A $p$-chain of $X_L \cup_f X_R$ with right representative $(z_L^\circ , z_R^\circ + z_R^\Sigma)$ is a $p$-cycle if:
\bea
\label{cycleconditHeeg}
\left\{ \begin{aligned}
& \partial^\circ_\circ z_L^\circ = 0 = \partial^\circ_\circ z_R^\circ \\
& \partial_\circ^\Sigma z_R^\circ + \partial_\Sigma^\Sigma z_R^\Sigma = f\left( \partial_\circ^\Sigma z_L^\circ \right)
\end{aligned} \right. \, .
\eea
The first line in the above relations means that $z_L^\circ$ and $z_R^\circ$ are relative $p$-cycles, whereas the second line implies that $z_R^\Sigma$ is not necessarily a cycle of $\Sigma_R$.  The set of $p$-cycles of $X_L \cup_f X_R$ is denoted by $Z_p(X_L \cup_f X_R)$. 

\noindent A $p$-cycle of $X_L \cup_f X_R$ with right representative $(b_L^\circ , b_R^\circ + b_R^\Sigma)$ is a $p$-boundary if there exists a $(p+1)$-chain with right representative $(c_L^\circ , c_R^\circ + c_R^\Sigma)$ such that:
\bea
\label{boundconditHeeg}
\left\{ \begin{aligned}
& b_L^\circ = \partial^\circ_\circ c_L^\circ \\
& b_R^\circ = \partial^\circ_\circ c_R^\circ \\
& b_R^\Sigma = \partial_\circ^\Sigma c_R^\circ - f\left( \partial_\circ^\Sigma c_L^\circ \right) + \partial_\Sigma^\Sigma c_R^\Sigma
\end{aligned} \right. \, .
\eea
The first two conditions simply means that $b_L^\circ$ and $b_R^\circ$ must be relative $p$-boundaries. According to the relations (\ref{nilpotdecomp}), it is quite obvious that a $p$-boundary is a $p$-cycle. The set of $p$-boundaries of $X_L \cup_f X_R$ is denoted by $B_p(X_L \cup_f X_R)$, the $p$-th homology group of $X_L \cup_f X_R$ thus being:
\bea
H_p(X_L \cup_f X_R) = \frac{Z_p(X_L \cup_f X_R)}{B_p(X_L \cup_f X_R)} \, .
\eea

As an exercise, let us compute $H_2(X_L \cup_f X_R)$. As already mentioned, the meridian disks $D_a$ form a set of generators for $H_2(X,\partial X)$. Hence, a couple of the form:
\bea
\label{2cyclesXcupX}
\left( \sum_{a=1}^g m^a_L D_a^L + \partial^\circ_\circ \chi_L , \sum_{a=1}^g m^a_R D_a^R + \partial^\circ_\circ \chi_R \right) \, ,
\eea
is a generator of $H_2(X_L \cup_f X_R)$ is and only if the two components are gluing properly along $\Sigma$, via the gluing mapping $f$, in order to form a $2$-cycle of $X_L \cup_f X_R$. This means that they fulfill:
\bea
\begin{aligned}
\sum_{a=1}^g (P \vec{m}_L)^a \lambda_a^{\Sigma_R} + & \sum_{a=1}^g (Q \vec{m}_L - \vec{m}_R)^a \mu_a^{\Sigma_R} = \\ & - \partial^\Sigma_\Sigma \left( \sum_{a=1}^g m_L^a \psi_a^{\Sigma_R}  + \partial^\Sigma_\circ \chi_R - f(\partial^\Sigma_\circ \chi_L) \right)  
\end{aligned} \, .
\eea
and hence that:
\bea
\left\{ \begin{aligned}
& P \vec{m}_L = 0 \\
& \vec{m}_R = Q \vec{m}_L
\end{aligned} \right. \, .
\eea
This yields the following property:

\begin{prop}\label{closedsurface}\leavevmode

Any $2$-cycle of $X_L \cup_f X_R$ have a right representative of the form:
\bea
\label{closedsurfacedef}
\left( \sum_{a=1}^g m^a D_a^L , \sum_{a=1}^g \left( Q \vec{m} \right)^a D_a^R - \sum_{a=1}^g m^a \psi_a^{\Sigma_R} \right) + \partial V \, .
\eea
with $\vec{m} \in \ker P$ and $V = (V_L^\circ,V_R^\circ)$ a $3$-chain of $X_L \cup_f X_R$. Hence:
\bea
H_2(X_L \cup_f X_R) \cong \ker P \, ,
\eea

\end{prop}

By using the same kind of arguments and exact sequence (\ref{Pexactsequence}) we can show that:

\begin{prop}\label{H1versuscokerP}\leavevmode

Any $1$-cycle of $X_L \cup_f X_R$ have a right representative of the form:
\bea
\label{1cycledecompalonglongitudes}
\left( 0 , \sum_a n^a \lambda_a^R \right) + \left( \partial_\circ^\circ c_L^\circ , \partial_\circ^\circ c_R^\circ + \partial_\circ^\Sigma c_R^\circ - f(\partial_\circ^\Sigma c_L^\circ) + \partial_\Sigma^\Sigma c_R^\Sigma \right) \, ,
\eea
with $\vec{n} \in \mathbb{Z}^g$ representing an element of $\coker P$. This cycle is homologically trivial if and only if $Q^\dag \vec{n} \in \Ima P^\dag$. Hence:
\bea
H_1(X_L \cup_f X_R) \cong \coker P \, ,
\eea

\end{prop}

\noindent A generator of $H_1(X_L \cup_f X_R)$ is necessarily a combination of the longitudes of $X_R$ (or equivalently of $X_L$), let say $z_{\vec{n}}^R = (0 , \sum_a n^a \lambda_a^R)$. Obviously, $z_{\vec{n}}^R$ is homologous to the $1$-cycle $z_{\vec{n}}^{\Sigma_R} = \sum_a n^a \lambda_a^{\Sigma_R}$, and if $\vec{n} = P \vec{m}$ then this last cycle is the boundary of $\left( \sum_{a=1}^g m^a j_{D_a^L} , \sum_{a=1}^g \left( Q \vec{m} \right)^a j_{D_a^R} - \sum_{a=1}^g m^a j_{\psi_a^{\Sigma_R}} \right)$. Hence, $z_{\vec{n}}^R$ is a boundary when $\vec{n} \in \Ima P$. Conversely, if $z_{\vec{n}}^R$ is a boundary then is is easy to check that $\vec{n} \in \Ima P$. The rest of the proof stems from Property \ref{trivclassandQ}. 

As a last check of the consistency of our definition of chains of $X_L \cup_f X_R$, let us consider $(X_L,X_R)$ which is nothing but $X_L \cup_f X_R$ itself and as such a $3$-cycle. Then, we have $\partial (X_L,X_R) = (\partial X_L , \partial X_R) = (\Sigma_L,\Sigma_R) = (0 , \Sigma_R - f(\Sigma_L) ) = (0,0)$, as intended.

\subsection{Forms, currents and linking number}

A differential $p$-form of $X_L \cup_f X_R$ is a couple $(\omega_L,\omega_R)  \in \Omega^p(X_L) \times \Omega^p(X_R)$ that fulfills the gluing condition:
\bea
\label{glueforms}
\omega_R^\Sigma = f_* \omega_L^\Sigma \, .
\eea
The de Rham differential of a $p$-form $(\omega_L,\omega_R)$ of $X_L \cup_f X_R$ is simply the $(p+1)$-form $(d\omega_L,d\omega_R)$. Hence, a $p$-form $(\omega_L,\omega_R)$ is closed if $\omega_L$ and $\omega_R$ are closed. If $(\omega_L,\omega_R)$ is exact then $\omega_L$ and $\omega_R$ are also exact. Conversely, if $\omega_L = d \eta_L$ and $\omega_R = d \eta_R$, then $(\omega_L,\omega_R)$ is exact if and only if $(\eta_L,\eta_R)$ is a $(p-1)$-form of 
$X_L \cup_f X_R$, that is to say if and only if it fulfills (\ref{glueforms}).

As among the closed forms of $X_L \cup_f X_R$ the closed $2$-forms with integral periods play a particular role in this article, we describe them in the following property.

\begin{prop}\label{closed2formsintgperiods}\leavevmode

A closed $1$-form with integral periods of $X_L \cup_f X_R$ is of the form:
\bea
\label{decompclosed1integperiods}
\alpha = \left( \sum_{a=1}^{g} N^a \, j_{D_a^L}^\infty , \sum_{a=1}^{g} (Q \vec{N})^a \, j_{D_a^R}^\infty \right) + d (\phi_L , \phi_R) \, ,
\eea
with $\vec{N} \in \ker P$ and $(\phi_R , \phi_L) \in \Omega^0(X_L) \times \Omega^0(X_R)$ such that:
\bea
\label{gluingphi}
\phi_R^\Sigma - f_* \phi_L^\Sigma = \sum_{a=1}^{g} N^a j_{\psi_a^R}^\infty   \, .
\eea
The de Rham cohomology class of $\alpha$ is $\vec{N} \in \ker P$ so that $\ker P \cong H_{dR}^1(X_L \cup_f X_R)$. 

\end{prop}

\noindent As meridian disks are generator of $H_2(X,\partial X)$ and $H^1(X) \cong H_2(X,\partial X)$ by Poincar\'e-Lefshetz duality, up to an exact term, every closed $1$-form of $X$ is a combination of the $1$-forms $j_{D_a}^\infty$. These forms are regularizations of the chain currents $j_{D_a}$. By restriction to $\Sigma$ they define regularization $j_{\mu_a}^\infty$ of the chain currents $j_{\mu_a}$. If we consider regularizations $j_{\lambda_a}^\infty$ and $j_{\psi_a^R}^\infty$ of $j_{\lambda_a}$ and $j_{\psi_a^R}$, respecitvely, then relations (\ref{images}) as well as the gluing condition (\ref{glueforms}) quite straightforwardly yield Property \ref{closed2formsintgperiods}. Note that the coefficients of $j_{D_a^L}^\infty$ and $j_{D_a^R}^\infty$ must be integers fort the closed form $\alpha$ to have integral periods. The fact that $\ker P \cong H_{dR}^1(X_L \cup_f X_R)$ is consistent with the following sequence of isomorphisms $H_{dR}^1(X_L \cup_f X_R) \cong H^1(X_L \cup_f X_R) \cong H_2(X_L \cup_f X_R) \cong \ker P$ which stem from Poincaré-Lefschetz duality.

In analogy with $p$-chains, the set of de Rham $p$-current of $X_L \cup_f X_R$ is the quotient:
\bea
\left( \Omega^p(X_L)^\diamond \times \Omega^p(X_L)^\diamond \right) / \sim_f \, ,
\eea
where the equivalence relation $\sim_f$ is defined by $(J^\Sigma , f(J^\Sigma)) \sim_f 0$. Every de Rham $p$-current of $X_L \cup_f X_R$  admits a right representative $(J_L^\circ , J_R^\circ + J_R^\Sigma)$, the boundary operator $d^\dag$ acting on them according to:
\bea
\label{boundaryJ}
d^\dag (J_L^\circ , J_R^\circ + J_R^\Sigma) = \left( d^{\dag \circ}_\circ J_L^\circ , d^{\dag \circ}_\circ J_R^\circ + d^{\dag \Sigma}_\circ J_R^\circ - f\left( d^{\dag \Sigma}_\circ J_L^\circ \right) + d^{\dag \Sigma}_\Sigma J_R^\Sigma \right) \, .
\eea
A de Rham $p$-current with right representative $(J_L^\circ , J_R^\circ + J_R^\Sigma)$ is closed if and only if:
\bea
\label{closeddRcurrent}
\left\{ \begin{aligned}
& d^{\dag \circ}_\circ J_L^\circ = 0 = d^{\dag \circ}_\circ J_R^\circ \\
& d^{\dag \Sigma}_\Sigma J_R^\Sigma + d^{\dag \Sigma}_\circ J_R^\circ - f\left( d^{\dag \Sigma}_\circ J_L^\circ \right) = 0
\end{aligned} \right. \, ,
\eea
and it is exact if and only if there exists a de Rham $(p+1)$-current with right representative $(S_L^\circ , S_R^\circ + S_R^\Sigma)$ such that:
\bea
\label{exactdRcurrent}
\left\{ \begin{aligned}
& J_L^\circ = \partial^\circ_\circ S_L^\circ \\
& J_R^\circ = \partial^\circ_\circ S_R^\circ \\
& J_R^\Sigma = \partial_\circ^\Sigma S_R^\circ - f\left( \partial_\circ^\Sigma S_L^\circ \right) + \partial_\Sigma^\Sigma S_R^\Sigma
\end{aligned} \right. \, .
\eea
When the de Rham current is the current of a chain, aka a chain current, the above relations are nothing but relations (\ref{cycleconditHeeg}) and (\ref{boundconditHeeg}). In the case of chain current, we denote by $j_c$ or $C$ the de Rham current of a chain $c$.

Let $z = \partial c$ and $\tilde{z} = \partial \tilde{c}$ be two homologically trivial $1$-cycles of a smooth closed $3$-manifold $M$ whose supports do not intersect and such that $z$ and $\tilde{c}$ on the one hand, and  $\tilde{z}$ and $c$ on the second hand, are in transverse position in $M$. The linking number of $z$ and $\tilde{z}$ is the integer defined as:
\bea
\label{linkingnumber}
\textit{lk}(z,\tilde{z}) = c \odot \tilde{z} = \tilde{c} \odot z = \textit{lk}(\tilde{z},z) \, .
\eea
In order to express the above expression of the  linking number with respect to the right representatives $c = (c_L^\circ , c_R^\circ + c_R^\Sigma)$ and $\tilde{z} = (z_L^\circ , z_R^\circ + z_R^\Sigma)$, all we have to do is identify the pointwise oriented intersections between the various components of $c$ and $\tilde{z}$, which yields:
\bea
\textit{lk}(z,\tilde{z}) = c_L^\circ \odot \tilde{z}_L^\circ + c_R^\circ \odot \tilde{z}_R^\circ + \partial_\circ^\Sigma c_R^\circ \odot \tilde{z}_R^\Sigma + c_R^\Sigma \odot \left( \partial_\circ^\Sigma \tilde{z}_R^\circ + \partial_\Sigma^\Sigma \tilde{z}_R^\Sigma \right) \, .
\eea
Let us point out that the first two terms in the above expression correspond to intersections in the 3D interior of $X_L$ and $X_R$ while the last two terms yield intersections in the 2D boundary $\Sigma _R$. Taking into account the second condition of (\ref{cycleconditHeeg}) as well as the first two of (\ref{boundconditHeeg}) we obtain:
\bea
\textit{lk}(z,\tilde{z}) = c_L^\circ \odot \partial_\circ^\circ \tilde{c}_L^\circ + c_R^\circ \odot \partial_\circ^\circ \tilde{c}_R^\circ + \partial_\circ^\Sigma c_R^\circ \odot \tilde{z}_R^\Sigma + c_R^\Sigma \odot f( \partial_\circ^\Sigma \tilde{z}_L^\circ) \, .
\eea
Finally, by using the third relation of (\ref{nilpotdecomp}), we deduce that:
\bea
\label{rightreplinking}
\textit{lk}(z,\tilde{z}) = c_L^\circ \odot \partial_\circ^\circ \tilde{c}_L^\circ + c_R^\circ \odot \partial_\circ^\circ \tilde{c}_R^\circ + \partial_\circ^\Sigma c_R^\circ \odot \tilde{z}_R^\Sigma - (\partial_\Sigma^\Sigma c_R^\Sigma) \odot f( \partial_\circ^\Sigma \tilde{c}_L^\circ) \, ,
\eea
with $\tilde{z}_R^\Sigma = \partial_\circ^\Sigma \tilde{c}_R^\circ - f(\partial_\circ^\Sigma \tilde{c}_L^\circ) +\partial_\Sigma^\Sigma \tilde{c}_R^\Sigma$. This expression can be written in term of de Rham currents as:
\bea
\label{linkingformula}
\begin{aligned}
	\textit{lk}(z,\tilde{z}) = & \left( j_{c_L^\circ}  \wedge d_\circ^{\dag \circ} j_{\tilde{c}_L^\circ} \right) \eval{1_L} + \left( j_{c_R^\circ}  \wedge d_\circ^{\dag \circ} j_{\tilde{c}_R^\circ} \right) \eval{1_R} + \\
	& + \left( d_\circ^{\dag \Sigma} j_{c_R^\circ} \wedge (\delta_{-1}^\star \tilde{\mathcal{C}})_R^\Sigma \right) \eval{1_{\Sigma_R}} - \left( d_\Sigma^{\dag \Sigma} j_{c_R^\Sigma} \wedge f(d_\circ^{\dag \Sigma} j_{\tilde{c}_L^\circ}) \right) \eval{1_{\Sigma_R}} 
\end{aligned} \, ,
\eea
with $(\delta_{-1}^\star \tilde{\mathcal{C}})_R^\Sigma = d_\circ^{\dag \Sigma} j_{\tilde{c}_R^\circ} - f(d_\circ^{\dag \Sigma} j_{\tilde{c}_L^\circ}) + d_\Sigma^{\dag \Sigma} j_{\tilde{c}_R^\Sigma} = j_{\tilde{z}_R^\Sigma}$. Some remarks must be made concerning this expression of the linking number. Normally, we must take $c$ (resp. $\tilde{c}$) in transverse position with $\tilde{z}$ (resp. $z$). So, if $\tilde{z}_R^\Sigma \neq 0$ (resp. $z_R^\Sigma \neq 0$) we may have to chose $c$ (resp. $\tilde{c}$) such that $c_R^\Sigma = 0$ (resp. $\tilde{c}_R^\Sigma = 0$). Thanks to the long exact sequence in homology (\ref{longexacseqhomo}) such a choice is always possible. The expression of the linking number then reduces to the first three terms of the right hand side of (\ref{linkingformula}) with the convention that $c = (c_L^\circ,c_R^\circ)$. However, the fourth term in the right hand side of the above expression of the linking number is well-defined when the two currents that make up it are in transverse position, thus making (\ref{linkingformula}) an acceptable generalized expression of the linking number. Last but not least, although this is not obvious when looking at (\ref{linkingformula}), the linking number fulfills $\textit{lk}(z,\tilde{z}) = \textit{lk}(\tilde{z},z)$.

\subsection{Smooth DB classes} 

A Deligne-Beilinson $1$-cocycle of $X_L \cup_f X_R$ is a triple $(A_L,A_R,\Lambda_{\Sigma_R})  \in \Omega^p(X_L) \times \Omega^p(X_R) \times H^1_D(\Sigma_R)$ that fulfills the DB-gluing condition:
\bea
\label{glueDBcocycles}
A_R^\Sigma - f_* A_L^\Sigma = \bar{d} \Lambda_{\Sigma_R} \, .
\eea
Two DB $1$-cocycles are said to be DB equivalent if their difference is of the form:
\bea
\label{1DBambig}
\left( \sum_{a=1}^g m_L^a j_{D_a^L}^\infty + dq_L , \sum_{a=1}^g m_R^a j_{D_a^R}^\infty + dq_R , \xi_{\vec{m}_L,\vec{m}_R}^\infty + \overline{q_R^\Sigma - f_* q_L^\Sigma} \right) \, ,
\eea
where $\xi_{\vec{m}_L,\vec{m}_R}^\infty \in H^0_D(\Sigma_R)$ is such that:
\bea
\bar{d} \xi_{\vec{m}_L,\vec{m}_R}^\infty = - \sum_{a=1}^g (P \vec{m}_L)^a j_{\lambda_a^{\Sigma_R}}^\infty + \sum_{a=1}^g (\vec{m}_R - Q \vec{m}_L)^a j_{\mu^{\Sigma_R}}^\infty - d j_{\psi_a^{\Sigma_R}}^\infty \, ,
\eea
and $\overline{q_R^\Sigma - f_* q_L^\Sigma}$ denotes the restriction to $\mathbb{R}/\mathbb{Z}$ of $q_R^\Sigma - f_* q_L^\Sigma$. This choice for ambiguities comes from the fact that $H_D^1(X) \cong \Omega^1(X) / \Omega^1_\mathbb{Z}(X)$ so that the elements of $\Omega^1_\mathbb{Z}(X)$ generate ambiguities. It turn out that this last space is generated by $j_{D_a^L}^\infty$, and when the gluing condition (\ref{glueDBcocycles}) is taken into account, we obtain the form (\ref{1DBambig}) for the DB ambiguities.

The set of DB classes of $X_L \cup_f X_R$ thus obtained is a $\mathbb{Z}$-module which can be embedded into the exact sequence \cite{T19}:
\bea
\label{exactsequenceHD}
0 \rightarrow H^1(X_L \cup_f X_R,\mathbb{R}/\mathbb{Z}) \rightarrow H^1_D(X_L \cup_f X_R) \xrightarrow{\bar{d}} \Omega^2_{\mathbb{Z}}(X_L \cup_f X_R) \rightarrow 0 \, .
\eea
The curvature of $A$ is defined as $\bar{d} A = \left( d A_L , d A_R \right)$. It is a $2$-form of $X_L \cup_f X_R$ since $d A_R^\Sigma = d (f_*A_L^\Sigma + \bar{d} \Lambda_{\Sigma_R}) = f_*(d A_L^\Sigma)$. This $2$-form is obviously closed. Moreover, it has integral periods as a consequence of Property \ref{closedsurface}.

A DB $3$-cocycle is a triple $(G_L,G_R,\upsilon_{\Sigma_R}) \in \Omega^3(X_L) \times \Omega^3(X_R) \times H_D^2(\Sigma_R)$. A DB $3$-cocycle of the form $(d F_L , d F_R , \overline{F_R^\Sigma - f_*(F_L^\Sigma)})$ is then a DB ambiguity, with $\overline{{^{^{\quad ^{}}}}}:\Omega^2(\Sigma_R) \rightarrow H_D^2(\Sigma_R)$ denoting the injection associated with the exact sequence $0 \rightarrow \Omega^2_\mathbb{Z}(\Sigma) \rightarrow \Omega^2(\Sigma) \rightarrow H_D^2(\Sigma) \rightarrow 0$. The set $H_D^3(X_L \cup_f X_R)$ of DB classes thus induced appears in the exact sequence:
\bea
\label{exactsequenceHD3}
0 \rightarrow H^3(X_L \cup_f X_R,\mathbb{R}/\mathbb{Z}) \rightarrow H^3_D(X_L \cup_f X_R) \xrightarrow{\bar{d}} \Omega^4_{\mathbb{Z}}(X_L \cup_f X_R) = 0 \rightarrow 0 \, ,
\eea
from which we deduce that $H^3_D(X_L \cup_f X_R) \cong H^3(X_L \cup_f X_R,\mathbb{R}/\mathbb{Z}) = \mathbb{R}/\mathbb{Z}$.

There is a pairing in $H_D^1$, the DB product $\star$, which is defined at the level of DB cocyles as follows:
\bea
\label{smoothDBproduct}
\left( A_L , A_R , \Lambda_{\Sigma_R} \right) \star \left( B_L , B_R , \Pi_{\Sigma_R} \right) = \left( A_L \wedge d B_L , A_R \wedge d B_R , \overline{B_R^\Sigma \wedge \bar{d} \Lambda_{\Sigma_R}} \right) \, .
\eea
It is obvious that $\left( A_L , A_R , \Lambda_{\Sigma_R} \right) \star \left( B_L , B_R , \Pi_{\Sigma_R} \right)$ is a $3$-cocycles of $X_L \cup_f X_R$. The above pairing is commutative and goes to DB classes \cite{T19}.

\section{Pontryagin duality on $X_L \cup_f X_R$}

With all the geometrical preliminaries having been done, let us now concentrate on the core of the article, Pontryagin duality. The construction starts with integration of DB classes along cycles which shows that $1$-cycles are elements of $H_D^1(X_L \cup_f X_R)^\star := \Hom \, (H_D^1(X_L \cup_f X_R), \mathbb{R}/\mathbb{Z})$, the Pontryagin dual of $H_D^1(X_L \cup_f X_R)$. This yields DB $1$-cycles which on their turn give rise to DB $1$-currents.

\subsection{Integration and DB $1$-cycles}

If $\omega = (\omega_L,\omega_R)$ is a $p$-form of $X_L \cup_f X_R$ and $c = (c_L,c_R)$ is a $p$-chain of $X_L \cup_f X_R$, then the integral of $\omega$ along $c$ is defined as:
\bea
\label{intformchain}
\int_c \omega = \int_{c_L} \omega_L + \int_{c_R} \omega_R \, .
\eea
This definition is consistent since the integral of $\omega = (\omega_L,\omega_R)$ along any $(u_\Sigma,f(u_\Sigma))$ is zero, these last couples representing the zero $p$-chain of $X_L \cup_f X_R$. Indeed, the gluing condition $\omega_R^\Sigma = f(\omega_L^\Sigma)$ fulfilled by $(\omega_L,\omega_R)$ straightforwardly yields this results. With respect to the right representative $(c_L^\circ , c_R^\circ + c_R^\Sigma)$ of $c$ the above integral takes the form:
\bea
\int_c \omega = \int_{c_L^\circ} \omega_L + \int_{c_R^\circ} \omega_R + \int_{c_R^\Sigma} \omega_R^\Sigma \, .
\eea 

Let $(z_L^\circ , z_R^\circ + z_R^\Sigma)$ be the right representative of a $1$-cycle $z$ of $X_L \cup_f X_R$. Since $H_1(X,\partial X) = 0$, we deduce that there exist relative $2$-chains $c_L^\circ$ and $c_R^\circ$ such that $z_L^\circ = \partial_\circ^\circ c_L^\circ$ and $z_R^\circ = \partial_\circ^\circ c_R^\circ$. Moreover, since $\partial_\circ^\circ c^\circ = \partial c^\circ - \partial_\circ^\Sigma c^\circ$ the integral of $\omega = (\omega_L,\omega_R)$ along $z$ can be written as:
\bea
\oint_z \omega = \int_{c_L^\circ} d \omega_L + \int_{c_R^\circ} d \omega_R - \int_{\partial_\circ^\Sigma c_L^\circ} \omega_L^\Sigma - \int_{\partial_\circ^\Sigma c_R^\circ} \omega_R^\Sigma + \int_{z_R^\Sigma} \omega_R^\Sigma \, .
\eea
By taking into account the gluing condition for $(\omega_L,\omega_R)$, we find that:
\bea
\label{intomegacyclebis}
\oint_z \omega = \int_{c_L^\circ} d \omega_L + \int_{c_R^\circ} d \omega_R + \oint_{\mathfrak{z}_{\Sigma_R}} \omega_R^\Sigma \, ,
\eea
where $\mathfrak{z}_{\Sigma_R}$ is the $1$-cycle of $\Sigma_R$ such that:
\bea
z_R^\Sigma = \partial_\circ^\Sigma c_R^\circ - f( \partial_\circ^\Sigma c_L^\circ) + \mathfrak{z}_{\Sigma_R} \, .
\eea
Note that $\partial_\Sigma^\Sigma \mathfrak{z}_{\Sigma_R} = 0$ is a consequence of (\ref{nilpotdecomp}), (\ref{cycleconditHeeg}) and of the fact that $z_L^\circ = \partial_\circ^\circ c_L^\circ$ and $z_R^\circ = \partial_\circ^\circ c_R^\circ$. All this suggests the following property:

\begin{prop}\label{preDBconstruction1}\leavevmode

With respect to integration of $1$-forms, any $1$-cycle $z$ of $X_L \cup_f X_R$ can be represented by a triple:
\bea
\label{triplerepcycle}
\left( c_L^\circ , c_R^\circ , \mathfrak{z}_{\Sigma_R} \right) \in C_2(X_L, \partial X_L) \times C_2(X_R, \partial X_R) \times Z_1(\Sigma_R) \, ,
\eea
so that:

 $\bullet$ the right representative of $z$ is:
\bea
\left( \partial_\circ^\circ c_L^\circ , \partial_\circ^\circ c_R^\circ + \partial_\circ^\Sigma c_R^\circ - f(\partial_\circ^\Sigma c_L^\circ) + \mathfrak{z}_{\Sigma_R} \right) \, , 
\eea

 $\bullet$ for any $\omega \in \Omega(X_L \cup_f X_R)$ we have:
\bea
\oint_z \omega = \int_{c_L^\circ} d \omega_L + \int_{c_R^\circ} d \omega_R + \oint_{\mathfrak{z}_{\Sigma_R}} \omega_R^\Sigma \, ,
\eea
A triple $\mathfrak{c} = \left( c_L^\circ , c_R^\circ , \mathfrak{z}_{\Sigma_R} \right)$ has above will be referred to as a \textbf{DB $1$-cycle} of $X_L \cup_f X_R$.

\end{prop}

\noindent To prove this property we must identify ambiguities in the DB $1$-cycles which represent a given $1$-cycle and show that the integrals along these ambiguities are zero. In other words we must identify the DB $1$-cycles which represent the trivial (i.e. zero) $1$-cycle of $X_L \cup_f X_R$. It is quite easy to check that these DB $1$-cycles are of the form:
\bea
\label{trivial1cycle}
\left( \mathfrak{z}_L^\circ , \mathfrak{z}_R^\circ , f(\partial_\circ^\Sigma \mathfrak{z}_L^\circ) - \partial_\circ^\Sigma \mathfrak{z}_R^\circ \right) \, ,
\eea
where $\mathfrak{z}_L^\circ$ and $\mathfrak{z}_R^\circ$ are relative $2$-cycles. Finally, it is also easy to check that a DB $1$-cycle $\left(c_L^\circ , c_R^\circ , \mathfrak{z}_{\Sigma_R} \right)$ represents a $1$-boundary of $X_L \cup_f X_R$ if and only if $\mathfrak{z}_{\Sigma_R}$ is a $1$-boundary of $\Sigma_R$.

Following the same logic, let us try to extend the above to DB cocycles and classes. We already know that in this case integration is an $\mathbb{R}/\mathbb{Z}$-valued linear functional.

\begin{defi}\label{integDBsalong1cycles}\leavevmode

The integral of a DB $1$-cocycle $A = (A_L,A_R,\Lambda_{\Sigma_R})$ along a $1$-cycles $z = \left( z_L , z_R  \right)$ is defined as:
\bea
\label{intAalong1cycle}
\oint_z A := \int_{z_L} A_L + \int_{z_R} A_R - \oint_{(\partial z_R)^\Sigma} \Lambda_{\Sigma_R} \, ,
\eea
with $(\partial z_R)^\Sigma = \partial_\circ^\Sigma z_R^\circ + \partial_\Sigma^\Sigma z_R^\Sigma$.

\end{defi}

\noindent The third term in the right-hand side of (\ref{intAalong1cycle}) is meaningful because $\partial \left( (\partial z_R)^\Sigma \right) = 0$, this term ensuring that the integral is defined modulo integers. Indeed, the integral of $A$ along any representative $(u_L^\Sigma , f(u_L^\Sigma))$ of the zero $1$-cycle of $X_L \cup_f X_R$ is:
\bea
\int_{f(u_L^\Sigma)} \bar{d} \Lambda_{\Sigma_R} - \oint_{\partial_\Sigma^\Sigma f(u^\Sigma)} \Lambda_{\Sigma_R} \, .
\eea
Since $\Lambda_R^\Sigma$ is a DB class, this number is an integer which not necessarily vanishes unlike the case of forms.

The next step would be to check that the integral along $z$ of a DB ambiguity like (\ref{1DBambig}) is an integer so that integration along $1$-cycles extends to DB classes as an $\mathbb{R}/\mathbb{Z}$-valued functional. We leave it as an exercise and will return to it in a moment. Like for $1$-forms we can use the fact that $z_L^\circ = \partial_\circ^\circ c_L^\circ$ and $z_R^\circ = \partial_\circ^\circ c_R^\circ$ for some $2$-chains $c_L^\circ$ and $c_R^\circ$ in order to rewrite the integral of a DB cocycle along a $z$. After some algebraic juggles, we find that:
\bea
\int_{c_L^\circ} d A_L + \int_{c_R^\circ} d A_R + \oint_{\mathfrak{z}_{\Sigma_R}} A_R^\Sigma \, .
\eea
where we also use the fact that the curvature $\bar{d} \Lambda_\Sigma$ has integral periods. Of course, the above expression must be considered in $\mathbb{R}/\mathbb{Z}$. Then, like for forms the integral of a DB $1$-cocycle along a $1$-cycle $z$ can be written with the help of a DB $1$-cycle $\left(c_L^\circ , c_R^\circ , \mathfrak{z}_{\Sigma_R} \right)$. Moreover, by applying the above integral formula to a DB $1$-cycle $\left( \mathfrak{z}_L^\circ , \mathfrak{z}_R^\circ , f\left( \partial_\circ^\Sigma \mathfrak{z}_L^\circ \right) - \partial_\circ^\Sigma \mathfrak{z}_R^\circ \right)$, known to represent the zero $1$-cycle, we obtain:
\bea
\int_{\mathfrak{z}_L^\circ} d A_L + \int_{\mathfrak{z}_R^\circ} d A_R + \oint_{f( \partial_\circ^\Sigma \mathfrak{z}_L^\circ) - \partial_\circ^\Sigma \mathfrak{z}_R^\circ } \! \! \! \! \! \! \! \! \! \! \! \! \! \! \! \! \! \! \! \! \! \! \!  A_R^\Sigma \qquad \; = \int_{f(\partial_\circ^\Sigma \mathfrak{z}_L^\circ)} \bar{d} \Lambda_{\Sigma_R} \egzz 0 \, ,
\eea
which correctly defines integration along a $1$-cycles as a $\mathbb{R}/\mathbb{Z}$-valued linear functional on the $\mathbb{Z}$-module of DB $1$-cocycles. Let us recall that: 1) $\partial_\circ^\circ \mathfrak{z}_L^\circ = 0 = \partial_\circ^\circ \mathfrak{z}_R^\circ$ by construction; 2) $f(\partial_\circ^\Sigma \mathfrak{z}_L^\circ)$ is a cycle because $\partial_\Sigma^\Sigma f(\partial_\circ^\Sigma \mathfrak{z}_L^\circ) = - f(\partial_\circ^\Sigma \partial_\circ^\circ \mathfrak{z}_L^\circ) = - f(0)$; 3) $\bar{d} \Lambda_{\Sigma_R}$ has integral periods. It is then a simple exercise to check that the integral of a DB ambiguity along a $1$-cycle is an integer. Property \ref{preDBconstruction1} thus extends to DB cocycles and classes as follows.

\begin{prop}\label{preDBconstruction2}\leavevmode

Let $z$ be a $1$-cycle of $X_L \cup_f X_R$ with $\mathfrak{c} = \left( c_L^\circ , c_R^\circ , \mathfrak{z}_{\Sigma_R} \right)$ a DB $1$-cycle representing $z$, and let $A = (A_L,A_R,\Lambda_{\Sigma_R})$ be a DB $1$-cocycle of $X_L \cup_f X_R$.

1) The evaluation of $\mathfrak{c}$ on $A$, defined as:
\bea
\label{intAalongzviatriple}
\mathfrak{c} \eval{A} := \int_{c_L^\circ} d A_L + \int_{c_R^\circ} d A_R + \oint_{\mathfrak{z}_{\Sigma_R}} A_R^\Sigma \, ,
\eea
is equal, modulo integers, to the integral of $A$ along $z$. In other words, we have: 
\bea
\mathfrak{c} \eval{A} \egzz \oint_z A \, .
\eea

2) With respect to integration of $1$-forms and evaluation of DB $1$-cocycles, two DB $1$-cycles $\left( c_L^\circ , c_R^\circ , \mathfrak{z}_{\Sigma_R} \right)$ and $\left( \tilde{c}^\circ , \tilde{c}_R^\circ , \tilde{\mathfrak{z}}_{\Sigma_R} \right)$ represent the same $1$-cycle if and only if they differ by a DB $1$-cycle of the form:
\bea
\left( \mathfrak{z}_L^\circ , \mathfrak{z}_R^\circ , f( \partial_\circ^\Sigma \mathfrak{z}_L^\circ ) - \partial_\circ^\Sigma \mathfrak{z}_R^\circ \right) \, ,
\eea
where $\mathfrak{z}_L^\circ$ and $\mathfrak{z}_R^\circ$ are \textbf{relative} $2$-cycles. Indeed, the evaluation of $A$ on such DB $1$-cycle as given by (\ref{intAalongzviatriple}) is zero in $\mathbb{R}/\mathbb{Z}$ and therefore correctly represents the integral of $A$ along the trivial (i.e. zero) $1$-cycle of $X_L \cup_f X_R$.

3) With respect to integration of $1$-forms and evaluation of DB $1$-cocycles, a DB $1$-cycle $\left(c_L^\circ , c_R^\circ , \mathfrak{z}_{\Sigma_R} \right)$ represents a $1$-boundary if and only if $\mathfrak{z}_{\Sigma_R} = \partial_\Sigma^\Sigma c_R^\Sigma$, the corresponding boundary  being then $ \partial \left( c_L^\circ , c_R^\circ + c_R^\Sigma \right)$. This gives rise to an equivalence relation on DB $1$-cycles representing $1$-cycles of $X_L \cup_f X_R$.

4)Integration of DB $1$-cocycles along $1$-cycles goes to classes. This allows to see any $1$ cycle of $X_L \cup_f X_R$ as an $\mathbb{R}/\mathbb{Z}$-valued functional over $H_D^1(X_L \cup_f X_R)$, which yields the well known inclusion: 
\bea
Z_1(X_L \cup_f X_R) \subset H_D^1(X_L \cup_f X_R)^\star \, .
\eea
Then, evaluation of DB $1$-cycles of $X_L \cup_f X_R$ on DB $1$-cocycles of $X_L \cup_f X_R$ also goes to classes, for both DB $1$-cycles and DB $1$-cocycles.

\end{prop}

For the sake of completeness, let us precise that the evaluation of a DB $3$-cocycle $G = \left( G_L , G_R , \upsilon_{\Sigma_R} \right)$ along $X_L \cup_f X_R$ is defined as:
\bea
\oint_{X_L \cup_f X_R} \hspace{-0.3cm} G := \int_{X_L} G_L +  \int_{X_R} G_R + \oint_{\Sigma_R} \upsilon_{\Sigma_R} \, .  
\eea
A simple computation shows that the integral along $X_L \cup_f X_R$ of an ambiguity is zero modulo integer. The above definition is very similar to (\ref{intAalong1cycle}) as $X_L$ and $X_R$ are relative $3$-cycles and $(\partial X_R)^\Sigma = \Sigma_R$. Furthermore, the $3$-cycles of $X_L \cup_f X_R$ are just $\left( n X_L  , n X_R \right)$.

\subsection{Pontryagin duality}

To make contact with the previous subsection, let us consider a DB $1$-cycle $\left(c_L^\circ , c_R^\circ , \mathfrak{z}_{\Sigma_R} \right)$ representing a $1$-cycle $z$ of $X_L \cup_f X_R$. Its first two components, $c_L^\circ$ and $c_R^\circ$, obviously define relative $2$-currents, i.e. elements of $\Omega^2(X,\partial X)^\diamond$, $j_{c_L^\circ}$ and $j_{c_R^\circ}$. The third component, $\mathfrak{z}_{\Sigma_R}$, defines an element of $\Omega^1(\Sigma)^\diamond_\mathbb{Z}$. Thanks to the exact sequence (\ref{sequenceSigmaPontDual}) we can see $\mathfrak{z}_{\Sigma_R}$ (or its de Rham current) as the generalized curvature of some $\varsigma_{\Sigma_R} \in H^1_D(\Sigma)^\star$ and replace the original triple $\left(c_L^\circ , c_R^\circ , \mathfrak{z}_{\Sigma_R} \right)$ by the triple $\left(j_{c_L^\circ} , j_{c_R^\circ} , \varsigma_{\Sigma_R} \right)$. Let us note that the same exact sequence tells us that $\varsigma_{\Sigma_R}$ is not unique. To confirm this idea, let us recall that integration provides an injection $Z_1(\Sigma) \rightarrow H_D^1(\Sigma)^\star$, so that $\varsigma_{\Sigma_R}$ is just the image of the $1$-cycle $\mathfrak{z}_{\Sigma_R}$ under this injection. However, $\varsigma_{\Sigma_R}$ must not be confused with the de Rham current of $\mathfrak{z}_{\Sigma_R}$ as this current is actually the generalized curvature $\delta_{-1}^\star \varsigma_{\Sigma_R}$ of $\varsigma_{\Sigma_R}$.

All these remarks suggest the following definition.

\begin{defi}\label{DB1currentandcurv}\leavevmode

1) A \textbf{DB $1$-current} of $X_L \cup_f X_R$ is a triple:
\bea
\label{defDB1current}
\mathcal{J} = \left( J_L^\circ , J_R^\circ , \varsigma_{\Sigma_R} \right) \in \Omega^2(X_L,\partial X_L)^\diamond \times \Omega^2(X_R,\partial X_R)^\diamond \times H_D^1(\Sigma_R)^\star \, .
\eea

2) The \textbf{generalized curvature} of a DB $1$-current $\mathcal{J} = \left( J_L^\circ , J_R^\circ , \varsigma_{\Sigma_R} \right)$ is the element of $\Omega^1(X_L \cup_f X_R)^\diamond_\mathbb{Z}$ defined as:
\bea
\label{gencurvJ}
\delta_{-1}^\star \mathcal{J} = ( d_\circ^{\dag ^\circ} J_L^\circ , d_\circ^{\dag ^\circ} J_R^\circ + d_\circ ^{\dag \Sigma} J_R^\circ  - f( d_\circ^{\dag \Sigma } J_L^\circ) + \delta_{-1}^* \varsigma_{\Sigma_R} ) \, ,
\eea 
$\delta_{-1}^\star \varsigma_{\Sigma_R} \in \Omega^1(\Sigma)^\diamond_\mathbb{Z}$ being the generalized curvature of $\varsigma_{\Sigma_R} \in H_D^1(\Sigma_R)^\star$. A DB $1$-current with zero genralized curvature is said to be \textbf{flat}, in analogy with the smooth case.

3) A \textbf{DB $1$-current ambiguity} is a DB $1$-current of the form:
\bea
 \mathfrak{j} = \left( \mathfrak{j}_L^\circ , \mathfrak{j}_R^\circ , \xi_{\Sigma_R} \right) \, ,
\eea
with $\mathfrak{j}_L^\circ \in \Omega^2(X_L,\partial X_L)^\diamond_\mathbb{Z}$, $\mathfrak{j}_R^\circ \in \Omega^2(X_R,\partial X_R)^\diamond_\mathbb{Z}$ and $\xi_R^\Sigma \in H_D^1(\Sigma)^\star$ such that:
\bea
\label{constDBambig}
\delta_{-1}^\star \xi_R^\Sigma = f( d_\circ^{\dag \Sigma } \mathfrak{j}_L^\circ) - d_\circ^{\dag \Sigma } \mathfrak{j}_R^\circ \, .
\eea

4) Two DB $1$-currents of $X_L \cup_f X_R$ that differ by a DB ambiguity are said to be DB equivalent, and the equivalent class of a DB $1$-current of $X_L \cup_f X_R$ is simply called a \textbf{generalized DB class} of $X_L \cup_f X_R$. The set of generalized DB classes of $X_L \cup_f X_R$ is the Pontryagin dual, $H_D^1(X_L \cup_f X_R)^\star$, of $H_D^1(X_L \cup_f X_R)$.

5) Let $\mathcal{J} = \left( J_L^\circ , J_R^\circ , \varsigma_{R_\Sigma} \right)$ a DB $1$-current and $A = (A_L , A_R , \Lambda_{R_\Sigma})$ a DB $1$-cocycle. The evaluation of $\mathcal{J}$ on $A$ is defined as:
\bea
\label{evalDBcurrentonDBcocycle}
\mathcal{J} \eval{A} := J_L^\circ \eval{d A_L} + J_R^\circ \eval{d A_R} + \left( \delta_{-1}^\star \varsigma_{R_\Sigma} \right) \eval{A_R^\Sigma} \, .
\eea
As an $\mathbb{R}/\mathbb{Z}$-valued linear functional this evaluation goes to DB classes, for both DB currents and DB cocycles, thus realizing Pontryagin duality.

\end{defi}

\noindent By construction, the de Rham $1$-current $\delta_{-1}^\star \mathcal{J}$ fulfills relations (\ref{closeddRcurrent}) and hence is closed. Moreover, it belongs to $\Omega^1(X_L \cup_f X_R)^\diamond_\mathbb{Z}$. Indeed, for any $\alpha \in \Omega^1_{\mathbb{Z}}(X_L \cup_f X_R)$ decomposed according to Property \ref{closed2formsintgperiods} we have:
\bea
\label{periodsingularcurvature}
\delta_{-1}^\star \mathcal{J} \eval{\alpha} = \sum_{a=1}^{g} (Q \vec{N})^a (\delta_{-1}^\star \varsigma_{\Sigma_R}) \eval{j_{\mu_a^{\Sigma_R}}^\infty} \, ,
\eea
and since $\delta_{-1}^\star \varsigma_{\Sigma_R} \in \Omega^1(\Sigma_R)^\diamond_\mathbb{Z}$, the above sum is an integer. We leave it as an exercise to check the similarity between the above formula and the integral of the curvature along a $2$-cycle of $X_L \cup_f X_R$. Let note that:
\bea
\delta_{-1}^\star \mathcal{J} = \left( 0 , \delta_{-1}^* \varsigma_{\Sigma_R} \right) + d^\dag \left( J_L^\circ , J_R^\circ \right) \, ,
\eea
which provides another way to see why generalized curvature are closed and $\mathbb{Z}$-valued on $\Omega^1_\mathbb{Z}(X_L \cup_f X_R)$. The choice for DB ambiguities is based on Property \ref{H1Dualisquotient}. Moreover, because of the constraint (\ref{constDBambig}) they must fulfill, DB ambiguities are flat: $\delta_{-1}^\star \mathfrak{j} = 0$. Hence, the generalized curvature of a generalized DB class $\mathfrak{J}$ is well-defined as the generalized curvature of any of its representatives $\mathcal{J}$: $\delta_{-1}^\star \mathfrak{J} = \delta_{-1}^\star \mathcal{J}$.

By comparing evaluations formulas (\ref{intAalongzviatriple}) and (\ref{evalDBcurrentonDBcocycle}) we deduce that a $1$-cycle $z$ represented by a DB $1$-cycle $\mathfrak{c} = \left( c_L^\circ , c_R^\circ , \mathfrak{z}_{\Sigma_R} \right)$ should define a DB $1$-current of $X_L \cup_f X_R$ and hence a generalized DB class. For this purpose, let us chose a DB class $\varsigma_{\Sigma_R} \in H_D^0(\Sigma_R)^\star$ such that $\delta_{-1}^\star \varsigma_{\Sigma_R} = j_{\mathfrak{z}_{\Sigma_R}}$. Then, $\mathcal{J}_z = (j_{c_L^\circ} , j_{c_R^\circ} , \varsigma_{\Sigma_R})$ is a DB $1$-current whose generalized curvature is $j_z$, and we have: 
\bea
\mathfrak{c} \eval{A} = \mathcal{J}_z \eval{A} \, ,
\eea
for any DB $1$-cocycle $A$ of $X_L \cup_f X_R$. The generalized DB class of $\mathcal{J}_z$ is said to be canonically associated with $z$ and as such usually identified with $z$ itself. Let us point out that the de Rham current $j_z$ of a $1$-cycle $z$ is the curvature of many inequivalent classes of DB $1$-currents, $\mathfrak{J}_z$ being one of these classes. So, we have the following property \cite{BGST05}: 

\begin{prop}\label{cyclemapprop}\leavevmode

There is a canonical injection:
\bea
Z_1(X_L \cup_f X_R) \to H_D^1(X_L \cup_f X_R)^\star \, ,
\eea 
usually referred to as a \textbf{cycle map}. The generalized DB class canonically associated with a $1$-cycle $z$ is denoted by $\mathfrak{J}_z$. A DB $1$-current in the class of $\mathfrak{J}_z$ is said to be \textbf{associated} with $z$.

\end{prop}

\noindent If $z$ is a homologically trivial $1$-cycle so that there exists a $2$-chain $c = \left( c_L^\circ , c_R^\circ + c_R^\Sigma \right)$ such that $z = \partial c$, then the  generalized DB class canonically associated with $z$ is the class of the DB $1$-current:
\bea
\label{trivcyclDBcurrent}
\mathcal{J}_z = \mathcal{J}_{\partial c} = \left( j_{c_L^\circ} , j_{c_R^\circ} , \overline{j_{c_R^\Sigma}} \right) \, ,
\eea
where $\overline{j_{c_R^\Sigma}} \in H_D^1(\Sigma)^\star$ is defined by:
\bea
\forall a_\Sigma \in H_D^1(\Sigma) , \quad \overline{j_{c_R^\Sigma}} \eval{a_\Sigma} := \overline{j_{c_R^\Sigma} \eval{\bar{d} a_\Sigma}}  \, .
\eea
This last relation obviously extend to all de Rham $2$-currents of $\Sigma$, thus defining a canonical mapping $\overline{^{^{\quad ^{}}}} : \Omega^2(\Sigma)^\diamond \to H_D^1(\Sigma)^\star$. The kernel of this mapping is obviously $\Omega^2(\Sigma)_\mathbb{Z}^\diamond$ so that $\overline{^{^{\quad ^{}}}} :  \Omega^2(\Sigma)^\diamond/\Omega^2(\Sigma)_\mathbb{Z}^\diamond \to H_D^1(\Sigma)^\star$ is injective. 

Let us complete these first remarks concerning DB $1$-currents and generalized DB classes with the following property about DB ambiguities which is a consequence of Property \ref{closedsurface}.

\begin{prop}\label{decompDBambig}\leavevmode

Let $\left( \mathfrak{j}_L^\circ , \mathfrak{j}_R^\circ , \xi_R^\Sigma \right)$ be a DB $1$-current ambiguity. We have:
\bea
\mathfrak{j}_L^\circ = \sum_{a=1}^g m_L^a j_{D_a^L} + d_\circ^{\dag ^\circ} q_L^\circ \qquad \text{and} \qquad \mathfrak{j}_R^\circ = \sum_{a=1}^g m_R^a j_{D_a^R} + d_\circ^{\dag ^\circ} q_L^\circ \, ,
\eea
where $m_L^a, m_R^a \in \mathbb{Z}$, $q_L^\circ \in \Omega^3(X_L,\partial X_L)^\diamond$ and $q_R^\circ \in \Omega^3(X_R,\partial X_R)^\diamond$, and such that:
\bea
\begin{aligned}
\delta_{-1}^\star \xi_R^\Sigma = - \sum_{a=1}^g (P \vec{m}_L)^a j_{\lambda_a^{\Sigma_R}} - & \sum_{a=1}^g (Q \vec{m}_L - \vec{m}_R)^a j_{\mu_a^{\Sigma_R}} + \\ 
& d^{\dag \Sigma} \left(  d_\circ^{\dag ^\Sigma} q_R^\circ - f(d_\circ^{\dag ^\Sigma} q_L^\circ) - \sum_{a=1}^g \vec{m}_L^a j_{\psi_a^{\Sigma_R}} \right) 
\end{aligned} \, .
\eea

\end{prop}

Now that we have identified the objects to work with, their ambiguities and the classes thus obtained, we can state the main result of this article:

\begin{prop}\label{DB1currenclassseq}\leavevmode

The set $H_D^1(X_L \cup_f X_R)^\star$ is a $\mathbb{Z}$-module which can be embedded into the following exact sequence:
\bea
\label{curvatureexactsequence}
0 \rightarrow H^1(X_L \cup_f X_R,\mathbb{R}/\mathbb{Z}) \rightarrow H_D^1(X_L \cup_f X_R)^\star \rightarrow \Omega^1(X_L \cup_f X_R)_\mathbb{Z}^\diamond  \rightarrow 0 \, .
\eea

\end{prop}

\noindent To prove the above property we start by recalling that a DB ambiguity $\mathfrak{j} = \left( \mathfrak{j}_L^\circ , \mathfrak{j}_R^\circ , \xi_{\Sigma_R} \right)$ is flat. Hence, the curvature operation defined on DB $1$-cocycles goes to DB classes thus yielding the morphism $\delta_{-1}^\star : H_D^1(X_L \cup_f X_R)^\star \rightarrow \Omega^1(X_L \cup_f X_R)_\mathbb{Z}^\diamond$. This morphism is surjective. Indeed, let $\mathfrak{F}$ be an element of $\Omega^1(X_L \cup_f X_R)_\mathbb{Z}^\diamond$. Since $H^2(X) \simeq H_1(X,\partial X)$ is generated by the longitudes $\lambda_a$ of $X$, we can always write $\mathfrak{F}$ as $\left( 0 , \sum_a n^a j_{\lambda_a^R} \right) + d^\dag S$ with $S = \left( S_L^\circ , S_R^\circ + S_R^\Sigma \right)$. The triple $\left( S^\circ_L , S^\circ_R + \sum_a n^a j_{A_a^R} , \varsigma_{\Sigma_R} + \overline{S_R^\Sigma} \right)$ such that $\delta_{-1}^\star \varsigma_{\Sigma_R} = \sum_a n^a j_{\lambda^{\Sigma_R}_a} + d^{\dag \Sigma}_\Sigma S^\Sigma_R$ is then a DB $1$-current whose generalized curvature is $\mathfrak{F}$. The existence of $\varsigma_{\Sigma_R} \in H^1_D(\Sigma)^\star$ is ensured by the exactness of the sequence (\ref{sequenceSigmaPontDual}) and the fact that $j_{\lambda^{\Sigma_R}_a}$ and $d^{\dag \Sigma}_\Sigma S^\Sigma_R$ both belong to $\Omega^1(\Sigma)^\diamond_\mathbb{Z}$. Hence, we obtain the exact sequence:
\bea
H_D^1(X_L \cup_f X_R)^\star \rightarrow \Omega^1(X_L \cup_f X_R)_\mathbb{Z}^\diamond  \rightarrow 0 \, .
\eea

Let us now consider a flat DB $1$-current $\mathcal{J} = \left( J_L^\circ , J_R^\circ , \varsigma_{\Sigma_R} \right)$ which is not in the trivial class. The curvature of $\mathcal{J}$ is then vanishing and we have:
\bea
\label{nontrivialflatDBcurrent}
d_\circ^{\dag ^\circ} J_L^\circ = 0 = d_\circ^{\dag ^\circ} J_R^\circ \, ,
\eea
which means that $J_L^\circ$ and $J_R^\circ$ are relatively closed. Since the meridian disks generate $H_2(X,\partial X)$ the relatively closed currents $J_L^\circ$ and $J_R^\circ$ are necessarily of the form:
\bea
J_L^\circ = \sum_{a=1}^g x_L^a j_{D_a} + d_\circ^{\dag ^\circ} q_L^\circ \qquad \text{and} \qquad J_R^\circ = \sum_{a=1}^g x_R^a j_{D_a} + d_\circ^{\dag ^\circ} q_R^\circ \, ,
\eea
the coefficients $x_L^a$ and $x_R^a$ being real numbers. Furthermore, the flatness of $\mathcal{J}$ implies that $\delta_{-1}^\star \varsigma_{\Sigma_R} = f\left( d_\circ^{\dag \Sigma } J_L^\circ  \right) - d_\circ ^{\dag \Sigma} J_R^\circ$ with $\delta_{-1}^\star \varsigma_{\Sigma_R} \in \Omega^1(\Sigma)^\diamond_\mathbb{Z}$, and since this last space is generated by the median and longitude cycles of $\Sigma$, we deduce that there exist $\vec{L} , \vec{M} \in \mathbb{Z}^g$ such that:
\bea
\label{constraintnontrivflat}
\sum_{a=1}^g (P \vec{x}_L)^a j_{\lambda_a^{\Sigma_R}} + \sum_{a=1}^g (Q \vec{x}_L - \vec{x}_R)^a j_{\mu_a^{\Sigma_R}}  = \sum_{a=1}^g L^a_\Sigma j_{\lambda_a^{\Sigma_R}} + \sum_{a=1}^g M^a_\Sigma j_{\mu_a^{\Sigma_R}} \, ,
\eea
and hence that:
\bea
\label{constraintsx}
\left\{ \begin{aligned}
& P \vec{x}_L = \vec{L} \\
& Q \vec{x}_L - \vec{x}_R = \vec{M}
\end{aligned} \right. \, .
\eea
We can add to our initial DB $1$-current any DB $1$-current ambiguity whose generic form is given in Property \ref{decompDBambig} which induces the shifts $\vec{x}_L \rightarrow \vec{x}_L + \vec{m}_L$, $\vec{x}_R \rightarrow \vec{x}_R + \vec{m}_R$ and $\vec{L} \rightarrow \vec{L} + P \vec{m}_L$. Hence, at the level of DB classes, we must consider $\vec{\theta}_L, \vec{\theta}_R \in (\mathbb{R} / \mathbb{Z})^g$ instead of $\vec{x}_L, \vec{x}_R \in \mathbb{R}^g$. Note that $\vec{L} \in (\mathbb{Z}^g / \Ima P) = \coker P$. The first of the above constraint can be written in term of the Pontryagin dual, $P^\star$, of $P$ as:
\bea
P^\star \vec{\theta}_L = \vec{0} \, ,
\eea
which means that $\vec{\theta}_L \in \ker P^\star$. Furthermore, since on the one hand $\ker P^\star = (\coker P)^\star$ and on the second hand $\coker P = H_1(X_L \cup_f X_R)$, we conclude that $\vec{\theta}_L \in H^1(X_L \cup_f X_R,(\mathbb{R}/\mathbb{Z})) = H_1(X_L \cup_f X_R)^\star$, which means that the set of DB classes of a flat DB $1$-currents is isomorphic to $H^1(X_L \cup_f X_R,(\mathbb{R}/\mathbb{Z}))$. This achieves the demonstration of Property \ref{DB1currenclassseq}. Note that the constraints (\ref{constraintsx}) are the same as those that yield the extension to the left of the exact sequence (\ref{exactsequenceHD}) as explained in \cite{T19}. Moreover, the second constraints in (\ref{constraintsx}) plays no role in the proof of Property \ref{DB1currenclassseq}. It just determines $\vec{\theta}_R$ in term of $\vec{\theta}_L$. In the particular case of the lens space $\mathcal{L}(p,q)$ with Heegaard splitting such that $\partial X = S^1 \times S^1$ we obtain $\theta_L = L/p$ and $\theta_R = q L / p$. A flat DB $1$-current of $\mathcal{L}(p,q)$ is then of the form $\left( \tfrac{L}{p} j_{D^L} , \tfrac{qL}{p} j_{D^R} , - \zeta_L - \tfrac{L}{p} j_{\psi^{\Sigma_R}} \right)$ with $\zeta_L \in H_D^0(S^1 \times S^1)^\star$ such that $\delta_{-1}^\star \zeta_L = L j_{\lambda^{\Sigma_R}}$.

Thanks to exact sequence (\ref{curvatureexactsequence}) we can see the $\mathbb{Z}$-module $H_D^1(X_L \cup_f X_R)^\star$ as a discrete fiber space over the space of generalized curvatures $\Omega^1(X_L \cup_f X_R)_\mathbb{Z}^\diamond$, the group $H^1(X_L \cup_f X_R,\mathbb{R}/\mathbb{Z})$ acting on fibers as a group of translation. As a closed de Rham $1$-current, any generalized curvature defines an element in $F_1(X_L \cup_f X_R)$. Hence, if we have chosen a set of de Rham $1$-currents, $j_k$ ($k=1, \cdots , b_1$), which generate  $F_1(X_L \cup_f X_R)$, then any generalized curvature $\mathfrak{F}$ can be written as:
\bea
\label{decompossegencurv}
\mathfrak{F} = \sum_{k=1}^{b_1} m^k j_k + d^\dag S \, ,
\eea
with $S \in \Omega^2(X_L \cup_f X_R)^\diamond$. To make this decomposition unique we must consider the class of $S$ in the quotient space $\Omega^2(X_L \cup_f X_R)^\diamond / \Omega^2(X_L \cup_f X_R)_0^\diamond$. This defines an isomorphism between $\Omega^1(X_L \cup_f X_R)_\mathbb{Z}^\diamond$ and $F_1(X_L \cup_f X_R) \oplus \left( \Omega^2(X_L \cup_f X_R)^\diamond / \Omega^2(X_L \cup_f X_R)_0^\diamond \right)$. Now extending what was done for $2$-chains to $2$-currents, we associate to the $2$-current $S = \left( S_L^\circ , S_R^\circ + S_R^\Sigma \right)$ the DB $1$-current $\mathcal{S} = \left( S_L^\circ , S_R^\circ , \overline{S_R^\Sigma} \right)$ which fulfills $\delta_{-1}^\star \mathcal{S} = d^\dag S$. However, the class of $S$ in $\Omega^2(X_L \cup_f X_R)^\diamond / \Omega^2(X_L \cup_f X_R)_0^\diamond$ must not be confused with the DB class of $\mathcal{S}$. Indeed, if we add to $S$ a closed $2$-current $T_0$, then the the DB $1$-current associate with $S + T_0$ is equivalent to $\mathcal{S}$ if and only if $T_0 \in \Omega^2(X_L \cup_f X_R)_\mathbb{Z}^\diamond$. In other words, if we see the DB class of $\mathcal{S}$   as an element of $\Omega^2(X_L \cup_f X_R)_0^\diamond / \Omega^2(X_L \cup_f X_R)_\mathbb{Z}^\diamond$, then the DB class of $\mathcal{T}_0$, where $T_0 \in \Omega^2(X_L \cup_f X_R)_0^\diamond$, is an element of $\Omega^2(X_L \cup_f X_R)_0^\diamond / \Omega^2(X_L \cup_f X_R)_\mathbb{Z}^\diamond$. The DB $1$-currents $\mathcal{S}$ and $\mathcal{S} + \mathcal{T}_0$ are generically inequivalent although they have the same generalized curvature: $d^\dag S$. To end this remark, let us point out that the situation is quite difference when $S$ is the current of a $2$-chain. Indeed, we saw that the DB class of the current of a $2$-chain $c$ is the class of DB $1$-current $\mathcal{J}_\mathfrak{c}$ defined in (\ref{trivcyclDBcurrent}). The integral nature of the $2-$chains is then a strong constraint on the associated DB classes, thus making this association into a canonical one for these particular $2$-currents. This is the cycle map.

\vspace{0.3cm}

Let us put all these results together into the following main property which extends the decomposition property of smooth DB classes discussed in \cite{T19}.

\begin{prop}\label{decompsingularDBclasses}\leavevmode

For each $\textbf{n}_f \in F_1(X_L \cup_f X_R)$ we pick a representative $\vec{n}_f \in \mathbb{Z}^g$,  with $\vec{n}_f = \vec{0}$ representing $\textbf{n} = \textbf{0}$. Similarly, for each $\boldsymbol{\theta} \in H_{P^\dag}^\star \cong H_1(X_L \cup_f X_R)^\star$ we pick a representative $\vec{\theta} = \vec{\theta}_f + \vec{\theta}_\tau$ as in Property \ref{pontdualcokerP}, with $\vec{\theta} = \vec{0}$ representing $\boldsymbol{\theta} = \textbf{0}$. 

Any $\mathfrak{J} \in H_D^1(X_L \cup_f X_R)^\star$ is represented in a unique way by a DB 1-current of the form:
\bea
\label{decompuniversal}
\mathcal{J}_{( \vec{n}_f , \vec{\theta} , S)} = \mathcal{J}_{\vec{n}_f} +  \mathcal{J}_{\vec{\theta}} + \mathcal{S} \, ,
\eea
where:
\bea
\label{defcurrentdecomp}
\left\{ \begin{aligned}
& \mathcal{J}_{\vec{n}_f} := \left( 0 , \sum_{a=1}^{g} n_f^a j_{A_a^R} , \, \zeta_{\vec{n}_f} \right) \\
& \mathcal{J}_{\vec{\theta}} := \left( \sum_{a=1}^{g} \theta^a j_{D_a^L} \, , \, \sum_{a=1}^{g} (Q \vec{\theta})^a j_{D_a^R} \, , \, - \zeta_{P \vec{\theta}} - \sum_{a=1}^{g} \theta^a \overline{j_{\psi^{\Sigma_R}_a}}  \right) 
\end{aligned} \right. \, ,
\eea
with:

 $\bullet$ $\mathcal{J}_{( \vec{0} , \vec{0} , 0)}$ is the representative of $\mathfrak{J} = 0$;
  
 $\bullet$ $\mathcal{S} \in \Omega^2(X_L \cup_f X_R)^\diamond$ parametrizes an element of $\Omega^2(X_L \cup_f X_R)^\diamond / \Omega^2(X_L \cup_f X_R)_\mathbb{Z}^\diamond$ whose curvature is $d^\dag S$, with $S = \left( S_L^\circ , S_R^\circ + S_R^\Sigma \right)$;
 
 $\bullet$ $\zeta_{\vec{n}} \in H_D^1(\Sigma_R)^\star$ such that: 
\bea
\label{zetacurv}
\delta_{-1}^\star \zeta_{\vec{n}} = \sum\limits_{a=1}^g n^a j_{\lambda_a^{\Sigma_R}} \, .
\eea
Hence, we can consider $( \vec{n}_f , \vec{\theta} , S)$ as a parametrization of $H_D^1(X_L \cup_f X_R)^\star$.  Moreover, we have the following generalized curvatures:
\bea
\label{GencurvcomponentsofJ}
\left\{ \begin{aligned}
& \delta_{-1}^\star \mathcal{J}_{\vec{n}_f} = \left( 0 , \sum_{a=1}^{g} n_f^a j_{\lambda_a^R} \right) \\
& \delta_{-1}^\star \mathcal{J}_{\vec{\theta}} = 0 \\
& \delta_{-1}^\star \mathcal{S} =  d_\circ ^{\dag \Sigma} S_R^\circ  - f\left( d_\circ^{\dag \Sigma } S_L^\circ  \right) + d^{\dag \Sigma} S_R^\Sigma
\end{aligned} \right. \, .
\eea

\end{prop}

\noindent With respect to exact sequence (\ref{curvatureexactsequence}), $\delta_{-1}^\star \mathcal{J}_{\left( \vec{n}_f , \vec{\theta} , S \right)}$ is the base point of the fiber on which the DB class of $\mathcal{J}_{\left( \vec{n}_f , \vec{\theta} , S \right)}$ stands. The DB class of $\mathcal{J}_{\vec{\theta}_\tau}$ plays the same role as the torsion origins introduced in the standard construction \cite{GT2014}, and the DB class of $\mathcal{J}_{\vec{\theta}_f}$ is a (free) translation along the corresponding fiber. To clarify the role played by the representatives $\mathcal{S}$ we can refer to the discussion preceding Property \ref{decompsingularDBclasses}. Hence, let us fix some $u \in \Omega^2(X_L \cup_f X_R)^\diamond / \Omega^2(X_L \cup_f X_R)_0^\diamond$. We consider two cases:

\begin{itemize}
\item if $u$ admits a representative which is the $2$-current of a $2$-chain $c = \left( c_L^\circ , c_R^\circ + c_R^\Sigma \right)$ then we \textbf{decide} to represent $u$ by $j_c = \left( j_{c_L^\circ} , j_{c_R^\circ} + j_{c_R^\Sigma} \right)$, and the DB $1$-current we associate to $j_c$ is $\mathcal{C} = \left( j_{c_L^\circ} , j_{c_R^\circ} , \overline{j_{c_R^\Sigma}} \right)$. Then, the set of DB classes whose generalized curvature is $d^\dag j_c = j_{\partial c}$ is generated by the DB $1$-currents $\mathcal{C} + \mathcal{J}_{\vec{\theta}_f}$ when $[\vec{\theta}_f]$ runs through $\ker \hat{P}$.

\item if there is no current of $2$-chains representing $u$, then we select randomly a representative $2$-current $S = \left( S_L^\circ , S_R^\circ + S_R^\Sigma \right)$ and the DB $1$-current we associate to $S$ is $\mathcal{S} = \left( S_L^\circ , S_R^\circ , \overline{S_R^\Sigma} \right)$. The set of DB classes whose generalized curvature is $d^\dag S$ is generated by the DB $1$-currents $\mathcal{S} + \mathcal{J}_{\vec{\theta}_f}$ when $\vec{\theta}_f$ runs through $\ker P$.

\end{itemize}

\noindent It should be noted that by clarifying the role played by $\mathcal{S}$ we have at the same time shed light on the role played by the free translations $\mathcal{J}_{\vec{\theta}_f}$. As an exercise the reader can check that up to an ambiguity we have $\mathcal{J}_{P \vec{m}} = \mathcal{C}$ with $\delta_{-1}^\star \mathcal{J}_{P \vec{m}} = d^\dag j_c = j_{\partial c}$ for some $2$-chain $c$. This is consistent with the fact that $P \vec{m}$ is zero in $H_1(X_L \cup_f X_R) \cong \coker P$ and justify the choice made in Property \ref{decompsingularDBclasses} to represent the trivial class in $H_1(X_L \cup_f X_R)$ by $\vec{n} = \vec{0}$.

We complete our description of $H_D^1(X_L \cup_f X_R)^\star$ with the following property

\begin{prop}\label{defcohomologyDB}\leavevmode

For $\mathcal{J}_{\left( \vec{n}_f , \vec{\theta} , S \right)}$ a DB 1-current, let us write $\epsilon_\star \mathcal{J}_{\left( \vec{n}_f , \vec{\theta} , S \right)}$ the class of $\vec{n}_f + P \vec{\theta} \in \mathbb{Z}^g$ in $\coker P \cong H_1(X_L \cup_f X_R)$. At the level of generalized DB classes, this defines a surjective homomorphism:
\bea
\epsilon_\star : H_D^1(X_L \cup_f X_R)^\star \to H_1(X_L \cup_f X_R) \, ,
\eea
which can be extended to the left to form the exact sequence:
\bea
0 \rightarrow \Omega_\mathbb{Z}^2(X_L \cup_f X_R)^\star \rightarrow H_D^1(X_L \cup_f X_R)^\star \to H_1(X_L \cup_f X_R) \rightarrow 0 \, .
\eea

\end{prop}
\noindent This property, the detailed proof of which is left as an exercise, is a consequence of decomposition (\ref{decompuniversal}) and of the discussion which follows Property \ref{decompsingularDBclasses}.

We would like to end this section by looking closer at the DB $1$-currents associated with $1$-cycles of $X_L \cup_f X_R$. 
Let us first point out that if $z$ is a $1$-cycle and $\mathcal{J}$ is a DB $1$-current whose generalized curvature is $j_z$ then the generalized curvature of $\mathcal{J} + \mathcal{J}_{\vec{\theta}}$ is $j_z$ too. Hence, with respect to decomposition (\ref{decompuniversal}) the identification of the DB $1$-currents which are associated with a given $1$-cycle of $X_L \cup_f X_R$, in the sense of Property \ref{cyclemapprop}, is not totally obvious. However, as this identification will be essential in the context of Chern-Simons and BF Quantum Field theories, we must exhibit it properly.

Let $\textbf{n}$ be an element of $\coker P$ and let $\vec{n} \in \mathbb{Z}^g$ be a representative of $\textbf{n}$. According to Property \ref{pontdualcokerP}, we can write $\vec{n} = \vec{n}_f + \vec{n}_\tau = \vec{n}_f + P \vec{\theta}_\tau$ with $\vec{\theta}_\tau = \vec{m}/p$ for some $p \in \mathbb{Z}$ and $\vec{m} \in \mathbb{Z}^g$. Let us introduce the DB $1$-current:
\bea
\label{torsionfracchain}
\mathcal{C}_{\vec{\theta}_\tau} = \left( \sum_{a=1}^{g} \theta_\tau^a j_{D_a^L} , \sum_{a=1}^{g} (P \vec{\theta}_\tau)^a j_{A_a^R} + (Q \vec{\theta}_\tau)^a j_{D_a^R} , - \sum_{a=1}^{g}  \overline{\theta_\tau^a j_{\psi_a^{\Sigma_R}}} \right) \, .
\eea
According to decomposition (\ref{decompuniversal}), this DB $1$-current is of type $\mathcal{S}$ and not of type $\mathcal{J}_{\vec{\theta}}$, the de Rham $2$-current associated with $\mathcal{C}_{\vec{\theta}_\tau}$ being:
\bea
C_{\vec{\theta}_\tau} = \left( \sum_{a=1}^{g} \theta_\tau^a j_{D_a^L} , \sum_{a=1}^{g} (P \vec{\theta}_\tau)^a j_{A_a^R} + (Q \vec{\theta}_\tau)^a j_{D_a^R} - \theta_\tau^a j_{\psi_a^{\Sigma_R}} \right),
\eea
and we have:
\bea
\left\{ \begin{aligned}
& \delta_{-1}^\star \mathcal{C}_{\vec{\theta}_\tau} = \left( 0 , \sum_{a=1}^{g} n_\tau^a j_{\lambda_a^R} \right) = d^\dag C_{\vec{\theta}_\tau} \\
& \epsilon_\star \mathcal{C}_{\vec{\theta}_\tau} = 0
\end{aligned} \right. \, .
\eea
The first equality indicates that the generalized curvature of $\mathcal{C}_{\vec{\theta}_\tau}$ is the de Rham $1$-current of the $1$-cycle $\mathfrak{z}_{\vec{n}_\tau} = \left( 0 , \sum_{a=1}^{g} n_\tau^a j_{\lambda_a^R} \right)$ and the boundary of the de Rham $2$-current $C_{\vec{\theta}_\tau}$ which implies that $\mathfrak{z}_{\vec{n}_\tau}$ is a torsion cycle. As a matter of fact, although $C_{\vec{\theta}_\tau}$ is not a chain current, the $2$-current $p \kern1pt C_{\vec{\theta}_\tau}$ is the de Rham current of the $2$-chain:
\bea
c_{\vec{m}} = \left( \sum_{a=1}^{g} m^a D_a^L , \sum_{a=1}^{g} (P \vec{m})^a A_a^R + (Q \vec{m})^a D_a^R - m^a \psi_a^{\Sigma_R} \right) \, ,
\eea
which fulfills $\partial c_{\vec{m}} = p \kern1pt \mathfrak{z}_{\vec{n}_\tau}$. This also explains why we decided to write $C_{\vec{\theta}_\tau}$ instead of $S_{\vec{\theta}_\tau}$, the nature of the index $\vec{\theta}_\tau$ preventing a possible confusion with a chain current.

Let us now consider the DB $1$-current:
\bea
\mathcal{J}_{(\vec{n}_f,-\vec{\theta}_\tau,C_{\vec{\theta}_\tau})} = \mathcal{J}_{\vec{n}_f} - \mathcal{J}_{\vec{\theta}_\tau} + \mathcal{C}_{\vec{\theta}_\tau} \, .
\eea
Then, for any DB $1$-cocycle $A = (A_L , A_R , \Lambda_{\Sigma_R})$ we have:
\bea
\mathcal{J}_{(\vec{n}_f,- \vec{\theta}_\tau,C_{\vec{\theta}_\tau})} \eval{A} = \oint_{\mathfrak{z}_{\vec{n}}} A_R^\Sigma \, ,
\eea
with:
\bea
\mathfrak{z}_{\vec{n}} = \left( 0 , \sum_{a=1}^{g} n^a \lambda_a^R \right) \, .
\eea
Hence, the DB $1$- current $\mathcal{J}_{(\vec{n}_f,-\vec{\theta}_\tau,C_{\vec{\theta}_\tau})}$ is associated with the $1$-cycle $\mathfrak{z}_{\vec{n}}^R$. Note that:
\bea
\left\{ \begin{aligned}
& \delta_{-1}^\star \mathcal{J}_{(\vec{n}_f,-\vec{\theta}_\tau,C_{\vec{\theta}_\tau})} = j_{\mathfrak{z}_{\vec{n}}} \\
& \epsilon_\star \mathcal{J}_{(\vec{n}_f,-\vec{\theta}_\tau,C_{\vec{\theta}_\tau})} = \textbf{n}
\end{aligned} \right. \, .
\eea
Any $1$-cycle in the homology class $\textbf{n}$ can be written as:
\bea
z_{\vec{n}} = \mathfrak{z}_{\vec{n}} + \partial c \, ,
\eea
for some $2$-chain $c = (c_L^\circ, c_R^\circ + c_R^\Sigma)$. Consequently, the DB $1$-current:
\bea
\mathcal{J}_{z_{\vec{n}}} = \mathcal{J}_{(\vec{n}_f,-\vec{\theta}_\tau,C_{\vec{\theta}_\tau})} + \mathcal{C} \, , 
\eea 
where $\mathcal{C} = ( j_{c_L^\circ} , j_{c_R^\circ} , \overline{j_{c_R^\Sigma}} )$, is associated with the $1$-cycle $z_{\vec{n}}$, and as such fulfills:
\bea
\mathcal{J}_{z_{\vec{n}}} \eval{A} = \oint_{_{z_{\vec{n}}}} A \, ,
\eea
for any DB $1$-cocycle $A = (A_L , A_R , \Lambda_{\Sigma_R})$. We gather this into the following property.

\begin{prop}\label{cyclesDBdecomp}\leavevmode

Let $\textbf{n}$ be an element of $\coker P \cong H_1(X_L \cup_f X_R)$. 

Let $\vec{n} = \vec{n}_f + P \vec{\theta}_\tau $, with $\vec{\theta}_\tau = \vec{m}/p$, be a representative of $\textbf{n}$ in $\mathbb{Z}^g$.  

Let $z_{\vec{n}}$ be a $1$-cycle whose de Rham current is:
\bea
j_{z_{\vec{n}}} = \left( 0 , \sum_{a = 1}^g n^a j_{\lambda_a^R} \right) + d^\dag C \, .
\eea

\noindent The generalized DB class canonically associated with $z_{\textbf{n}}$ is then the class of the DB $1$-current:
\bea
\label{cyclesDBdecompdef}
\mathcal{J}_{z_{\vec{n}}} = \mathcal{J}_{{\vec{n}_f}} - \mathcal{J}_{\vec{\theta}_\tau} + \mathcal{C}_{\vec{\theta}_\tau} + \mathcal{C} = \mathcal{J}_{(\vec{n}_f , - \vec{\theta}_\tau , C_{\vec{\theta}_\tau} + C )} \, ,
\eea
where $\mathcal{C}$ is the DB $1$-current associated with the chain $2$-current $C$. 

\end{prop}
\noindent Let us point out that once a representative of type $\mathcal{J}_{(\vec{n}_f,\vec{\theta},S)}$ has been chosen in each generalized DB class, the DB $1$-current (\ref{cyclesDBdecompdef}) is the unique one associated with the $1$-cycle $z_{\vec{n}}$. Moreover, still with this choice made, $\mathcal{J}_{z_{\vec{n}}}$ is the only DB $1$-current with generalized curvature $j_{z_{\vec{n}}}$ and homology class the class of $\vec{n}$. These two constraints thus provide another way to specify the DB $1$-current associated with a given $1$-cycle of $X_L \cup_f X_R$. If $z = \partial c$, then we have $\mathcal{J}_z = \mathcal{J}_{\partial c} = \mathcal{C} = \left( j_{c_L^\circ} , j_{c_R^\circ} ,  \overline{j_{c_R^\Sigma}} \right) = \mathcal{J}_{(\vec{0} , \vec{0} , C )}$, with $C = j_c$.

\subsection{Generalized DB product}

In this subsection we shall see how the DB product between DB $1$-cocycles can be extend firstly to a product between DB $1$-currents and DB $1$-cocycles, and secondly, with some precautions, to a product between DB $1$-currents. The construction is very similar to what happen for de Rham currents and forms with respect to the exterior product: the exterior product of two forms is well-defined and so is the exterior product of a de Rham current with a form. The product of two de Rham currents whose total dimension is the manifold's dimension may be defined when these currents fulfill certain conditions. Our interest in trying to extend the DB product to generalized DB classes is twofold. On the one hand, the Chern-Simons and BF actions are both defined using the DB product. On the second hand, since Chern-Simons and BF observables all rely on $1$-cycles which are singular DB classes, it seems natural to consider $H_D^1(X_L \cup_f X_R)^\star$, which contains $Z_1(X_L \cup_f X_R)^\star$, as the space of fields for these two theories. These considerations bring us to wonder whether it is possible to extend the standard BD product to generalized DB classes. 

As a first step, let us associate a DB $1$-current to a DB $1$-cocycle so that we can naturally pair singular and regular DB classes. Let $A = (A_L , A_R , \Lambda_{R_\Sigma})$ be a DB $1$-cocycle of $X_L \cup_f X_R$. We set:
\bea
\mathcal{J}_{A} := \int_{X_L \cup_f X_R} \hspace{-1cm} A \star \bullet \quad .
\eea
This defines a functional which acts linearly (and continuously) on any BD $1$-cocycle $B = (B_L , B_R , \Pi_{R_\Sigma})$ according to:
\bea
\label{evaluationDBalongDB}
\mathcal{J}_{A} \eval{B} =  \int_{X_L} \! \! A_L \wedge d B_L + \int_{X_R} \! \! A_R \wedge d B_R + \oint_{\Sigma_R} \! \! \bar{d} \Lambda_{R_\Sigma} \wedge B_R^\Sigma \, .
\eea
To check that $\mathcal{J}_{A}$ is a DB $1$-current we need to identify its components according to Definition \ref{DB1currentandcurv}. The injection $\Omega^1(X) \xrightarrow{J} \Omega^2(X,\partial X)^\diamond$, Property \ref{decompderivregular} and the comparison of expressions (\ref{evaluationDBalongDB}) and (\ref{intAalongzviatriple}) bring us to set:
\bea
\label{smoothtosingularDB}
\left\{ \begin{aligned}
& \mathcal{A}_L^\circ = J_{A_L} \\
& \mathcal{A}_R^\circ = J_{A_R} 
\end{aligned} \right. \, ,
\eea
with:
\bea
\label{derivsmoothDB}
\left\{ \begin{aligned}
& d_\circ^{\dag \circ} \mathcal{A}_L^\circ = J_{d A_L} \\
& d_\circ^{\dag \circ} \mathcal{A}_R^\circ = J_{d A_R} 
\end{aligned} \right. \qquad , \qquad
\left\{ \begin{aligned}
& d_\circ^{\dag \Sigma} \mathcal{A}_L^\circ = - J_{A_L^\Sigma} \\
& d_\circ^{\dag \Sigma} \mathcal{A}_R^\circ = - J_{A_R^\Sigma} 
\end{aligned} \right. 
\eea
After noticing that:
\bea
f ( J_{A_R^\Sigma} ) = J_{f_* A_R^\Sigma} \, ,
\eea
we conclude that:
\bea
d_\circ^{\dag \Sigma} \mathcal{A}_R^\circ - f(d_\circ^{\dag \Sigma} \mathcal{A}_L^\circ) = - J_{A_R^\Sigma} + J_{f_* A_L^\Sigma} = - J_{A_R^\Sigma - f_* A_L^\Sigma} = - J_{\bar{d} \Lambda_{\Sigma_R}} \, .
\eea
The de Rham $1$-current  $J_{\bar{d} \Lambda_{\Sigma_R}}$ is obviously an element of $\Omega^1(\Sigma)^\diamond_\mathbb{Z}$. Exact sequence (\ref{sequenceSigmaPontDual}) ensures us that there exists $\xi_{\Sigma_R} \in H^1_D(\Sigma)^\star$ such that $\delta_{-1}^\star \xi_{\Sigma_R} = J_{\bar{d} \Lambda_{\Sigma_R}}$. More precisely, thanks to the canonical injection $H^0_D(\Sigma) \xrightarrow{\mathfrak{J}} H^1_D(\Sigma)^\star$ already mentioned in the end of subsection 2.1 we can set:
\bea
\xi_{\Sigma_R} = \mathfrak{J}_{\Lambda_{\Sigma_R}} \, ,
\eea
so that:
\bea
\label{gencurvfromsmoothcurv}
\delta_{-1}^\star \xi_{\Sigma_R} = J_{\bar{d} \Lambda_{R_\Sigma}} \, .
\eea
Hence, we can see $\mathcal{J}_{A}$ as the DB $1$-current $\left( \mathcal{A}_L^\circ , \mathcal{A}_R^\circ , \xi_{\Sigma_R} \right)$, with:
\bea
\mathcal{J}_{A} \eval{B} =  \mathcal{A}_L^\circ \eval{d B_L} + \mathcal{A}_R^\circ \eval{d B_R} + (\delta_{-1}^\star \xi_{\Sigma_R} ) \eval{B_R^\Sigma} \, ,
\eea
which perfectly matches evaluation formula (\ref{evalDBcurrentonDBcocycle}). Since evaluation goes to classes, we have  the following property \cite{HL01}.

\begin{prop}\label{includeHinHstar}\leavevmode

There is a canonical injective homomorphism:
\bea
\label{canonicalinclusion}
H_D^1(X_L \cup_f X_R) \xrightarrow{\mathfrak{J}} H_D^1(X_L \cup_f X_R)^\star \, ,
\eea
also called the canonical inclusion of $H_D^1(X_L \cup_f X_R)$ in $H_D^1(X_L \cup_f X_R)^\star$. By analogy with de Rham currents, the elements of $\Ima \mathfrak{J}$ are said to be \textbf{regular}, and a DB $1$-current is said to be regular if it is of the form $\mathcal{J}_{A}$ for some DB $1$-cocycle $A$.

\end{prop}

\noindent Let us note that in a regular class there is always at least one regular representative. However not all representatives are regular. Conversely, in a singular class there is no regular representative.

Thanks to the previous discussion, we see that evaluation of DB $1$-current on DB $1$-cocycles can be considered as a first extension of the DB product if we set:
\bea
\label{singstarreg}
(\mathcal{J} \star \mathcal{J}_B) \eval{1} := \mathcal{J} \eval{B} \, ,
\eea
for any DB $1$-current $\mathcal{J}$ and any \emph{regular} DB $1$-current $\mathcal{J}_B$. In particular, for regular DB $1$-currents $\mathcal{J}_A$ and $\mathcal{J}_B$ we find that:
\bea
(\mathcal{J}_A \star \mathcal{J}_B) \eval{1} = \int_{X_L \cup_f X_R} \hspace{-1cm} A \star B \, ,
\eea
as it must be. This extended DB product goes to classes provided that evaluations are taken in $\mathbb{R} / \mathbb{Z}$.

Before going on, let us comment the use of the $1$ which appears in the above evaluations. The DB product $A \star B$ defines a DB $3$-cocycles of $X_L \cup_f X_R$ whose DB class is an element of $H_D^3(X_L \cup_f X_R)$. If we follow the logic of de Rham currents, we should have written:
\bea
(\mathcal{J} \star \mathcal{J}_B) \eval{1} := \mathcal{J} \eval{B \star 1} \, .
\eea
By comparing this relation with (\ref{singstarreg}) we deduce that the $1$ appearing here behaves like an identity for the star product with $B$. The DB class of $1$ cannot be an element of $H _D^1(X_L \cup_f X_R)$ as otherwise the product $B \star 1$ would belong to $H _D^3(X_L \cup_f X_R)$, and since $\mathcal{J} \in H_D^1(X_L \cup_f X_R)^\star = \Hom \, (H_D^1(X_L \cup_f X_R) , \mathbb{R}/\mathbb{Z})$ the evaluation $\mathcal{J} \eval{B \star 1}$ would be meaningless. However, the DB product is more generally a pairing $\star : H _D^p(M) \times H _D^q(M)  \to H _D^{p+q+1}(M)$ \cite{HLZ}. Hence, $B \star 1$ belongs to $H_D^1(X_L \cup_f X_R)$ if and only if $1 \in H_D^{-1}(X_L \cup_f X_R)$, and since by convention $H_D^{-1}(X_L \cup_f X_R) = \mathbb{Z}$ \cite{HLZ}, we conclude that the DB class $1$ appearing in the DB product $B \star 1$ is simply the integer $1$ considered as an element of $H_D^{-1}(X_L \cup_f X_R)$.

To go one step further, let us compare more carefully regular and singular DB $1$-currents. Let $\mathcal{J}_A = \left( \mathcal{A}_L^\circ , \mathcal{A}_R^\circ , \xi_{\Sigma_R} \right)$ be a regular DB $1$-current and let $\mathcal{J} = \left( J_L^\circ , J_R^\circ , \varsigma_{\Sigma_R} \right)$ be a singular one. Their generalized curvatures are:
\bea
\left\{ \begin{aligned}
& \delta_{-1}^\star \mathcal{J}_A = \left( d_\circ^{\dag \circ} \mathcal{A}_L^\circ , d_\circ^{\dag \circ} \mathcal{A}_R^\circ + d_\circ^{\dag \Sigma} \mathcal{A}_R^\circ - f(d_\circ^{\dag \Sigma} \mathcal{A}_L^\circ) + \delta_{-1}^\star \xi_{\Sigma_R} \right) \\
& \delta_{-1}^\star \mathcal{J} = \left( d_\circ^{\dag \circ} J_L^\circ , d_\circ^{\dag \circ} J_R^\circ + d_\circ^{\dag \Sigma} J_R^\circ - f(d_\circ^{\dag \Sigma} J_L^\circ) + \delta_{-1}^\star \varsigma_{\Sigma_R} \right)
\end{aligned} \right. \, .
\eea
Furthermore, since the generalized curvature of $\xi_{\Sigma_R}$ is:
\bea
\delta_{-1}^\star \xi_{\Sigma_R} = J_{\bar{d} \Lambda_{\Sigma_R}} = f(d_\circ^{\dag \Sigma} \mathcal{A}_L^\circ) - d_\circ^{\dag \Sigma} \mathcal{A}_R^\circ \, ,
\eea
we deduce that:
\bea
\label{regularcurvconstraint}
\delta_{-1}^\star \mathcal{J}_A = \left( d_\circ^{\dag \circ} \mathcal{A}_L^\circ , d_\circ^{\dag \circ} \mathcal{A}_R^\circ \right) = J_{\bar{d} A} \, .
\eea
Hence, if $( \delta_{-1}^\star \mathcal{J} )_R^\Sigma$ denotes the $\Sigma_R$ component of the generalized curvature of a DB $1$-current $\mathcal{J}$, the regular current $\mathcal{J}_A$ fulfills:
\bea
\label{sigmacompregul}
( \delta_{-1}^\star \mathcal{J}_A )_R^\Sigma = d_\circ^{\dag \Sigma} \mathcal{A}_R^\circ - f(d_\circ^{\dag \Sigma} \mathcal{A}_L^\circ) + \delta_{-1}^\star \xi_{\Sigma_R} = 0 \, .
\eea
Let $\mathcal{J} \eval{A}$ be given as in (\ref{evalDBcurrentonDBcocycle}). Integrating by parts this evaluation yields:
\bea
\mathcal{J} \eval{A} = (d^\dag  J_L^\circ) \eval{A_L} + (d^\dag  J_R^\circ) \eval{A_R} + (\delta_{-1}^\star \varsigma_{\Sigma_R}) \eval{A_R^\Sigma} \, .
\eea
Since $d^\dag = d_\circ^{\dag \circ} + d_\circ^{\dag \Sigma}$, we can rewrite this evaluation as:
\bea
(d_\circ^{\dag \circ}  J_L^\circ) \eval{A_L} + (d_\circ^{\dag \circ}  J_R^\circ) \eval{A_R} + ( d_\circ^{\dag \Sigma} J_L^\circ) \eval{A_L^\Sigma} + ( d_\circ^{\dag \Sigma} J_R^\circ + \delta_{-1}^\star \varsigma_{\Sigma_R}) \eval{A_R^\Sigma} \, ,
\eea
and as:
\bea
( d_\circ^{\dag \Sigma} J_L^\circ) \eval{A_L^\Sigma} = - \left( f(d_\circ^{\dag \Sigma} J_L^\circ) \right) \eval{f_* A_L^\Sigma} = - \left( f(d_\circ^{\dag \Sigma} J_L^\circ) \right) \eval{A_R^\Sigma - \bar{d} \Lambda_{\Sigma_R}} \, .
\eea
we find that:
\bea
\begin{aligned}
\mathcal{J} \eval{A} = (d_\circ^{\dag \circ}  J_L^\circ) \eval{A_L} + (d_\circ^{\dag \circ}  J_R^\circ) \eval{A_R} + \qquad \qquad \qquad \qquad \qquad \qquad \qquad \\ + ( d_\circ^{\dag \Sigma} J_R^\circ -  f(d_\circ^{\dag \Sigma} J_L^\circ) + \delta_{-1}^\star \varsigma_{\Sigma_R}) \eval{A_R^\Sigma} +  f(d_\circ^{\dag \Sigma} J_L^\circ) \eval{\bar{d} \Lambda_{\Sigma_R}}
\end{aligned} \, .
\eea
Note the similitude with formula (\ref{intAalong1cycle}). By introducing the $\Sigma_R$ component of $\mathcal{J}$ and replacing evaluation by formal exterior product we get:
\bea
\begin{aligned}
\mathcal{J} \eval{A} = (d_\circ^{\dag \circ}  J_L^\circ \wedge J_{A_L}) \eval{1_L} + (d_\circ^{\dag \circ}  J_R^\circ \wedge J_{A_R}) \eval{1_R} + \qquad \qquad \qquad \qquad \\ + \left( (\delta_{-1}^\star \mathcal{J})_R^\Sigma \wedge J_{A_R^\Sigma} \right) \eval{1_{\Sigma_R}} + \left( f(d_\circ^{\dag \Sigma} J_L^\circ) \wedge J_{\bar{d} \Lambda_{\Sigma_R}} \right) \eval{1_{\Sigma_R}} 
\end{aligned} \, ,
\eea
and by using relations (\ref{derivsmoothDB}) and (\ref{gencurvfromsmoothcurv}) we obtain:
\bea
\begin{aligned}
\mathcal{J} \eval{A} = (d_\circ^{\dag \circ}  J_L^\circ \wedge \mathcal{A}_L^\circ ) \eval{1_L} + (d_\circ^{\dag \circ}  J_R^\circ \wedge \mathcal{A}_R^\circ ) \eval{1_R} + \qquad \qquad \qquad \qquad \\ - ( \left( \delta_{-1}^\star \mathcal{J} \right)_R^\Sigma \wedge d_\circ^{\dag \Sigma} \mathcal{A}_R^\circ ) \eval{1_{\Sigma_R}} + ( f(d_\circ^{\dag \Sigma} J_L^\circ) \wedge \delta_{-1}^\star \xi_{\Sigma_R})  \eval{1_{\Sigma_R}}
\end{aligned} \, ,
\eea
and finally:
\bea
\begin{aligned}
\label{firstresultsingregDB}
\mathcal{J} \eval{A} = ( \mathcal{A}_L^\circ \wedge d_\circ^{\dag \circ}  J_L^\circ ) \eval{1_L} + ( \mathcal{A}_R^\circ \wedge d_\circ^{\dag \circ}  J_R^\circ ) \eval{1_R} + \qquad \qquad \qquad \qquad \\ + ( d_\circ^{\dag \Sigma} \mathcal{A}_R^\circ \wedge \left( \delta_{-1}^\star \mathcal{J} \right)_R^\Sigma ) \eval{1_{\Sigma_R}} - ( \delta_{-1}^\star \xi_{\Sigma_R} \wedge f(d_\circ^{\dag \Sigma} J_L^\circ) )  \eval{1_{\Sigma_R}}
\end{aligned} \, .
\eea
All the terms in this last expression are well defined because the currents $\mathcal{A}_L^\circ$, $\mathcal{A}_R^\circ$, $d_\circ^{\dag \Sigma} \mathcal{A}_R^\circ$ and $\delta_{-1}^\star \xi_{\Sigma_R}$ are regular \cite{dR55}. By comparing the above expression with the one of $\int A \star B$ and remembering that we started with an integration by parts, it seems more logical to consider that the former defines $\left( \mathcal{J}_A \star \mathcal{J} \right) \eval{1}$ rather then $\left( \mathcal{J} \star \mathcal{J}_A \right) \eval{1}$ so that:
\bea
\begin{aligned}
\left( \mathcal{J}_A \star \mathcal{J}_B \right) \eval{1} = \, & ( \mathcal{A}_L^\circ \wedge d_\circ^{\dag \circ}  \mathcal{B}_L^\circ ) \eval{1_L} + ( \mathcal{A}_R^\circ \wedge d_\circ^{\dag \circ}  \mathcal{B}_R^\circ ) \eval{1_R} +  \\ & + ( d_\circ^{\dag \Sigma} \mathcal{A}_R^\circ \wedge (\delta_{-1}^\star \mathcal{J}_B )_R^\Sigma ) \eval{1_{\Sigma_R}} + ( \delta_{-1}^\star \xi_{\Sigma_R} \wedge f(d_\circ^{\dag \Sigma} \mathcal{B}_L^\circ ) \eval{1_{\Sigma_R}} \\
= \, & ( \mathcal{A}_L^\circ \wedge d_\circ^{\dag \circ}  \mathcal{B}_L^\circ ) \eval{1_L} + ( \mathcal{A}_R^\circ \wedge d_\circ^{\dag \circ}  \mathcal{B}_R^\circ ) \eval{1_R} + \\
& - ( J_{\bar{d} \Lambda_{R_\Sigma}} \wedge d_\circ^{\dag \Sigma} \mathcal{B}_R^\circ ) \eval{1_{\Sigma_R}} 
\end{aligned} \, ,
\eea
which agrees with (\ref{evaluationDBalongDB}). Note that we use the fact that $(\delta_{-1}^\star \mathcal{J}_B )_R^\Sigma = 0$.

All this yields the first generalization of the DB product according to the following property.

\begin{prop}\label{extendDBsingreg}\leavevmode

1) A DB $1$-current $\left( \mathcal{A}_L^\circ , \mathcal{A}_R^\circ , \xi_{\Sigma_R} \right)$ is regular if and only if there exist a DB $1$-cocycle $\left( A_L , A_R , \Lambda_{\Sigma_R} \right)$ such that:
\bea
\mathcal{A}_L^\circ = J_{A_L} \quad , \quad \mathcal{A}_R^\circ = J_{A_R} \quad , \quad \xi_{\Sigma_R} = \mathfrak{J}_{\Lambda_{\Sigma_R}} \, ,
\eea
with $H^0_D(\Sigma) \xrightarrow{\mathfrak{J}} H^1_D(\Sigma)^\star$ the canonical injection. The generalized curvature of a regular DB $1$-current $\mathcal{J}_A$ is the image of the curvature of $A$ under the injection $\Omega_\mathbb{Z}^2(X_L \cup_f X_R) \to \Omega^1(X_L \cup_f X_R)_\mathbb{Z}^\diamond$ and as such is a regular de Rham current. 

2) Let $\mathcal{J} = \left( J_L^\circ , J_R^\circ , \varsigma_{\Sigma_R} \right)$ and $\mathcal{J}_A = \left( \mathcal{A}_L^\circ , \mathcal{A}_R^\circ , \xi_{\Sigma_R} \right)$ be two DB $1$-currents, the first being singular and the second regular. The generalized DB product of $\mathcal{J}_A$ with $\mathcal{J}$ is: 
\bea
\label{regularstarsingular}
\begin{aligned}
\left( \mathcal{J}_A \star \mathcal{J} \right) \eval{1} = ( \mathcal{A}_L^\circ \wedge d_\circ^{\dag \circ}  J_L^\circ  ) \eval{1_L} + ( \mathcal{A}_R^\circ \wedge d_\circ^{\dag \circ}  J_R^\circ ) \eval{1_R} + \qquad \qquad \qquad \qquad \\ + ( d_\circ^{\dag \Sigma} \mathcal{A}_R^\circ \wedge \left( \delta_{-1}^\star \mathcal{J} \right)_R^\Sigma ) \eval{1_{\Sigma_R}} - ( \delta_{-1}^\star \xi_{\Sigma_R} \wedge f(d_\circ^{\dag \Sigma} J_L^\circ) )  \eval{1_{\Sigma_R}} \, ,
\end{aligned} 
\eea
and fulfills:
\bea
\left( \mathcal{J}_A \star \mathcal{J} \right) \eval{1} \egzz \mathcal{J} \eval{A} = \left( \mathcal{J} \star \mathcal{J}_A \right) \eval{1} \, ,
\eea
with $A$ the DB $1$-cocycle defining $\mathcal{J}_A$. In particular, if $\mathcal{J}_z$ is associated with a $1$-cycle $z$ we have:
\bea
\left( \mathcal{J}_z \star \mathcal{J}_A \right) \eval{1} = \mathcal{J}_z \eval{A} = \oint_z A \, .
\eea

\end{prop}

\noindent From the first of the above properties, we deduce that a generalized DB class is regular if and only if it is a smooth Pontryagin dual in the sense of \cite{HL01}. In other words, a generalized DB class is regular if and only if its generalized curvature is regular.

To further extend the generalized DB product the first idea is to replace the regular DB $1$-current $\mathcal{J}_A$ in formula (\ref{regularstarsingular}) with a general DB $1$-current $\mathcal{K}$ which yields:
\bea
\begin{aligned}
\left( \mathcal{J} \star \mathcal{K} \right) \eval{1} = \, & (J_L^\circ \wedge d_\circ^{\dag \circ}  K_L^\circ ) \eval{1_L} + (J_R^\circ \wedge d_\circ^{\dag \circ} K_R^\circ ) \eval{1_R} + \\ & + ( d_\circ^{\dag \Sigma} J_R^\circ \wedge ( \delta_{-1}^\star \mathcal{K} )_R^\Sigma ) \eval{1_{\Sigma_R}} - ( \delta_{-1}^\star \varsigma_{\Sigma_R} \wedge f(d_\circ^{\dag \Sigma} K_L^\circ) ) \eval{1_{\Sigma_R}} \, .
\end{aligned} 
\eea
Before wondering whether this expression is meaningful, let us see if it gives rise to a commutative product or not. To this end, we consider the difference $\Delta(\mathcal{J},\mathcal{K}) = \left( \mathcal{J} \star \mathcal{K} \right) \eval{1} - \left( \mathcal{K} \star \mathcal{J} \right) \eval{1}$. We have:
\bea
\begin{aligned}
\Delta(\mathcal{J},\mathcal{K}) = & - \left( d_\circ^{\dag \circ} ( J_L^\circ \wedge K_L^\circ) \right) \eval{1_L} - \left( d_\circ^{\dag \circ} ( J_R^\circ \wedge K_R^\circ) \right) \eval{1_L} \\
& + ( d_\circ^{\dag \Sigma} J_R^\circ \wedge ( \delta_{-1}^\star \mathcal{K} )_R^\Sigma ) \eval{1_{\Sigma_R}} - ( d_\circ^{\dag \Sigma} K_R^\circ \wedge ( \delta_{-1}^\star \mathcal{J} )_R^\Sigma ) \eval{1_{\Sigma_R}} \\ 
& - ( \delta_{-1}^\star \varsigma_{\Sigma_R} \wedge f(d_\circ^{\dag \Sigma} K_L^\circ) ) \eval{1_{\Sigma_R}} + ( \delta_{-1}^\star \kappa_{\Sigma_R} \wedge f(d_\circ^{\dag \Sigma} J_L^\circ) ) \eval{1_{\Sigma_R}} \, ,
\end{aligned}
\eea
which after some algebraic juggles yields:
\bea
\Delta(\mathcal{J},\mathcal{K}) = \left( ( \delta_{-1}^\star \mathcal{J} )_R^\Sigma ) \wedge ( \delta_{-1}^\star \mathcal{K} )_R^\Sigma ) \right) \eval{1_{\Sigma_R}} - \left( \delta_{-1}^\star \varsigma_{\Sigma_R} \wedge \delta_{-1}^\star \kappa_{\Sigma_R} \right) \eval{1_{\Sigma_R}} \, .
\eea
Since $\delta_{-1}^\star \varsigma_{\Sigma_R}$ and $\delta_{-1}^\star \kappa_{\Sigma_R}$ are generalized curvature (on $\Sigma_R$) they are closed which implies that the quantity $\left( \delta_{-1}^\star \varsigma_{\Sigma_R} \wedge \delta_{-1}^\star \kappa_{\Sigma_R} \right) \eval{1_{\Sigma_R}}$ is firstly meaningful \cite{dR55} and secondly an integer. So, the term $\left( \delta_{-1}^\star \varsigma_{\Sigma_R} \wedge \delta_{-1}^\star \kappa_{\Sigma_R} \right) \eval{1_{\Sigma_R}}$ is not an issue if commutativity is considered modulo integers, i.e. in $\mathbb{R}/\mathbb{Z}$. Hence, it is the first term in the right-hand side of the above equation which breaks commutativity. To ensure commutativity, we must add to the original expression of $\left( \mathcal{J} \star \mathcal{K} \right) \eval{1}$ the term:
\bea
\label{additionnalcontrib}
\label{commutnecessterm}
- \frac{1}{2} \left( ( \delta_{-1}^\star \mathcal{J} )_R^\Sigma \wedge ( \delta_{-1}^\star \mathcal{K} )_R^\Sigma \right) \eval{1_{\Sigma_R}} = \frac{1}{2} \left( ( \delta_{-1}^\star \mathcal{K} )_R^\Sigma \wedge ( \delta_{-1}^\star \mathcal{J} )_R^\Sigma \right) \eval{1_{\Sigma_R}} \, .
\eea

We can now set the following definition:

\begin{defi}\label{defgenDBproductsing}\leavevmode
	
	Let $\mathcal{J} = \left( J_L^\circ , J_R^\circ , \varsigma_{\Sigma_R} \right)$ and $\mathcal{K} = \left( K_L^\circ , K_R^\circ , \kappa_{\Sigma_R} \right)$ be two DB $1$-currents of $X_L \cup_f X_R$. When it exists, their generalized DB product is:
	\bea
	\begin{aligned}
		\left( \mathcal{J} \star \mathcal{K} \right) \eval{1} = \, & (J_L^\circ \wedge d_\circ^{\dag \circ}  K_L^\circ ) \eval{1_L} + (J_R^\circ \wedge d_\circ^{\dag \circ} K_R^\circ ) \eval{1_R} + \\ & + ( \Delta_\mathcal{J}^{\Sigma_R} \wedge ( \delta_{-1}^\star \mathcal{K} )_R^\Sigma \eval{1_{\Sigma_R}} - ( \delta_{-1}^\star \varsigma_{\Sigma_R} \wedge f(d_\circ^{\dag \Sigma} K_L^\circ) ) \eval{1_{\Sigma_R}} \, ,
	\end{aligned} 
	\eea
	with:
	\bea
	\Delta_\mathcal{J}^{\Sigma_R} = d_\circ^{\dag \Sigma} J_R^\circ - \frac{1}{2} ( \delta_{-1}^\star \mathcal{J} )_R^\Sigma = \frac{1}{2} \left(d_\circ^{\dag \Sigma} J_R^\circ + f(d_\circ^{\dag \Sigma} J_L^\circ) - \delta_{-1}^\star \varsigma_{\Sigma_R} \right) \, .
	\eea 
\end{defi}

\noindent To see the consistency of this definition we must check that the generalized DB product with a DB $1$-current ambiguity is zero (modulo integers). We leave this as an exercise. Of course we find that $\left( \mathcal{J}_A \star \mathcal{J}_B \right) \eval{1} = \int_M A \star B$ is well-defined. If one of the two DB $1$-currents is regular all the evaluations appearing in the above definition are well-defined \cite{dR55}, the additional contribution (\ref{additionnalcontrib}) vanishes and as expected we have:
\bea
\left( \mathcal{J} \star \mathcal{J}_A \right) \eval{1} = \left( \mathcal{J}_A \star \mathcal{J} \right) \eval{1} = J \eval{A} \, .
\eea
When both DB $1$-currents are singular some or all of the evaluations appearing in the definition of $\left( \mathcal{J} \star \mathcal{K} \right) \eval{1}$ may be ill-defined. Fortunately, the situation is not as bad as it seems. To see this, let us give some fundamental generalized DB products relying on decomposition (\ref{decompuniversal}). After some algebraic juggles, we find:
\bea
\label{basicgenDBproducts}
\begin{aligned}
& \bullet \quad \left( \mathcal{J}_{\vec{n}_f} \star \mathcal{J}_{\vec{n}'_f} \right) \eval{1} \egzz 0 \, ,\\
& \bullet \quad \left( \mathcal{J}_{\vec{n}_f} \star \mathcal{J}_{\vec{\theta}} \right) \eval{1} =  \left\langle Q \vec{\theta} ,\vec{n}_f \right\rangle \, , \\
& \bullet \quad \left( \mathcal{J}_{\vec{n}_f} \star \mathcal{S} \right) \eval{1} =  \sum_{a = 1}^g \left( n_f^a j_{\lambda_a^R} \wedge S_R^\circ \right) \eval{1_{\Sigma_R}} \, , \\
& \bullet \quad  \left( \mathcal{J}_{\vec{\theta}} \star \mathcal{J}_{\vec{\theta}'} \right) \eval{1} = - \left\langle P \vec{\theta} , Q \vec{\theta}' \right\rangle = - \left\langle P \vec{\theta}_\tau , Q \vec{\theta}'_\tau \right\rangle \, , \\
& \bullet \quad \left( \mathcal{J}_{\vec{\theta}} \star \mathcal{S} \right) \eval{1} = 0 \, ,\\
& \bullet \quad \left( \mathcal{C}_{\vec{\theta}_\tau} \star \mathcal{C}_{\vec{\theta}'_\tau} \right) \eval{1} \egzz \left\langle Q \vec{\theta}_\tau , P \vec{\theta}'_\tau \right\rangle \, .
\end{aligned}
\eea

\noindent In the computation of the first and last of the above relations we used $(j_{A_a^R} \wedge j_{\lambda_b^R}) \eval{1_R} \egzz 0$ which derives from the following regularization. First, we chose the support of the annuli $A_a^R$ in such a way that $(j_{A_a^R} \wedge j_{\lambda_b^R}) \eval{1_R} =0$ when $a \neq b$. Then, when $a = b$ we consider a framing $\breve{\lambda}_a^R$ of $\lambda_a^R$ and set $(j_{A_a^R} \wedge j_{\lambda_a^R}) \eval{1_R} := (j_{A_a^R} \wedge j_{\breve{\lambda}_a^R}) \eval{1_R}$. This last term is a Kronecker index and as such an integer, that is to say zero modulo integers. This is an example of what we will call the zero regularization procedure in the next property. The third generalized product in the above list might be ill-defined. However, if the singular support of $\delta_{-1}^\star \mathcal{S}$ does not intersect the support of any of the $j_{\lambda_a^R}$, then the corresponding generalized DB product is well-defined. In fact, it is even sufficient to require that the singular support of $(\delta_{-1}^\star \mathcal{S})_R^\circ = d_\circ^{\dag \circ} S_R^\circ$ does not intersect the support of any of the $j_{\lambda_a^R}$. Thus, this third term is not hard to manage, for instance by putting some simple support restriction on the set of $S$ fields. The last kind of generalized product which can be ill-defined are those of the form $(\mathcal{S} \star \mathcal{S}') \eval{1}$. Identifying the ``good fields" to consider seems a hard task. Fortunately, we will see in the last section that finding a precise set of ``good fields" is a rather irrelevant question in the context of Chern-Simons and BF abelian theories. Nonetheless, because they will play a role in these theories, the more specific case of DB $1$-currents associated with $1$-cycles requires some attention. 

Let $\mathcal{J}_{\partial c} = \left( j_{c_L^\circ} , j_{c_R^\circ} ,  \overline{j_{c_R^\Sigma}} \right)$ and $\mathcal{J}_{\partial c'} = \left( j_{{c'_L}{\kern-0.25em ^\circ}} , j_{{c'_R}{\kern-0.28em ^\circ}} , \overline{j_{{c'_R}{\kern-0.25em ^\Sigma}}} \right)$ be DB $1$-currents associated with two homologically trivial $1$-cycles $\partial c$ and $\partial c'$, respectively. Then, we have:
\bea
\label{DBvslinking}
\begin{aligned}
\left( \mathcal{J}_{\partial c} \star \mathcal{J}_{\partial c'} \right) \eval{1} = & ( j_{c_L^\circ} \wedge d_\circ^{\dag \circ} j_{{c'_L}{\kern-0.25em ^\circ}} ) \eval{1_L} + ( j_{c_R^\circ} \wedge d_\circ^{\dag \circ} j_{{c'_R}{\kern-0.29em ^\circ}} ) \eval{1_R} + \\ 
& + (\Delta_{\mathcal{C}}^{\Sigma_R} \wedge ( \delta_{-1}^\star \mathcal{C}' )_R^\Sigma ) \eval{1_{\Sigma_R}} - ( d^{\dag \Sigma} j_{c_R^\Sigma} \wedge f(d_\circ^{\dag \Sigma} j_{{c'_L}{\kern-0.25em ^\circ}}) )  \eval{1_{\Sigma_R}} \, .
\end{aligned}
\eea
If the supports of $\partial c$ and $\partial c'$ do not intersect then $( \delta_{-1}^\star \mathcal{C} )_R^\Sigma \wedge ( \delta_{-1}^\star \mathcal{C}' )_R^\Sigma = 0$ so that the right-hand side of the above expression coincides with expression (\ref{linkingformula}) of the linking number of $\partial c$ with $\partial c'$ and as such is an integer. Thus, we have:
\bea
\label{DBcurrentlinking}
\left( \mathcal{J}_{\partial c} \star \mathcal{J}_{\partial c'} \right) \eval{1} = \textit{lk} (\partial c,\partial c') \egzz 0 \, .
\eea
Of course, it might be necessary to chose $c$ and $c'$ carefully for the above generalized DB product to be well-defined, a precaution which must also be taken when the linking number is written as a Kronecker index. 

Let $z$ and $z'$ be two homologically non-trivial $1$-cycles. There exist $p,p' \in \mathbb{Z}$ and two $2$-chains $c$ and $c'$ such that $p z = \partial c$ and $p' z' = \partial c'$. We already seen that the DB $1$-currents associated with $z$ (resp. $z'$) is $\mathcal{J}_z = \mathcal{C}_{\vec{\theta}_\tau} - \mathcal{J}_{\vec{\theta}_\tau}$ (resp. $\mathcal{J}_{z'} = \mathcal{C}_{\vec{\theta}_\tau} - \mathcal{J}_{\vec{\theta}_\tau}$) where $\vec{\theta}_\tau$ (resp. $\vec{\theta}'_\tau$) is such that $P \vec{\theta}_\tau$ (resp. $P \vec{\theta}'_\tau$) represents the homology class of $z$ (resp. $z'$). Now, from (\ref{basicgenDBproducts}) we deduce that:
\bea
\left( \mathcal{J}_z \star \mathcal{J}_{z'} \right) \eval{1} = \left( \mathcal{C}_{\vec{\theta}_\tau} \star \mathcal{C}_{\vec{\theta}'_\tau} \right) \eval{1} + \left( \mathcal{J}_{\vec{\theta}_\tau} \star \mathcal{J}_{\vec{\theta}'_\tau} \right) \eval{1} \egzz 0 \, .
\eea
In addition, we know that there is  a bilinear pairing $\Gamma : T_1(X_L \cup_f X_R) \times T_1(X_L \cup_f X_R) \to \mathbb{Q}/\mathbb{Z}$, referred to as the linking form of $X_L \cup_f X_R$ and which is defined as follows. Let $z_\tau$ and $z'_\tau$ be the two non-trivial torsion $1$-cycles we have just introduced. Following \cite{CT80}, we set:
\bea
\Gamma ( z_\tau , z'_\tau ) = \frac{1}{p} (c \odot z'_\tau) \, .
\eea
It is not hard to check that this definition is independent of the choice of $c$. Since $p' z' = \partial c'$, we can also write:
\bea
\Gamma ( z_\tau , z'_\tau ) = \frac{1}{p p'} (c \odot \partial c') =\frac{1}{p p'} \textit{lk} (\partial c , \partial c') \, .
\eea
This bilinear pairing becomes an $\mathbb{Q}/\mathbb{Z}$-valued bilinear pairing at the level of classes. Now, since we have $j_z = d^\dag \mathcal{C}_{\vec{\theta}_\tau}$ and $j_{z'} = d^\dag \mathcal{C}_{\vec{\theta}'_\tau}$, we deduce that:
\bea
\Gamma ( z_\tau , z'_\tau ) = \left( \mathcal{C}_{\vec{\theta}_\tau} \star \mathcal{C}_{\vec{\theta}'_\tau} \right) \eval{1} \egzz \left\langle Q \vec{\theta}_\tau , P \vec{\theta}'_\tau \right\rangle \, . 
\eea
\noindent See \cite{CFH18} for the particular case of a rational homology sphere. Let us point out that, the linking number of $z_\tau$ and $z'_\tau$, now defined as $\Gamma ( z_\tau , z'_\tau )$, is not an integer and hence not zero in $\mathbb{Q}/\mathbb{Z}$. Furthermore, this rational linking number is not $\left( \mathcal{J}_{z_\tau} \star \mathcal{J}_{z'_\tau} \right) \eval{1}$ since this generalized DB product is zero. Of course, the linking form $\Gamma$ is nothing but the linking form introduced in Property \ref{pontdualcokerP}.

Finally, let $z$ and $z'$ be two homologically non-trivial free $1$-cycles with associated DB $1$-currents $\mathcal{J}_z = \mathcal{J}_{\vec{n}_f} + \mathcal{C}$ and $\mathcal{J}_{z'} = \mathcal{J}_{\vec{n}'_f} + \mathcal{C}'$, respectively. By combining the fundamental evaluations (\ref{basicgenDBproducts}) together with property (\ref{DBcurrentlinking}) we deduce that:
\bea
\left( \mathcal{J}_z \star \mathcal{J}_{z'} \right) \eval{1} \egzz 0 \, .
\eea
We end with the following property.

\begin{prop}\label{zeroregul}\leavevmode
	
	1) Let $z$ and $z'$ be two $1$-cycles of $X_L \cup_f X_R$ and let $\mathcal{J}_z$ and $\mathcal{J}_{z'}$ be two DB $1$-current associated with $z$ and $z'$, respectively. Then we have:
	\bea
	\label{zeroregul1}
	\left( \mathcal{J}_z \star \mathcal{J}_{z'} \right) \eval{1} \egzz 0 \, .
	\eea
	Moreover, we define the \textbf{zero regularization} procedure by setting:
	\bea
	\label{zeroregul2}
	\left( \mathcal{J}_z \star \mathcal{J}_{z} \right)\eval{1} = \left( \mathcal{J}_z \star \mathcal{J}_{\breve{z}} \right)\eval{1} \egzz 0 \, ,
	\eea
	where $\breve{z}$ is any framing of $z$.
	
	2) The linking form of $X_L \cup_f X_R$ is defined with the help of generalized DB products according to:
	\bea
	\label{linkingformDBprod}
	\Gamma ( z_\tau , z'_\tau ) = \left( \mathcal{C}_{\vec{\theta}_\tau} \star \mathcal{C}_{\vec{\theta}'_\tau} \right) \eval{1} \egzz \left\langle Q \vec{\theta}_\tau , P \vec{\theta}'_\tau \right\rangle \, ,
	\eea
	where $z_\tau$ and $z'_\tau$ are torsion $1$-cycles whose homology classes are given by $P \vec{\theta}_\tau$ and $P \vec{\theta}'_\tau$, respectively.

\end{prop}

\noindent For homologically trivial $1$-cycles, the zero regularization procedure can be identified with the framing procedure which allows to define the linking of an homologically trivial $1$-cycle $z$ with itself, aka the self-linking of $z$. The other cases are an extension of this framing procedure to generalized DB products.

\section{Chern-Simons and BF theories}

\subsection{Actions}

Let us start with the BF theory as the Chern-Simons action can be derived from the BF one. We already know \cite{MT1,MT2} that on a closed $3$-manifold $M$ the the $U(1)$ BF action can be defined as the functional:
\bea
S_{BF,k} \left( \bar{A} , \bar{B} \right) = k \int_{\mathbb{R}^3} \bar{A} \star \bar{B} \, ,
\eea
where $\bar{A}$ and $\bar{B}$ are elements of $H_D^1(M)$ and $k$ is a coupling constant. This action cannot be considered as classical since it is $\mathbb{R} / \mathbb{Z}$-valued. Nevertheless, this choice of definition for the BF action is twofold. Firstly it goes straightforwardly to the quantum world and secondly it automatically implies that the coupling constant $k$ is an integer, i.e. is quantized. Following the same path, the Chern-Simons action is defined by:
\bea
S_{CS,k} \left( \bar{A} \right) = S_{BF,k} \left( \bar{A} , \bar{A} \right) = k \int_{\mathbb{R}^3} \bar{A} \star \bar{A} \, ,
\eea
with $\bar{A} \in H_D^1(M)$ and $k \in \mathbb{Z}$.

These actions where studied in various contexts \cite{T19,GT2014,MT1}. Our purpose here is to find expressions of the CS and BF actions in the context of a Heegaard spliting $X_L \cup_f X_R$ when $H_D^1(M)$ is replaced by $H_D^1(M)^\star$, that is to say when we deal with generalized DB classes instead of smooth ones. Let us recall that we want to do that in order to include links into the set of fields. Indeed, a link can be see as a $1$-cycle and hence as a generalized DB class.

We shall concentrate on the BF action and try to find its expression in term of the variables which appear in the decomposition formula (\ref{decompuniversal}). In other words, we want to compute:
\bea
S_{BF,k} ( \vec{n}_f , \vec{\theta} , S , \vec{n}'_f , \vec{\theta}' , S') = k \left( \mathcal{J}_{( \vec{n}_f , \vec{\theta} , S)} \star \mathcal{J}_{( \vec{n}'_f , \vec{\theta}' , S')} \right) \eval{1} \, ,
\eea
where the generalized DB product was given in Definition \ref{defgenDBproductsing}. By using the sets of fundamental generalized DB products (\ref{basicgenDBproducts}) we straightforwardly get:
\bea
\begin{aligned}
	\label{BFactiondone}
S_{BF,k} ( \vec{n}_f , \vec{\theta} , S , \vec{n}'_f , \vec{\theta}' , S') & = \; k ( \mathcal{S} \star \mathcal{S}') \eval{1} + \\ 
   &  + k \sum_{a = 1}^g \left( n_f^a j_{\lambda_a^{\Sigma_R}} \wedge {S'_R}{\kern-4pt ^\circ} +  {n'_f}{\! ^a} j_{\lambda_a^{\Sigma_R}} \wedge S_R^\circ \right) \eval{1_R} + \\
   &  + k \left\langle Q \vec{\theta} , \vec{n}'_f \right\rangle  + k \left\langle Q \vec{\theta}' ,  \vec{n}_f \right\rangle  - k \left\langle P \vec{\theta}_\tau , Q \vec{\theta}'_\tau \right\rangle  
\end{aligned} \, .
\eea

The Chern-Simons action is then:
\bea
\label{CSactiondone}
\begin{aligned}
S_{CS,k} ( \vec{n}_f , \vec{\theta} , S ) = & \; k (\mathcal{S} \star \mathcal{S}) \eval{1} + 2k \sum_{a = 1}^g \left( n_f^a j_{\lambda_a^{\Sigma_R}} \wedge S_R^\circ \right) \eval{1_R} \\ 
& \qquad + 2k \left\langle Q \vec{\theta} , \vec{n}_f \right\rangle - k \left\langle P \vec{\theta}_\tau , Q \vec{\theta}'_\tau \right\rangle 
\end{aligned} \, .
\eea
Of course, these expressions are formal because the various product appearing in them might be ill-defined. These expressions, although formal, are very close to the expressions obtained in the smooth case in \cite{T19}. In particular the inner products are the same.

\subsection{Functional measures}

Let's have a look at the functional measures which are involved in the determination of the CS and BF partition function as well as of the expectation values of $U(1)$ Wilson loops. These observables will be discussed in the last subsection. Decomposition (\ref{decompuniversal}) provides us with the most natural expression of the functional measures:
\bea
\label{CSfuncmeasure}
d \mu _{CS,k} ( \vec{n}_f , \vec{\theta} , S) = D \! S \sum_{\vec{n}_f} \sum_{\vec{\theta}_\tau } d^{b_1} \vec{\theta }_f \, e^{2i\pi S_{CS,k} ( \vec{n}_f , \vec{\theta} , S) } \, . 
\eea
All integrals and sums are taken over classes, or rather over the representatives of Property \ref{decompsingularDBclasses}. Note that the only infinite dimensional, and so formal, measure is the one dealing with the variables $\mathcal{S}$. All the others are finite dimensional. Of course the Bf measure is:
\bea
\label{BFfuncmeasure}
d \mu _{BF,k} ( \vec{n}_f , \vec{\theta} , S , \vec{n}'_f , \vec{\theta}' , S') = D \! S  D \! S'  \sum_{\vec{n}_f,\vec{n}'_f} \sum_{\vec{\theta}_\tau , \vec{\theta}'_\tau} d^{b_1} \vec{\theta }_f d^{b_1} \vec{\theta }'_f\, e^{2i\pi S_{BF,k} ( \vec{n}_f , \vec{\theta} , S , \vec{n}'_f , \vec{\theta}' , S') } \, . 
\eea

Let us consider the shift:
\bea
\label{freeshift}
\vec{n}_f \to \vec{n}_f + \vec{t}_f = \vec{u}_f \, ,
\eea
where $\vec{t}_f$ is fixed. Since $T_1(X_L \cup_f X_R)$ is a free abelian group, this shift does not change the sums over $\vec{n}_f$ in the functional measure $d \mu _{CS,k}$. However, under this shift the CS action (\ref{CSactiondone}) changes according to:
\bea
\label{translationalCM}
S_{CS,k} ( \vec{u}_f - \vec{t}_f , \vec{\theta} , S ) = S_{CS,k} ( \vec{u}_f , \vec{\theta} , S ) - 2k \left( \mathcal{J}_{ ( \vec{u}_f , \vec{\theta} , S)} \star \mathcal{J}_{\vec{t}_f} \right) \eval{1} \, ,
\eea
with:
\bea
\mathcal{J}_{\vec{t}_f} = \left( 0 , \sum_{a=1}^g t_f^a j_{A_a^R} , \zeta_{\vec{t}_f} \right) \, .
\eea
This is a simple consequence of the relation:
\bea
\begin{aligned}
\left(  \mathcal{J}_{ ( \vec{u}_f , \vec{\theta} , S)} \star \mathcal{J}_{\vec{t}_f} \right) \eval{1} = & \sum_{a=1}^g \left( t_f^a j_{\lambda_a^R} \wedge 
S_R^\circ  \right) \eval{1_R} + \left\langle Q \vec{\theta} , \vec{t}_f \right\rangle 
\\ = & \left(  \mathcal{J}_{ ( \vec{n}_f , \vec{\theta} , S)} \star \mathcal{J}_{\vec{t}_f} \right) \eval{1}
\end{aligned} \, ,
\eea
which itself can be deduced from (\ref{BFactiondone}). The second line follows from the fact that the right-hand side of the first equality is independent of $\vec{u}_f$. The zero regularization assumption implying that $\left( \mathcal{J}_{\vec{t}_f} \star \mathcal{J}_{\vec{t}_f} \right) \eval{1} = 0$, we finally find:
\bea
S_{CS,k} ( \vec{u}_f - \vec{t}_f , \vec{\theta} , S ) = k \left( (\mathcal{J}_{ ( \vec{u}_f , \vec{\theta} , S)} - \mathcal{J}_{\vec{t}_f}) \star (\mathcal{J}_{ ( \vec{u}_f , \vec{\theta} , S)} - \mathcal{J}_{\vec{t}_f}) \right) \eval{1} \, .
\eea 
Hence, since the sum over $\vec{n}_f$ is invariant under the shift (\ref{freeshift}), the CS measure is invariant by a shift of the form:
\bea
\mathcal{J}_{ ( \vec{u}_f , \vec{\theta} , S)} \rightarrow \mathcal{J}_{ ( \vec{u}_f , \vec{\theta} , S)} - \mathcal{J}_{\vec{t}_f} \, ,
\eea
Conversely, if we add to the CS action a term like the second term appearing in the right hand side of (\ref{translationalCM}) then this contribution can be totally reabsorbed in the CS measure thanks to the shift invariance. This property is referred to as the \textbf{color} $\textbf{2k}$\textbf{-periodicity} of the CS theory with coupling constant $k$. Let us point out that the color periodicity of the Chern-Simons theory is related to the Cameron-Martin property of the action. An equivalent result is obtained for the BF action when considering shifts of the form:
\bea
\left\{
\begin{aligned}
& \vec{n}_f \to \vec{n}_f + \vec{t}_f = \vec{u}_f \\
& \vec{n}'_f \to \vec{n}'_f + \vec{t}'_f = \vec{u}'_f
\end{aligned} \right. \, ,
\eea
which infers the \textbf{color} $\textbf{k}$\textbf{-periodicity} of the BF theory with coupling constant $k$.

A dual symmetry comes from the term $2k \left\langle Q \vec{\theta}_f , \vec{n}_f \right\rangle$ in the CS action. For any $\vec{m} \in \mathbb{Z}^g$ and any $\vec{n}_f$ we have:
\bea
2k \left\langle Q \left( \frac{\vec{m}}{2k} \right) , \vec{n}_f \right\rangle = \left\langle Q \vec{m} , \vec{n}_f \right\rangle \egzz 0 \, ,
\eea
which means that the CS action is invariant with respect to the shift:
\bea
\label{freeangleshift}
\vec{\theta}_f \rightarrow \vec{\theta}_f + \frac{\vec{m}}{2k} \, .
\eea
Then, since the finite measure $d^{b_1} \vec{\theta }_f$ is obviously invariant under this shift, we conclude that the CS functional measure is invariant under (\ref{freeangleshift}). The directions $\vec{m}/2k$ generating this symmetry are called \textbf{zero modes} of the CS theory with coupling constant $k$. Similarly, the BF functional measure is invariant under
\bea
\label{2freeangleshift}
\left\{
\begin{aligned}
& \vec{\theta}_f \rightarrow \vec{\theta}_f + \frac{\vec{m}}{k} \\
& \vec{\theta}'_f \rightarrow \vec{\theta}'_f + \frac{\vec{m}'}{k}
\end{aligned} \right. \, .
\eea
The directions $\vec{m}/k$ generating this symmetry are called the \textbf{zero modes} of the BF theory with coupling constant $k$. In the next subsections we shall see how the symmetry on color and the dual one provided by zero modes dual symmetries couple in the functional integration procedure which yield partition functions and expectation values of observables.

\subsection{Partition function}

Let us start again with the BF theory. The BF partition function is defined by:
\bea
\label{BFpartition}
\begin{aligned}
\mathcal{Z}_{BF,k} = & \frac{1}{\mathcal{N}_{BF}} \int d \mu _{BF,k} ( \vec{n}_f , \vec{\theta} , S , \vec{n}'_f , \vec{\theta}' , S') \\
= & \frac{1}{\mathcal{N}_{BF}} \int D \! S  D \! S'  \sum_{\vec{n}_f,\vec{n}'_f} \sum_{\vec{\theta}_\tau , \vec{\theta}'_\tau} d^{b_1} \vec{\theta }_f d^{b_1} \vec{\theta }'_f\, e^{2i\pi S_{BF,k} ( \vec{n}_f , \vec{\theta} , S , \vec{n}'_f , \vec{\theta}' , S') }
\end{aligned} \, ,
\eea
where $\mathcal{N}_{BF}$ is a normalization factor supposed to take care of the divergences that may occur due to the infinite dimensional part in the functional integration. It expression will be chosen almost at the end of the computation.

The integration is formally performed over $H_D^1(X_L \cup_f X_R)^\star \times H_D^1(X_L \cup_f X_R)^\star$. However, the BF action is not well-defined on this configuration space. Hence, we should restrict the integration to a (maximal) subspace $\mathcal{E}$ on which the action is well-defined. Nevertheless, it turns out that only the finite integrals will provide us with quantity related to the $U(1)$ Turaev-Viro invariant.

Let us first perform the integration over the circular variables $\vec{\theta }_f$ and $\vec{\theta }'_f$. By looking at the expression (\ref{BFactiondone}) of the BF action it is quite obvious that this integration yields the factor:
\bea
\delta_{Q^\dag \vec{n}_f , \vec{0}} \, \delta_{Q^\dag \vec{n}'_f , \vec{0}} \, ,
\eea
in the remaining sums and integrals. Thanks to Property \ref{trivclassandQ} and its corollary, this factor can be replaced by the factor:
\bea
\delta_{\vec{n}_f , \vec{0}} \, \delta_{\vec{n}'_f , \vec{0}} \, ,
\eea
Let us recall that we have assumed that the trivial class is represented by $\vec{n} = \vec{0}$. Now, taking into account this factor when performing the sum of $\vec{n}_f$ and $\vec{n}'_f$ then gives:
\bea
\mathcal{Z}_{BF,k} = \frac{1}{\mathcal{N}_{BF}} \int D \! S  D \! S' \sum_{\vec{\theta}_\tau , \vec{\theta}'_\tau} e^{2i\pi S_{BF,k} ( \vec{0} , \vec{\theta}_\tau , S , \vec{0} , \vec{\theta}'_\tau , S') } \, ,
\eea
with:
\bea
S_{BF,k} ( \vec{0} , \vec{\theta}_\tau , S , \vec{0} , \vec{\theta}'_\tau , S') = & k ( S \star S') \eval{1} - k \left\langle P \vec{\theta}_\tau , Q \vec{\theta}'_\tau \right\rangle \, .
\eea
The two terms in this expression of the action are totally decoupled so that we have:
\bea
\mathcal{Z}_{BF,k} = \frac{1}{\mathcal{N}_{BF}} \left( \int D \! S  D \! S'  e^{2i\pi k ( S \star S') \eval{1} } \right) \left( \sum_{\vec{\theta}_\tau , \vec{\theta}'_\tau} e^{- 2i\pi k \left\langle P \vec{\theta}_\tau , Q \vec{\theta}'_\tau \right\rangle } \right) \, .
\eea
We now decide to set:
\bea
\mathcal{N}_{BF} = \int D \! S  D \! S'  e^{2i\pi k ( S \star S') \eval{1} }  \, ,
\eea
and we conclude that:
\bea
\mathcal{Z}_{BF,k} = \sum_{\vec{\theta}_\tau , \vec{\theta}'_\tau} e^{- 2i\pi k \left\langle P \vec{\theta}_\tau , Q \vec{\theta}'_\tau \right\rangle } \, .
\eea
Recalling that $\left\langle P \vec{\theta}_\tau , Q \vec{\theta}'_\tau \right\rangle = \Gamma (\vec{\theta}_\tau,\vec{\theta}'_\tau)$ where $\Gamma: T_1 \times T_1 \to \mathbb{R}/\mathbb{Z}$ induces the linking form of $X_L \cup_f X_R$ according to Property \ref{pontdualcokerP}, we recover the standard partition function of the $U(1)$ BF theory which is related to a Turaev-Viro invariant \cite{MT1}.

With exactly the same steps, we obtain the following expression for the CS partition function:
\bea
\mathcal{Z}_{CS,k} = \frac{1}{\mathcal{N}_{BF}} \left( \int D \! S  e^{2i\pi k ( S \star S) \eval{1} } \right) \left( \sum_{\vec{\theta}_\tau } e^{- 2i\pi k \left\langle P \vec{\theta}_\tau , Q \vec{\theta}_\tau \right\rangle } \right) \, ,
\eea
and by setting:
\bea
\mathcal{N}_{CS} = \int D \! S e^{2i\pi k ( S \star S) \eval{1} }  \, ,
\eea
 we finally get:
\bea
\mathcal{Z}_{CS,k} = \sum_{\vec{\theta}_\tau} e^{- 2i\pi k \left\langle P \vec{\theta}_\tau , Q \vec{\theta}_\tau \right\rangle } \, .
\eea
This is the standard expression for the CS partition function which is related to a Reshetikhin-Turaev invariant \cite{GT2013,GT2014,MT1}.

Let us stress out that thanks to the form we have chosen for the normalization factors $\mathcal{N}_{BF}$ and $\mathcal{N}_{CS}$ the partition functions are obtained from finite integrals. In particular, this means that beside the zero regularization assumption, we didn't need to check if the various generalized DB product are well defined.

\subsection{Observables and their expectation value}

In the CS theory, the \textbf{reduced} expectation value of an observable $\mathcal{O}_{CS}$ is:
\bea
\label{expectobservCS}
\left\langle \left\langle \mathcal{O}_{CS} \right\rangle \right\rangle  = \frac{1}{\mathcal{N}_{CS}} \int \mathcal{O}_{CS} \, d \mu _{CS,k} ( \vec{n}_f , \vec{\theta} , S) \, ,
\eea
and in the BF theory, for an observable $\mathcal{O}_{BF}$, by:
\bea
\label{expectobservBF}
\left\langle \left\langle \mathcal{O}_{BF} \right\rangle \right\rangle  = \frac{1}{\mathcal{N}_{BF}} \int \mathcal{O}_{BF} \, d \mu _{BF,k} ( \vec{n}_f , \vec{\theta} , S , \vec{n}'_f , \vec{\theta}' , S') \, .
\eea

As already noticed, the observables in the CS and BF theories are Wilson loop, that is to say $U(1)$ holonomies. Let us consider a framed link $\mathcal{L} = L_1 \cup L_2 \cup \cdots \cup L_n$ in $X_L \cup_f X_R$. We represent this link as a $1$-cycle $z_\mathcal{L}$ and denote by $\vec{N} = \vec{N}_f + \vec{N}_\tau$ its homology class, with $\vec{N}_\tau = \hat{P} \vec{\Theta}_\tau$ and $\vec{\Theta}_\tau = \vec{M} / p$. The observable defined by $\mathcal{L}$ is then:
\bea
W_{CS}(\mathcal{L},\mathcal{J}) = e^{ 2i\pi \int_\mathcal{L} \mathcal{J} } \, .
\eea
As a $1$-cycle of $X_L \cup_f X_R$, $z_\mathcal{L}$ defines a generalized DB class $\mathcal{J}_\mathcal{L}$ so that we can rewrite the above Wilson loop observable as:
\bea
W_{CS}(\mathcal{L},\mathcal{J}) = e^{ 2i\pi ( \mathcal{J} \star \mathcal{J}_\mathcal{L} ) \eval{1} } \, .
\eea
As usual, the quantity $( \mathcal{J} \star \mathcal{J}_\mathcal{L} ) \eval{1}$ is not always well-defined so we have to assume that the space of configuration we are working with only contains fields for which this generalized DB product is well-defined. In fact, we must chose our link $\mathcal{L}$ properly because the space of configuration have already been chosen for the action to be well-defined. In reality, all these subtleties will prove quite irrelevant since, as with the partition function, it is the finite part of the functional integral that will yield a link invariant.

In order to compute $\left\langle \left\langle W_{CS}(\mathcal{L},\mathcal{J}) \right\rangle \right\rangle$ we must compute
the generalized DB product $\mathcal{J} \star \mathcal{J}_\mathcal{L}$. We also want to do that in the variables $( \vec{n}_f , \vec{\theta} , S)$ used to write the CS functional measure. This means that we must decompose the DB $1$-current representing $\mathcal{J}_\mathcal{L}$ according to (\ref{decompuniversal}). Thanks to Property \ref{cyclesDBdecomp} we can write:
\bea
\mathcal{J}_\mathcal{L} = \mathcal{J}_{{\vec{N}_f}} - \mathcal{J}_{\Theta_\tau} +  \mathcal{C}_{\Theta_\tau} + \mathcal{C} = \mathcal{J}_{ ( \vec{N}_f , - \Theta_\tau , C_T)} \, ,
\eea
with $C_T = C_{\Theta_\tau} + C$. Then, by noticing that $( \mathcal{J} \star \mathcal{J}_\mathcal{L} ) \eval{1}$ is very close from the BF action, we can rely on expression (\ref{BFactiondone}) to obtain:
\bea
\begin{aligned}
( \mathcal{J} \star \mathcal{J}_\mathcal{L} ) \eval{1} = & ( \mathcal{J}_{ ( \vec{n}_f , \vec{\theta} , S)} \star \mathcal{J}_{ ( \vec{N}_f , - \Theta_\tau , C_T)} ) \eval{1} \\
= & ( \mathcal{S} \star \mathcal{C}_T ) \eval{1} + \sum_{a=1}^g ( n_f^a j_{\lambda_a^R} \wedge C_{T,R}^\circ  + N_f^a j_{\lambda_a^R} \wedge S_R^\circ ) \eval{1_R} + \\
& + \left\langle Q \vec{\theta} , \vec{N}_f \right\rangle - \left\langle Q \vec{\Theta}_\tau , \vec{n}_f \right\rangle + \left\langle P \vec{\theta}_\tau , Q \vec{\Theta}_\tau \right\rangle \\
= & ( \mathcal{S} \star \mathcal{C}_T ) \eval{1} + \sum_{a=1}^g ( N_f^a j_{\lambda_a^R} \wedge S_R^\circ ) \eval{1_R} + \left\langle Q \vec{\theta} , \vec{N}_f \right\rangle + \left\langle P \vec{\theta}_\tau , Q \vec{\Theta}_\tau \right\rangle \, ,
\end{aligned}
\eea
where $\mathcal{C} = ( C_L^\circ , C_R^\circ , \overline{C_R^\Sigma} )$ and $\mathcal{C}_T = \mathcal{C}_{\Theta_\tau} + \mathcal{C}$. We used the fact that:
\bea
\sum_{a=1}^g ( n_f^a j_{\lambda_a^R} \wedge C_{T,R}^\circ ) \eval{1_R} = \sum_{a=1}^g ( n_f^a j_{\lambda_a^R} \wedge (C_{\vec{\Theta}_\tau} + C)_R^\circ ) \eval{1_R} \egzz \left\langle Q \vec{\Theta}_\tau , \vec{n}_f \right\rangle \, ,
\eea
in order to compensate the term $- \left\langle Q \vec{\Theta}_\tau , \vec{n}_f \right\rangle$ in the expression of $( \mathcal{J} \star \mathcal{J}_\mathcal{L} ) \eval{1}$. Note that $\sum_{a=1}^g \left( n_f^a j_{\lambda_a^R} \wedge C_R^\circ \right) \eval{1_R} \egzz 0$ because $C_R^\circ$ is a $2$-chain. 

Now, we add the above result to the expression (\ref{CSactiondone}) of the CS action thus obtaining:
\bea
\label{expobservCS}
\begin{gathered}
 k ( \mathcal{J}_{ ( \vec{n}_f , \vec{\theta} , S)} \star \mathcal{J}_{ ( \vec{n}_f , \vec{\theta} , S)} ) \eval{1} + ( \mathcal{J}_{ ( \vec{n}_f , \vec{\theta} , S)} \star \mathcal{J}_{ ( \vec{N}_f , - \vec{\Theta}_\tau , C_T)} ) \eval{1}  = \qquad \qquad \qquad \qquad \qquad \\
\qquad \qquad k ( \mathcal{S} \star \mathcal{S}) \eval{1} + 2k \left\langle Q \vec{\theta}_\tau , \vec{n}_f \right\rangle - k \left\langle P \vec{\theta}_\tau , Q \vec{\theta}_\tau \right\rangle + \\ 
\qquad \qquad  + 2k \sum_{a=1}^g  \left( n_f^a j_{\lambda_a^R} \wedge S_R^\circ \right) \eval{1_R} + 2k \left\langle Q \vec{\theta}_f , \vec{n}_f \right\rangle + \\
\qquad \qquad  + ( \mathcal{S} \star \mathcal{C}_T ) \eval{1} + \left\langle Q \vec{\theta}_\tau , \vec{N}_f \right\rangle  + \left\langle P \vec{\theta}_\tau , Q \vec{\Theta}_\tau \right\rangle + \ \\
\qquad \qquad + \sum_{a=1}^g ( N_f^a j_{\lambda_a^R} \wedge S_R^\circ ) \eval{1_R} + \left\langle Q \vec{\theta}_f , \vec{N}_f \right\rangle \, .
\end{gathered}
\eea
The terms containing $Q \vec{\theta}_f$ can be gathered into:
\bea
\left\langle Q \vec{\theta}_f , 2k \vec{n}_f + \vec{N}_f \right\rangle \, .
\eea
Then, as in the computation of the partition function, we first perform the integration over the circular variables $\vec{\theta}_f$ which yields, thanks to Property \ref{trivclassandQ} and its corollary, the constraint $2k \vec{n}_f + \vec{N}_f = \vec{0}$. Hence, instead of (\ref{expobservCS}) we must consider:
\bea
\begin{gathered}
 k ( \mathcal{J}_{ ( \vec{n}_f , \vec{\theta} , S)} \star \mathcal{J}_{ ( \vec{n}_f , \vec{\theta} , S)} ) \eval{1} +  ( \mathcal{J}_{ ( \vec{n}_f , \vec{\theta} , \mathfrak{s})} \star \mathcal{J}_{ ( \vec{N}_f , - \frac{\vec{M}}{p} , C_T)} ) \eval{1} |_{2k \vec{n}_f + \vec{N}_f = \vec{0}} = \qquad \qquad \qquad \\
\qquad \qquad \qquad \qquad k ( \mathcal{S} \star \mathcal{S}) \eval{1} + ( \mathcal{S} \star \mathcal{C}_T ) \eval{1} - k \left\langle P \vec{\theta}_\tau , Q \vec{\theta}_\tau \right\rangle + \left\langle P \vec{\theta}_\tau , Q \vec{\Theta}_\tau \right\rangle \, ,
\end{gathered}
\eea
and limit the sum over the free variables $\vec{n}_f$ to one contribution. Hence, we obtain the following expression of the involved expectation value: 
\bea
\left\langle \left\langle W_{CS}(\mathcal{L},\mathcal{J}) \right\rangle \right\rangle  = \frac{\delta_{2k \vec{n}_f + \vec{N}_f}}{\mathcal{N}_{CS}} \int D \mathcal{S} \sum_{\vec{\theta}_\tau} e^{2i\pi \left(k \mathcal{S} \star \mathcal{S} + \mathcal{S} \star \mathcal{C}_T \right) \eval{1} } \, e^{ - 2i\pi \left(k \left\langle P \vec{\theta}_\tau , Q \vec{\theta}_\tau \right\rangle - \left\langle P \vec{\theta}_\tau , Q \vec{\Theta}_\tau \right\rangle \right)} \, .
\eea
The integrals once more decouple thus yielding:
\bea
\begin{aligned}
\left\langle \left\langle W_{CS}(\mathcal{L},\mathcal{J}) \right\rangle \right\rangle = \frac{\delta_{2k \vec{n}_f + \vec{N}_f}}{\mathcal{N}_{CS}} & \left( \int  D \mathcal{S} \, e^{2i\pi \left(k \mathcal{S} \star \mathcal{S} + \mathcal{S} \star \mathcal{C}_T \right) \eval{1} } \right) \times \\
& \qquad \qquad \times \left( \sum_{\vec{\theta}_\tau} e^{ - 2i\pi (k \left\langle P \vec{\theta}_\tau , Q \vec{\theta}_\tau \right\rangle - \left\langle P \vec{\theta}_\tau , Q \vec{\Theta}_\tau \right\rangle )} \right) \, .
\end{aligned}
\eea
The last step is to deal with the argument of the first exponential accoridng to:
\bea
\left(k \mathcal{S} \star \mathcal{S} + \mathcal{S} \star \mathcal{C}_T \right) \eval{1}  = k \left( \left( \mathcal{S} + \frac{\mathcal{C}_T}{2k} \right) \star  \left( \mathcal{S} + \frac{\mathcal{C}_T}{2k} \right) \right) \eval{1} - k \left( \frac{\mathcal{C}_T}{2k} \star \frac{\mathcal{C}_T}{2k} \right) \eval{1} \, .
\eea
Since by construction $\mathcal{C}_T$ is a DB $1$-current of type $\mathcal{S}$ the division by $2k$ is allowed and we can make the following change of variables:
\bea
\mathcal{U} = \mathcal{S} + \frac{\mathcal{C}_T}{2k} \, .
\eea
By referring to the discussion made right after Property \ref{decompsingularDBclasses} this redefinition of the "infinite dimensional" variables involve a redefinition of the free angular variables $\vec{\theta}_f$. However, we already integrated over these variables so a change in the angular variable is irrelevant at this stage. If we assume that the formal measure $D \mathcal{s}$ is invariant under the above shift, then we obtain:
\bea
\begin{aligned}
\left\langle \left\langle W_{CS}(\mathcal{L},\mathcal{J}) \right\rangle \right\rangle =  \frac{\delta_{\vec{N}_f , \vec{0}}^{[2k]}}{\mathcal{N}_{CS}} & \left( \int D \mathcal{U} \, e^{2i\pi k ( \mathcal{U} \star \mathcal{U} ) \eval{1} } \right) \times \\ 
\times & e^{- 2i\pi k (\frac{\mathcal{C}_T}{2k} \star \frac{\mathcal{C}_T}{2k}) \eval{1} } \left( \sum_{\vec{\theta}_\tau} e^{ - 2i\pi (k \left\langle P \vec{\theta}_\tau , Q \vec{\theta}_\tau \right\rangle - \left\langle P \vec{\theta}_\tau , Q \vec{\Theta}_\tau \right\rangle )} \right) \, .
\end{aligned}
\eea
Taking into account the expression of normalization factor $\mathcal{N}_{CS}$ we get:
\bea
\label{expectvalueresult}
\begin{aligned}
\left\langle \left\langle W_{CS}(\mathcal{L},\mathcal{J}) \right\rangle \right\rangle = \delta_{\vec{N}_f , \vec{0}}^{[2k]} \; e^{- 2i\pi k (\frac{\mathcal{C}_T}{2k} \star \frac{\mathcal{C}_T}{2k}) \eval{1} } \left( \sum_{\vec{\theta}_\tau} e^{ - 2i\pi (k \left\langle P \vec{\theta}_\tau , Q \vec{\theta}_\tau \right\rangle - \left\langle P \vec{\theta}_\tau , Q \vec{\Theta}_\tau \right\rangle )} \right) \, .
\end{aligned}
\eea
From what we saw in the previous section, we have:
\bea
k (\frac{\mathcal{C}_T}{2k} \star \frac{\mathcal{C}_T}{2k}) \eval{1} \egzz \frac{1}{4k p^2} \textit{lk}(\mathfrak{Z}^p , \breve{\mathfrak{Z}}^p) \, ,
\eea
with $\mathfrak{Z}^p$ being the $p$-fold of $\mathfrak{Z}$ and $\breve{\mathfrak{Z}}^p$ a framing of this $p$-fold, the cycle $\mathfrak{Z}$ being the generalized curvature of $\mathcal{C}_T = \mathcal{C}_{\vec{\Theta}_\tau} + \mathcal{C}$, where $\vec{\Theta}_\tau = \vec{M} / p$. If we inject this last relation in (\ref{expectvalueresult}), the resulting expression is:
\bea
\label{expectvaluefinal}
\begin{aligned}
\left\langle \left\langle W_{CS}(\mathcal{L},\mathcal{J}) \right\rangle \right\rangle = \delta_{\vec{N}_f , \vec{0}}^{[2k]} \; e^{- \frac{2i\pi}{4k p^2} \textit{lk}(\mathfrak{Z}^p , \breve{\mathfrak{Z}}^p) } \left( \sum_{\vec{\theta}_\tau} e^{ - 2i\pi (k \left\langle P \vec{\theta}_\tau , Q \vec{\theta}_\tau \right\rangle - \left\langle P \vec{\theta}_\tau , Q \vec{\Theta}_\tau \right\rangle )} \right)
\end{aligned} \, ,
\eea
in full agreement with formula (90) of \cite{GT2014}. Up to the factor $-2i\pi$, we can write the argument of the first exponential in the above expression as:
\bea
\label{pertubativepart}
\frac{1}{4k} \left( \mathcal{C} \star  \mathcal{C} \right) \eval{1} + \frac{1}{2kp} \sum_{a=1}^{g} (P \vec{M})^a \left( j_{\lambda_a^R} \wedge j_{c_R^\circ} \right) \eval{1_R} + \frac{1}{4kp^2} \left\langle P \vec{M}_ , Q \vec{M} \right\rangle \, .
\eea
Due to the presence of the factor $1/k$, which is the inverse of the coupling constant $k$, this contributions is referred to as the perturbative component of the reduced expectation value \cite{GT2014}.

In the case of BF observables, we consider:
\bea
W_{BF}(\mathcal{L}_1,\mathcal{L}_2,\mathcal{J},\mathcal{J}') = e^{2 i \pi \oint_{\mathcal{L}_1} \mathcal{J}} \cdot e^{2 i \pi \oint_{\mathcal{L}_2} \mathcal{J}'} \, ,
\eea
for two links $\mathcal{L}_1$ and $\mathcal{L}_2$. We then consider the DB $1$-currents:
\bea
\left\{
\begin{aligned}
	\mathcal{J}_{\mathcal{L}_1} = \mathcal{J}_{{\vec{N}_f^1}} - \mathcal{J}_{\Theta_\tau^1} +  \mathcal{C}_{\Theta_\tau^1} + \mathcal{C}^1 = \mathcal{J}_{ ( \vec{N}_f^1 , - \Theta_\tau^1 , C_T^1)} \\
	\mathcal{J}_{\mathcal{L}_2} = \mathcal{J}_{{\vec{N}_f^2}} - \mathcal{J}_{\Theta_\tau^2} +  \mathcal{C}_{\Theta_\tau^2} + \mathcal{C}^2 = \mathcal{J}_{ ( \vec{N}_f^2 , - \Theta_\tau^2 , C_T^2)}
\end{aligned} \right. \, ,
\eea
associated with $\mathcal{L}_1$ and $\mathcal{L}_2$, respectively. By following the same procedure as in the Chern-Simons case, and with obvious notations, we end with the reduced expectation value:
\bea
\begin{aligned}
\left\langle \left\langle W_{BF}(\mathcal{L}_1,\mathcal{L}_2,\mathcal{J},\mathcal{J}') \right\rangle \right\rangle = & \delta_{\vec{N}_f^1 , \vec{0}}^{[k]} \; \delta_{\vec{N}_f^2 , \vec{0}}^{[k]} \; e^{- \frac{2i\pi}{k p_1 p_2} \textit{lk}(\mathfrak{Z}_1^{p_1} , \mathfrak{Z}_2^{p_2}) } \\ 
& \left( \sum_{\vec{\theta}_\tau , \vec{\theta}'_\tau} e^{ - 2i\pi (k \left\langle P \vec{\theta}_\tau , Q \vec{\theta}'_\tau \right\rangle - \left\langle P \vec{\theta}_\tau , Q \vec{\Theta}_\tau^1 \right\rangle - \left\langle P \vec{\theta}'_\tau , Q \vec{\Theta}_\tau^2 \right\rangle)} \right) \, .
\end{aligned}
\eea
This expression coincides with formula (2.53) of \cite{MT2}. Of course, if the two links are equal then the linking number in the first exponential is actually a self-linking and thus is defined with the help of a framing as in the Chern-Simons case. We leave it to the reader to find the equivalent of the perturbative component (\ref{pertubativepart}) in the BF case.

\section{Conclusion}

In this second article dedicated to the use of a Heegaard splitting in the determination of the partition functions and expectation values in $U(1)$ CS and BF theories we saw how the use of DB $1$-currents yield to results which were obtained by different approaches. However, although the results are already known, the use of a Heegaard splitting allows to explicitly see the role played by the Riemann surface on which the spitting is based. All the degree of freedom which enter in the final results actually come from Property \ref{pontdualcokerP} and are referring to the gluing diffeomorphism $f:\Sigma_L \to \Sigma_R$ which defines the Heegaard splitting $X_L \cup_f X_R$ thought the endomorphism $P:\mathbb{Z}^g \to \mathbb{Z}^g$.

Another virtue of the Heegaard splitting approach is that it shows how the exact sequence based on (generalized) curvatures can be used in the context of CS and BF theories. Usually \cite{GT2013,MT1}, it is the exact sequence based on the second cohomology of $M$ which is used. It turns out that a Heegaard splitting also provides a nice realization of the so-called cycle map \cite{BGST05} which states that an $1$-cycle $z$ canonically defines a singular DB class $\eta_z$ thanks to which integration of a smooth DB class $A$ along $z$ takes the form $(\mathcal{J}_A \star \eta_z)\eval{1}$.

The so-called standard approach can be extended to $U(1)$ CS and BF theories on a smooth closed $(4n+3)$-dimensional manifold which also yield invariants \cite{MT1}. It is therefore natural to wonder whether there exists some equivalent of the Heegaard splitting approach for such higher dimensional manifolds. Unfortunately, a smooth closed $(4n+3)$-dimensional manifold ($n > 0$) cannot always be decomposed according to $V \cup_f W$ with $V$ and $W$ diffeomorphic manifolds and $f:\partial V \to \partial W$ a diffeomorphism. However, as the $U(1)$ CS and BF invariants discussed in this article all rely on $T_{2n+1}(M)$ and the linking form $\Gamma_M$ we could consider the following trick: first, we look for $V$, $W$ and $f$ as above and in such a way that $T_{2n+1}(M') = T_{2n+1}(M)$ and $\Gamma_M' = \Gamma_M$, where $M' = V \cup_f W$; then, we apply the Heegaard splitting approach to $M'$. Another interesting point would be to see if it is possible to extend the Heegaard construction to a decomposition $V \cup_f W$ with the only constraint that $f:\partial V \to \partial W$ is a diffeomorphism. More generally, as any smooth manifold admits a handle decomposition, we investigate how to adapt the Heegaard splitting approach to such a decomposition. A fourth and last point of interest concern the Kirk and Klassen construction in which a BF type action interpolates between non-abelian flat CS actions \cite{KK93}. We can wonder what remains of this in the $U(1)$ context where the space of equivalence classes of $U(1)$-connections is not connected.

The use of a Heegaard splitting can also be seen as a first step in order to understand how CS and BF theories could be considered as QFT on a closed $3$-manifold, whereas QFT in its standard approach, which is perturbative by essence, only deal with $\mathbb{R}^3$. These topological models are of particular interest in such a search since we know the results concerning their partition functions and the expectation values of some of their observables. Of course, the identification of the relevant degrees of freedom entering the final expression of the partition functions and expectation values drives us to the questions of these degrees of freedom in the non-abelian case. In that case too all should come from the gluing diffeomorphism $f$ which defines the Heegaard splitting. Unfortunately, these degrees of freedom still seem mysterious. Nevertheless, the abelian degrees of freedom should be somewhere among them as they are for instance in the case of lens spaces \cite{G93}.

Let us conclude this article with a questioning. It is pretty amazing that $U(1)$ CS and BF theories are almost absent from the usual lectures on QFT in $\mathbb{R}^3$ (or even $\mathbb{R}^{4n+3}$.) Indeed, the propagators of these theories have nice geometrical interpretations for a large class of gauge fixing procedure \cite{GMPT18}, while partition functions and expectation values of holonomies provide manifold invariants. As to the computation of these quantities they also rely on geometry \cite{GPT13,GPT15}. It is not usual at all for propagators, partition functions and expectation values to have such nice interpretations in QFT.

\vspace{0.5cm}

\textbf{Acknowledgments:} The author would like to thank Eric Pilon for fruitful discussions concerning currents and distributions.

\vfill\eject


\begin{thebibliography}{T19}


\bibitem{T19}
F. Thuillier, {\it 3D Topological Models and Heegaard Splitting I: Partition Function}, J. Math. Phys. 60 (2019), 32

\bibitem{GT2013}
E. Guadagnini and F. Thuillier, {\it Three-manifold invariant from functional integration}, J. Math. Phys. {\bf 54}, 082302 (2013).

\bibitem{GT2014}
E. Guadagnini and F. Thuillier, {\it Path-integral invariants in abelian Chern-Simons theory}, Nucl. Phys. B {\bf 882}, 450--484 (2014).

\bibitem{MT1}
P. Mathieu and F. Thuillier, {\it Abelian BF theory and Turaev-Viro invariant}, J. Math. Phy. {\bf 57}, 022306 (2016); doi: 10.1063/1.4942046.

\bibitem{MT2}
P. Mathieu and F. Thuillier, {\it A reciprocity formula from abelian BF and Turaev-Viro theories}, published in ``Eulogy for Raymond", Nucl. Phys. B {\bf 912},  327--353 (2016).

\bibitem{RT91}
N. Y. Reshetikhin and V. G. Turaev, {\it Invariants of 3-manifolds via link polynomials and quantum groups}, Invent. Math. {\bf 103}, 547--597 (1991).

\bibitem{TV}
V. G. Turaev, O. Yu. Viro, {\it State Sum Invariants of 3-Manifolds and Quantum 6j-symbols},
Topology {\bf 31}, 865--902 (1992).

\bibitem{MOO92}
H. Murakami, T. Ohtsuki and M. Okada, {\it Invariants of three-manifolds derived from linking matrices and framed links}, Osaka J. Math. {\bf 29} (1992), 545--572.

\bibitem{BGST05}
M. Bauer, G. Girardi, R. Stora and F. Thuillier, {\it A class of topological actions}, J. High Energy Phys. 2005 (2005), no. 8, 027, 35 pages.

\bibitem{dR55}
G. de Rham, {\it Vari\'et\'es Diff\'erentiables, Formes, Courants, Formes Harmoniques}, Hermann, Paris, 1955.

\bibitem{CFH18}
Conway A., Friedl S. and Herrmann G., {\it Linking forms revisited}, arXiv:1708.03754 v3 Feb. 2018.

\bibitem{HLZ}
Harvey, R., Lawson, B. and Zweck, J., {\it The de Rham--Federer theory of differential characters and character duality}, Amer. J. Math. {\bf 125} (2003), 791--847.

\bibitem{D72}
A. Dold, {\it Lectures on algebraic topology}, in Classics in Mathematics (Springer, 1972).

\bibitem{HL01}
F. Reese Harvey and H. Blaine Lawson, {\it Lefschetz-Pontrjagin Duality for Differential Characters}, Anais da Academia Brasileira de Ciencias, vol 73, no. 2(2001), 145-160.

\bibitem{CT80}
Seifert H. and Threlfall W., {\it Seifert and Threlfall: a textbook of topology}, volume 89 of Pure and
Applied Mathematics. Academic Press, Inc. [Harcourt Brace Jovanovich, Publishers], New York-London, 1980.
Translated from the German edition of 1934 by Michael A. Goldman.

\bibitem{KK93}
P. Kirk and E. Klassen, {\it Chern-Simons invariants of 3-manifolds decomposed along tori and the circle bundle over the representation space of $T^2$}, Communications in Mathematical Physics volume 153, 521-557, 1993.

\bibitem{G93}
E. Guadagnini, {\it The link invariants of the Chern-Simons field theory}. New developments in topological quantum field theory, de Gruyter Expositions in Mathematics, Vol. 10, Walter de Gruyter and Co., Berlin, 1993.

\bibitem{GMPT18}
L. Gallot, P. Mathieu, É. Pilon and F. Thuillier, {\it Geometric aspects of interpolating gauge-fixing in Chern-Simons theory}, Modern Physics Letters A, 2018.

\bibitem{GPT13}
L. Gallot, E. Pilon and F. Thuillier, {\it Higher dimensional abelian Chern-Simons theories and their link invariants}, Journal of Mathematical Physics, 2013.

\bibitem{GPT15}
L. Gallot, E. Pilon and F. Thuillier, {\it Topological gauge fixing II: A homotopy formulation}, Modern Physics Letters A, 2015.



\end{thebibliography}
\end{document}